\documentclass[aps,prx,twocolumn]{revtex4-2}
\usepackage{graphicx}
\usepackage{bm,amsmath}
\usepackage{xcolor}
\usepackage[percent]{overpic}
\usepackage{tikz}

\usepackage{array}

\begin{document}

\title{Flexibility Controls Active-Filament Transport in Crowded Landscapes}

\author{Qingyi Di}
\author{Mohammad Fazelzadeh}
\author{Sara Jabbari-Farouji}
\affiliation{Institute for Theoretical Physics, University of Amsterdam,
Science Park 904, 1098XH Amsterdam, The Netherlands}

\begin{abstract}

Active filaments, ranging from motor-driven biopolymers to elongated bacteria and worms, are paradigmatic examples of deformable active matter. How filament flexibility interacts with environmental heterogeneity to control their transport in crowded environments, however, remains poorly understood. Here, we perform large-scale Brownian dynamics simulations of tangentially driven active polymers moving through \emph{ordered} and \emph{disordered} obstacle arrays to map the long-time diffusion as a function of obstacle density and filament flexibility. We find that flexibility can either enhance or hinder transport depending on the structure of the medium. In disordered environments, transport is optimized at intermediate filament flexibility, whereas both highly flexible and semiflexible filaments diffuse more slowly. In contrast, dense ordered arrays enhance the mobility of semiflexible filaments by promoting directed motion along periodic channels. We identify three distinct transport regimes: (i) tortuosity-controlled diffusion of highly flexible filaments, characterized by trapping-and-hopping dynamics; (ii) confinement-assisted transport of moderately flexible filaments, which enhances diffusion in dense media; and (iii) persistence-controlled transport of semiflexible filaments, which facilitates diffusion in dense ordered media, but suppresses it in disordered media. Combining theory and simulations, we show that long-time diffusion is governed by confinement-induced changes in filament conformation and reorientation dynamics. Our work uncovers general transport principles for deformable active agents in heterogeneous environments and provides a predictive framework for active-filament navigation in complex porous landscapes.

\end{abstract}

\maketitle

\section{Introduction}

From living cells to engineered robotic worms, \textit{active filaments} are ubiquitous across scales. Biological examples of these elongated, self-propelled entities include motor-driven cytoskeletal polymers such as actin filaments and microtubules~\cite{wen2011polymer}, filamentous bacteria such as \textit{Thioploca} spp.~\cite{teske2006genera} and \textit{Chloroflexus aggregans}~\cite{fukushima2016gliding}, and motile multicellular organisms such as \textit{Caenorhabditis elegans}~\cite{juarez2010motility} and \textit{Tubifex tubifex}~\cite{sinaasappel2025locomotion}. Synthetic counterparts include actuated colloidal~\cite{kuei2017strings,wei2025autonomous} and robotic chains~\cite{nakagaki2016chainform}.

Across these biological and technological systems, active filaments often move through crowded and structurally heterogeneous environments, where transport is shaped by interactions with surrounding confinement~\cite{martinez2023active}. For example, worms and nematodes burrow through soils and sediments with widely varying particle densities and pore geometries~\cite{kudrolli2019burrowing,sreepadmanabh2025physical}. Motile parasites provide an example at cellular scales: \textit{Trypanosoma brucei} adjusts its morphology and swimming mode in the vertebrate bloodstream~\cite{heddergott2012trypanosome}, and \textit{Plasmodium} sporozoites navigate microstructured tissue by adapting their curvature to surrounding obstacles~\cite{muthinja2017microstructured}. At sub-cellular level, motor-driven cytoskeletal filaments move through the viscoelastic cytoplasm, where their dynamics is constrained by cross-linked networks of actin filaments, microtubules, and intermediate filaments~\cite{cabrales2017multivalent}.

This prevalence of transport through complex media has motivated extensive studies of active agents in obstacle-filled environments, showing that confinement and pore geometry can strongly alter motility and dispersal~\cite{bechinger2016active,brown2016swimming,zeitz2017active,bertrand2018optimized,alonso2019transport,wu2021wechanisms,frangipane2019invariance,makarchuk2019enhanced,bhattacharjee2019bacterial,Dehkharghani2023self,mattingly2025coarse,zhang2026bacterial,mokhtari2019dynamics,wu2022facilitated,theeyancheri2023active,yan2023conformation,fazelzadeh2023active,sinaasappel2025locomotion,majmudar2012experiments,licata2016diffusion,volpe2017topography,kurzthaler2021geometric,van2022role,pietrangeli2025universal,sreepadmanabh2025physical,wurthner2026geometry}. 
Many of these studies, however, have focused on isotropic active particles~\cite{brown2016swimming,zeitz2017active,bertrand2018optimized,alonso2019transport,wu2021wechanisms} or rigid microbial swimmers~\cite{frangipane2019invariance,makarchuk2019enhanced,bhattacharjee2019bacterial,Dehkharghani2023self,mattingly2025coarse,zhang2026bacterial}.
Active filaments differ fundamentally from such agents because their internal conformational degrees of freedom couple shape, flexibility, and motility~\cite{winkler2020physics}. Their transport is therefore controlled not only by self-propulsion, but also by how filament deformations interact with the geometry of the surrounding medium.
The impact of this interplay has been observed both in experiments on locomotion through porous media~\cite{majmudar2012experiments,sreepadmanabh2025physical,sinaasappel2025locomotion} and in simulations of active filaments in crowded environments~\cite{mokhtari2019dynamics,wu2022facilitated,theeyancheri2023active,yan2023conformation,fazelzadeh2023active}. However, a systematic framework predicting how filament flexibility maps to transport across different geometrical features of porous media remains lacking.

To address this gap, we focus on tangentially driven active polymers, a minimal model that captures key features of motor-driven biofilaments in motility assays~\cite{hiratsuka2001controlling}, \textit{T. tubifex} worms crawling through obstacle arrays~\cite{sinaasappel2025locomotion}, and active colloidal chains~\cite{wei2025autonomous}. In free space, the conformations and dynamics of this model are governed by the interplay between flexibility, activity, and inertia~\cite{isele2015self,peterson2020statistical,philipps2022tangentially,fazelzadeh2023effects,karan2024inertia}.
In porous media, these dynamics are further constrained by the density and spatial arrangement of obstacles~\cite{mokhtari2019dynamics,wu2022facilitated,fazelzadeh2023active,sinaasappel2025locomotion}. Previous studies, however, have typically varied either obstacle arrangement or packing fraction over a limited range of filament flexibility~\cite{sinaasappel2025locomotion}, or degree of flexibility within a fixed porous configuration~\cite{mokhtari2019dynamics,fazelzadeh2023active}. This leaves the joint effects of flexibility, obstacle density, and pore-space geometry unresolved; a gap emphasized by recent experiments on worms where increasing obstacle density enhanced diffusion in disordered arrays but suppressed it in ordered media~\cite{sinaasappel2025locomotion}. These counter-intuitive observations raise a central question: how does the interplay between filament flexibility and porous-medium geometry shape transport, and under what conditions is the diffusion optimized?

\begin{figure}
\centering
\begin{tikzpicture}
\node[anchor=south west, inner sep=0] (img1)
    {\includegraphics[width=0.48\linewidth]{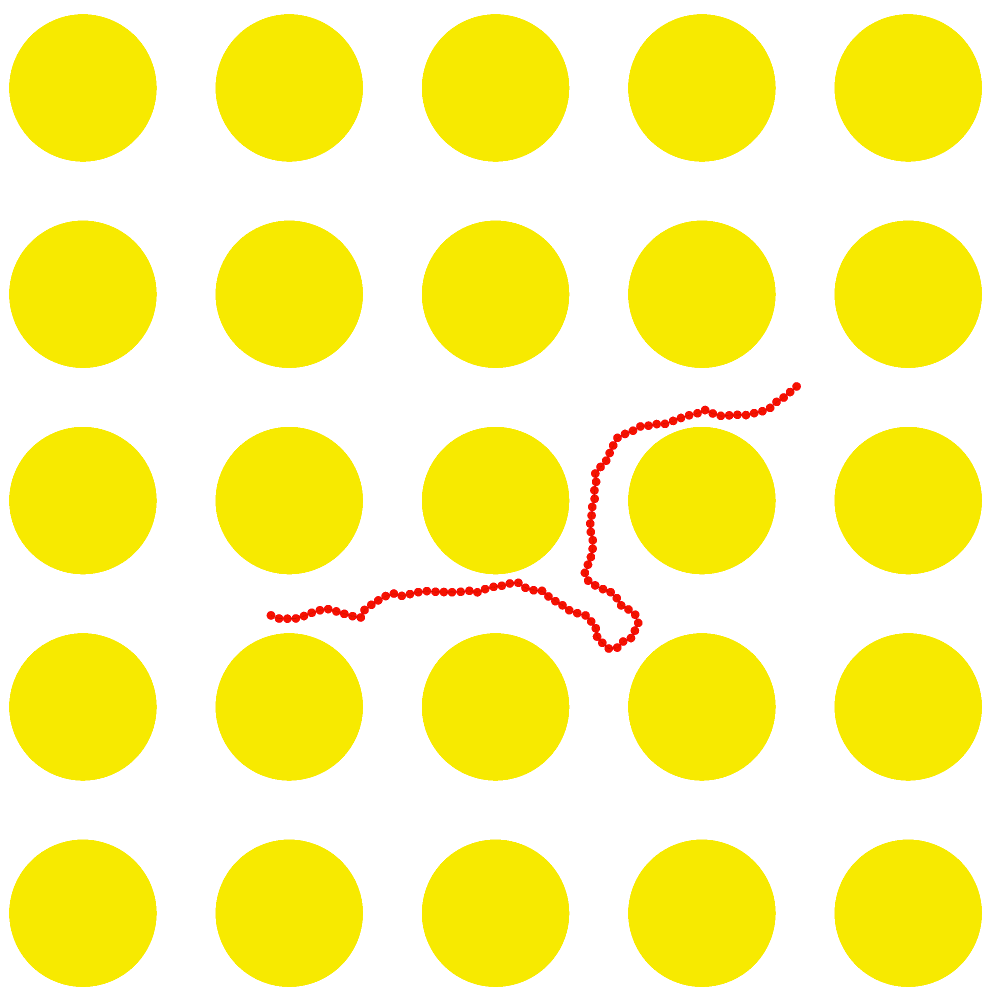}};
\begin{scope}[x={(img1.south east)}, y={(img1.north west)}]
    \node[anchor=north west] at (0.02,0.98) {\textbf{(a)}};
    \draw[->, thick, draw=brown] (0.3,0.35) -- (0.4,0.36);
    \draw[->, thick, draw=brown] (0.51,0.38) -- (0.59,0.32);
    \draw[->, thick, draw=brown] (0.62,0.43) -- (0.63,0.53);
    \draw[->, thick, draw=brown] (0.7,0.555) -- (0.8,0.575);
    \draw[dashed, thick, draw=gray!60] (0.16,0.0) -- (0.16,1.0);
    \draw[dashed, thick, draw=gray!60] (0.215,0.0) -- (0.215,1.0);
    \draw[dashed, thick, draw=gray!60] (0.0,0.165) -- (1.0,0.165);
    \draw[dashed, thick, draw=gray!60] (0.0,0.22) -- (1.0,0.22);
    \draw[<->, thick] (0.805,0.16) -- (0.805,0.22);
    \draw[<->, thick] (0.345,0.755) -- (0.446,0.856);
    \draw[<->, thick] (0.695,0.905) -- (0.915,0.905);
    \node [text=brown] at (0.5,0.3) {$\mathbf{f}^{\mathrm{a}}$};
    \node at (0.805,0.11) {$\xi_\text{sq}$};
    \node at (0.3,0.84) {$d_{\text{cage}}$};
    \node at (0.805,0.86) {$a$};
\end{scope}
\end{tikzpicture}
\hfill
\begin{tikzpicture}
\node[anchor=south west, inner sep=0] (img2)
    {\includegraphics[width=0.48\linewidth]{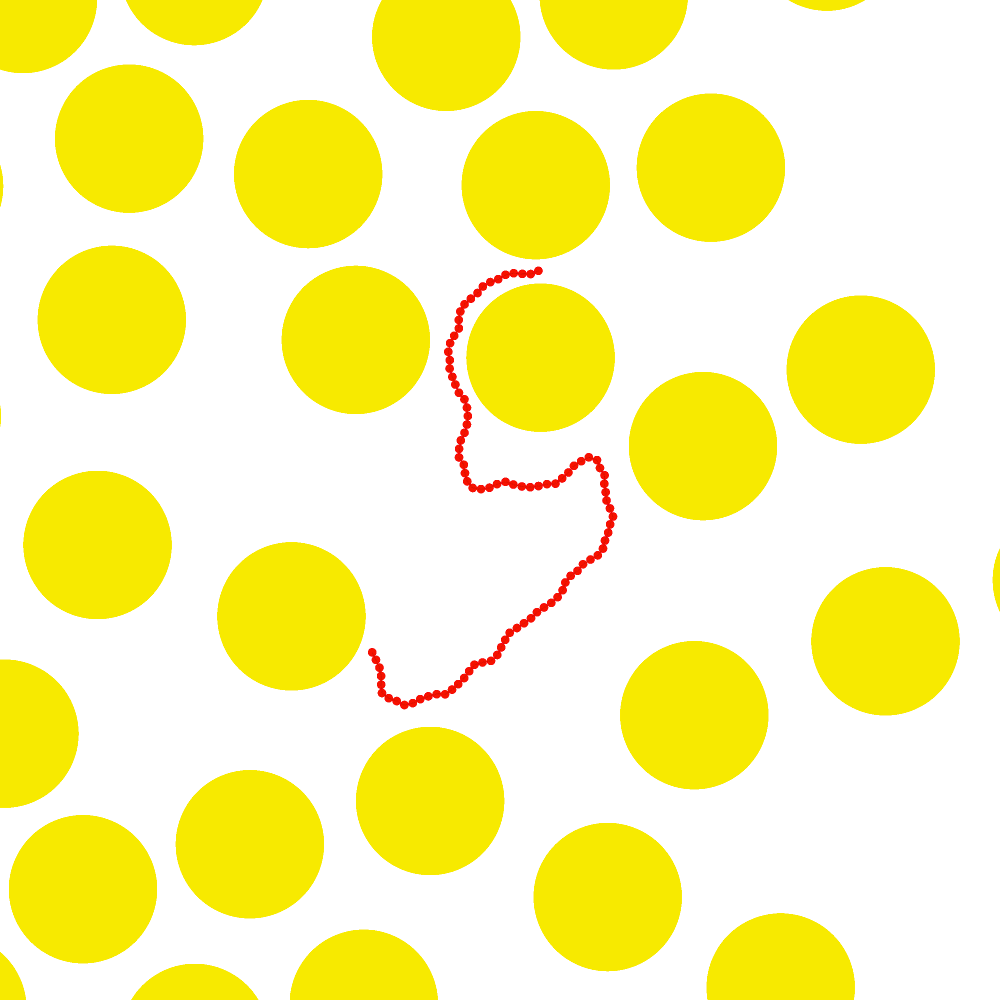}};
\begin{scope}[x={(img2.south east)}, y={(img2.north west)}]    
    \node[anchor=north west] at (0.02,0.98) {\textbf{(b)}};
    \draw[->, thick] (0.08,0.105) -- (0.14,0.16);
    \node[below] at (0.12,0.1) {$R_{\text{o}}$};
    \draw[->, thick,draw=teal!60] (0.37,0.35) -- (0.54,0.73);
    \node [text=teal] at (0.38,0.55) {$\mathbf{R}_{\mathrm{e}}$};
    \draw[<->, thick, draw=black!60] (0.15,0.5) -- (0.26,0.79);
    \draw[<->, thick, draw=black!60] (0.83,0.41) -- (0.75,0.5);
    \draw[<->, thick, draw=black!60] (0.51,0.935) -- (0.56,0.96);
    \node at (0.15,0.65) {$\ell_{\text{c,1}}$};
    \node at (0.75,0.42) {$\ell_{\text{c,2}}$};
    \node at (0.55,0.88) {$\ell_{\text{c,3}}$};
\end{scope}
\end{tikzpicture}
\caption{Schematic of a tangentially driven active polymer in obstacle arrays. (a) Ordered square-lattice arrangement, illustrating the active force \(\mathbf{f}^\text{a}\) direction, lattice spacing \(a\), pore size \(d_{\text{cage}}\), and channel width \(\xi_\text{sq}\). (b) Disordered obstacle arrangement, showing the polymer end-to-end vector \(\mathbf{R}_{\text{e}}\), obstacle radius \(R_{\text{o}}\), and representative chord lengths \(\ell_{\text{c,}i}\).}
\label{fig:schematics}
\end{figure}
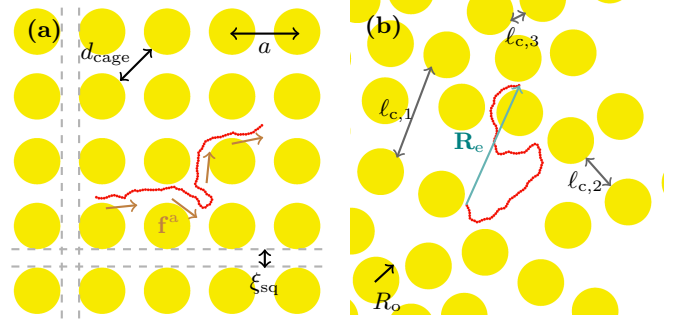

To answer this question, we perform overdamped Langevin dynamics simulations of tangentially driven bead-and-spring polymers moving through two-dimensional arrays of fixed circular obstacles. We systematically vary the polymer bending stiffness, obstacle packing fraction, and obstacle arrangement, considering both random and square-lattice geometries, see Fig.~\ref{fig:schematics}. The obstacles are approximately one order of magnitude larger than the polymer width. The polymers are phantom chains that can cross themselves, representing quasi-two-dimensional projections of systems in which three-dimensional filaments are loosely confined to a substrate plane, such as aquatic worms crawling under gravity~\cite{sinaasappel2025locomotion}. We focus on the low-activity regime, since previous work has shown that strongly driven filaments can largely overcome obstacle confinement~\cite{mokhtari2019dynamics,fazelzadeh2023active}.

Our results reveal that active-filament transport is governed by a subtle interplay between flexibility and pore geometry. Rather than a single optimal strategy, we find distinct transport regimes in ordered and disordered media. Highly flexible filaments undergo tortuosity-controlled trapping-and-hopping dynamics, which slows transport as crowding increases. Moderately flexible filaments benefit from confinement-induced shape fluctuations and persistent reorientation, leading to enhanced diffusion in dense media. Semiflexible filaments behave differently in ordered and disordered media: in dense ordered arrays, periodic channels promote persistent directed motion and efficient transport, whereas in disordered environments, their extended conformations are less able to adapt to tortuous pathways between obstacles, suppressing their transport. Combining simulations with a simple theoretical description, we show that long-time diffusion is controlled by confinement-induced changes in filament conformation and reorientation dynamics. These findings establish filament flexibility together with the structure of the porous medium as key design parameters for optimizing active transport in crowded environments.

The rest of the paper is organised as follows. In Sec.~\ref{sec:methodology}, we introduce the simulation models and parameters. In Sec.~\ref{sec:map}, we give an overview of the long-time diffusion dynamics of active polymers in square and random media. In Sec.~\ref{sec:results}, we investigate the combined effects of polymer flexibility and obstacle packing fraction on conformational properties and reorientational dynamics of active polymers. In Sec.~\ref{sec:theory}, we outline a theoretical framework for estimating the long-time centre-of-mass diffusion coefficient of active filaments in obstacle arrays. Finally, we discuss our main findings and summarise them in Sec.~\ref{sec:discussion-conclusion}.

\section{Methodology}
\label{sec:methodology}
In this section, we introduce the tangentially driven active polymer model, describe the construction of ordered and disordered obstacle arrays, and quantify the resulting pore-space geometry.

\subsection{Active filament model}
\label{subsec:polymer}
The active filament (polymer) is modelled as a linear chain of $N$ beads (monomers) connected by harmonic springs, subject to a bending potential that sets its persistence length $\ell_\text{p}^0$. Each bead is subject to stochastic forces $\mathbf{f}^r$ of thermal or biological origin and an active force which is tangential to the backbone. The stochastic force is modelled as white noise with the following properties:
\begin{equation}
\label{eq:randomforce}
\begin{cases} 
    \langle \mathbf{f}^{\text{r}}(t) \rangle = \mathbf{0} & \text{zero mean}, \\
    \langle f^{\text{r}}_{\alpha}(t) f^{\text{r}}_{\beta}(t')  \rangle = 2 \gamma^2 D_0 
    \delta_{\alpha\beta} \delta (t-t') & \text{\(\delta\)-correlation.}
\end{cases}
\end{equation}
Here, subscripts \(\alpha\) and \(\beta\) denote the spatial components of the vector \(\mathbf{f}^{\text{r}}\). \(D_0\) denotes the diffusivity of an individual bead, which becomes equal to \(k_{\mathrm{B}}T/\gamma\) when the noise is thermal in origin. 

The overdamped equation of motion for bead $i$, subject to conservative force \(-\nabla U\), active force \( \mathbf{f}^{\text{a}}_{i} \) and stochastic force \( \mathbf{f}^{\text{r}}_{i} \) is given by:
\begin{equation}
\label{eq:eom}
    \gamma \dot{\mathbf{r}}_i = - \sum_{k} \nabla_{\mathbf{r}_i} U(\mathbf{r}_i,\mathbf{r}_{i\pm 1}, \mathbf{R}_k) + \mathbf{f}^{\text{a}}_{i}+\mathbf{f}^{\text{r}}_{i},
\end{equation}
where \( \mathbf{r}_i \) denotes the position of bead \(i\) in the chain and \(\mathbf{R}_k\) refers to the position of the \(k\)\textsuperscript{th} obstacle. \( \gamma \) is the damping coefficient. 

The overall configuration potential \(U\) has three components:
\begin{equation}
\label{eq:configurationpotential}
    U = U_{\text{harmonic}} + U_{\text{bend}} +U_{\text{excl}}.
\end{equation}
The first two terms define the bond forces. We denote the bond vector from the \( (i-1) \)\textsuperscript{th} to \( i\)\textsuperscript{th} bead as \(\mathbf{b}_{i-1,i} = \mathbf{r}_i - \mathbf{r}_{i-1} \). Their elasticity is given by a harmonic spring potential:
\begin{equation}
\label{eq:harmonic}
    U_{\text{harmonic}}(\mathbf{b}_{i-1,i})=\frac{k_{\text{s}}}{2}(|\mathbf{b}_{i-1,i}|-\ell_0)^2,
\end{equation}
where \( k_{\text{s}} \) is the spring constant and $\ell_0$ is the equilibrium bond length. 
The stiffness of the active polymer is controlled via a bending potential:
\begin{equation}
\label{eq:bending}
    U_{\text{bend}}(\theta_i)=\kappa(1-\cos{\theta_i}),
\end{equation}
where \( \theta_i \) is the bond angle that satisfies \(\cos{\theta_i} = \hat{\mathbf{b}}_{i-1,i} \cdot \hat{\mathbf{b}}_{i,i+1}\), and \( \kappa \) is the bending stiffness. The bending force vanishes for perfectly aligned bonds. The bending stiffness sets the intrinsic persistence length $\ell_{\text{p}}^0=\frac{2\kappa}{D_0 \gamma}$ for thermal active polymers. 

The last term in Eq.~\eqref{eq:configurationpotential} specifies the excluded volume interactions between monomers and surrounding obstacles, given by a short-ranged repulsive Weeks-Chandler-Andersen (WCA) potential:
\begin{equation}
\label{eq:WCA}
    U_{\text{excl}}(r)=
\begin{cases}    
    4  \epsilon \left[(\frac{\sigma_{\text{w}}}{r})^{12} -(\frac{\sigma_{\text{w}}}{r})^6+\frac{1}{4}\right] & \text{for } 0 < r < r_{\text{cut}}, \\
    0 & \text{for } r \ge r_{\text{cut}},
\end{cases}
\end{equation}
where \(r\) is the distance between the two particles of interest, \( \epsilon \) is the strength of the interaction, \( \sigma_{\text{w}} \) is the interaction width, and \(r_{\text{cut}} = 2^{1/6} \sigma_{\text{w}} \) is the cutoff radius. The interaction width is \(\sigma_{\text{w}}=\frac{1}{2} \sigma+R_{\text{o}}\), where \(\sigma\) is the diameter of the monomer and \( R_{\text{o}} \) is the radius of the obstacle. 

In our model, we use a uniform active force density along the polymer chain backbone, where the force per unit length has a constant magnitude and is tangential to the bonds. The active force associated with each bond is equally distributed to the two neighbouring monomer beads in the following way:
\begin{equation}
\label{eq:activity}
\begin{cases} 
    \mathbf{f}^{\text{a}}_1=\frac{f^{\text{a}}}{2\ell_0 } \mathbf{b}_{1,2}  & \text{for the head bead,} \\
    \mathbf{f}^{\text{a}}_i=\frac{f^{\text{a}}}{2 \ell_0 } (\mathbf{b}_{i-1,i}+\mathbf{b}_{i,i+1})
 & \text{for beads in the middle,} \\
    \mathbf{f}^{\text{a}}_N=\frac{f^{\text{a}}}{2\ell_0 } \mathbf{b}_{N-1,N} & \text{for the tail bead,}
\end{cases}
\end{equation}
where \( f^\text{a} \) is the magnitude of the active force, such that the total active force acting on a straight polymer is \( (N-1) f^\text{a} \).

\subsubsection*{{\bf Simulation parameters}}
For our simulations, we choose the monomer diameter \(l_\text{u}=\sigma\), thermal energy \(E_\text{u}=D_0 \gamma \), and \(\tau_\text{u}= \frac{l_\text{u}^2}{D_0}\) as the units of length, energy, and time. The bead diffusion and damping coefficient are fixed to \(D_0=1\) and \(\gamma=1\).
Each polymer chain contains \(N=100\) monomers. We set the equilibrium bond length \(\ell_0=1\) matching the size of monomers, with spring constant \(k_\text{s}=5000\) for rigid bonds. The corresponding contour length is therefore \(L=(N-1) \ell_0+\sigma=100\). The polymers are under low active force \(f^{\text{a}}=0.1\). The strength of WCA interaction is set to \(\epsilon=1\). We vary the polymer bending stiffness from flexible to semiflexible regimes \(\kappa\in[1,100]\). The simulation results for each parameter set are obtained from time and ensemble averaging of 150--380 independent polymers. The long-time diffusion coefficient for a passive polymer in free space is given by \(D^\text{Free}_\text{Passive}=D_0/N=0.01\), where \(D_0\) is the diffusion coefficient of a single Brownian monomer in free space.

\subsection{Porous media model and pore space characterisation}
\label{subsec:media}

We model the porous media as non-overlapping fixed circular obstacles in a 2D square box with periodic boundary conditions. The obstacles are monodisperse in size and exert short-ranged volume-exclusion repulsive forces (through WCA potential in Eq.~\eqref{eq:WCA}) on monomers that come into contact with them.

For the majority of our simulations, the obstacles have radius \(R_\text{o}=8.7 \sigma\) substantially larger than the monomers, matching previous experiments involving worms~\cite{sinaasappel2025locomotion}. Moreover, some simulations with $R_\text{o}=5\sigma$ and \(15 \sigma\) for active filaments in random media are performed. 

The obstacle packing fraction is defined as the fraction of space filled by the obstacles. For non-overlapping circular obstacles, it is given by:
\begin{equation}
\label{eq:packingfraction}
    \phi = \frac{N_{\text{o}}\pi {R_{\text{o}}^2}}{L_{\text{box}}^2},
\end{equation}
where \(N_{\text{o}}\) is the number of obstacles, \( R_{\text{o}} \) is the obstacle radius, and \( L_{\text{box}} \) is the side length of the 2D simulation box. The porosity of the medium is obtained as \(\phi_{\text{pore}} = 1 - \phi\).
We study two types of obstacle configurations (random and square lattice), and vary the obstacle packing fraction in the range \(\phi\in[0,0.6]\).

\subsubsection{Generation of porous media with different configurations}

\noindent a. Square Lattice

We place obstacles on the lattice sites of 2D square lattices. The simulation box has approximate side length $L_\text{box}\approx120 \sigma$ with periodic boundary conditions. The lattice constant for a prescribed obstacle packing fraction is \(a=\sqrt{\pi/\phi}R_{\text{o}}\). 
The corresponding channel width is \(\xi=(\sqrt{\pi/\phi}-2)R_{\text{o}}\), and the approximate pore size is \(d_{\text{cage}}=(\sqrt{2\pi/\phi}-2)R_{\text{o}}\), as illustrated in Fig.~\ref{fig:schematics}(a).\\

\noindent b. Random Packing

To generate disordered configurations, we initially place the obstacles independently and randomly in the simulation box of size $1000\times1000 \, \sigma^2$, with no correlation between positions. We then apply short-ranged repulsive dissipative-particle-dynamics (DPD) potential~\cite{phillips2011pseudo} consisting of a repulsive part and a pairwise drag to remove overlaps. The relaxed equilibrium configuration with fixed obstacle positions is then taken as the model porous medium. Figure~\ref{fig:schematics} (b) shows the schematics of an active polymer in a disordered obstacle array.
We enforce a minimum gap between obstacles to prevent the formation of extremely narrow bottlenecks and ensure that polymers can pass through the gap between any two obstacles.
The minimum surface-to-surface obstacle separation is set to \(d_{\mathrm{gap}}=2.84\,\sigma\), corresponding to a DPD cutoff radius \(10\%\) larger than the geometric contact distance, \(R_{\mathrm{o}}+0.5\sigma\). This choice is also consistent with the monomer--obstacle WCA interaction, for which the repulsive force becomes comparable to the active driving force (\(f^{\mathrm a}=0.1\)) at a gap width of roughly \(3\sigma\).

For each parameter set, observables are obtained by ensemble averaging over 4--7 independent realizations of the disordered obstacle configuration.

\subsubsection {Characteristic length scales of porous media}

Here, we discuss and characterize the relevant geometric length scales of the porous medium. A key measure is the chord length, defined as the length of a straight line segment that lies entirely within the pore phase and connects two points on the obstacle surface, see Fig.~\ref{fig:schematics}(b). It provides a geometric proxy for the mean free path of a persistently moving particle. The chord-length distribution, $P(\ell)$, gives the probability density of finding a pore-space chord with length between $\ell$ and $\ell+d\ell$. For randomly positioned overlapping circular obstacles, this distribution decays exponentially, $p(\ell)=\frac{1}{\ell_\text{c}^0}\exp\left(-\frac{\ell}{\ell_\text{c}^0}\right)$, where $\ell_\text{c}^0$ is both the decay length and the mean chord length~\cite{lu1993chord}. When the disks become impenetrable, the chord length distribution (CLD) deviates from exponential at small length scales. For a two-dimensional medium of randomly positioned impenetrable disks, the mean chord length $\ell_\text{c} = \langle \ell \rangle = \int \ell P(\ell) d\ell$ is known analytically~\cite{lu1993chord}. It is given in terms of the obstacle radius \(R_\text{o}\) as \(\ell_\text{c} =\pi R_\text{o}\frac{1-\phi}{2\phi}\).
\begin{figure}
    \centering
    \begin{overpic}[width=0.32\linewidth]{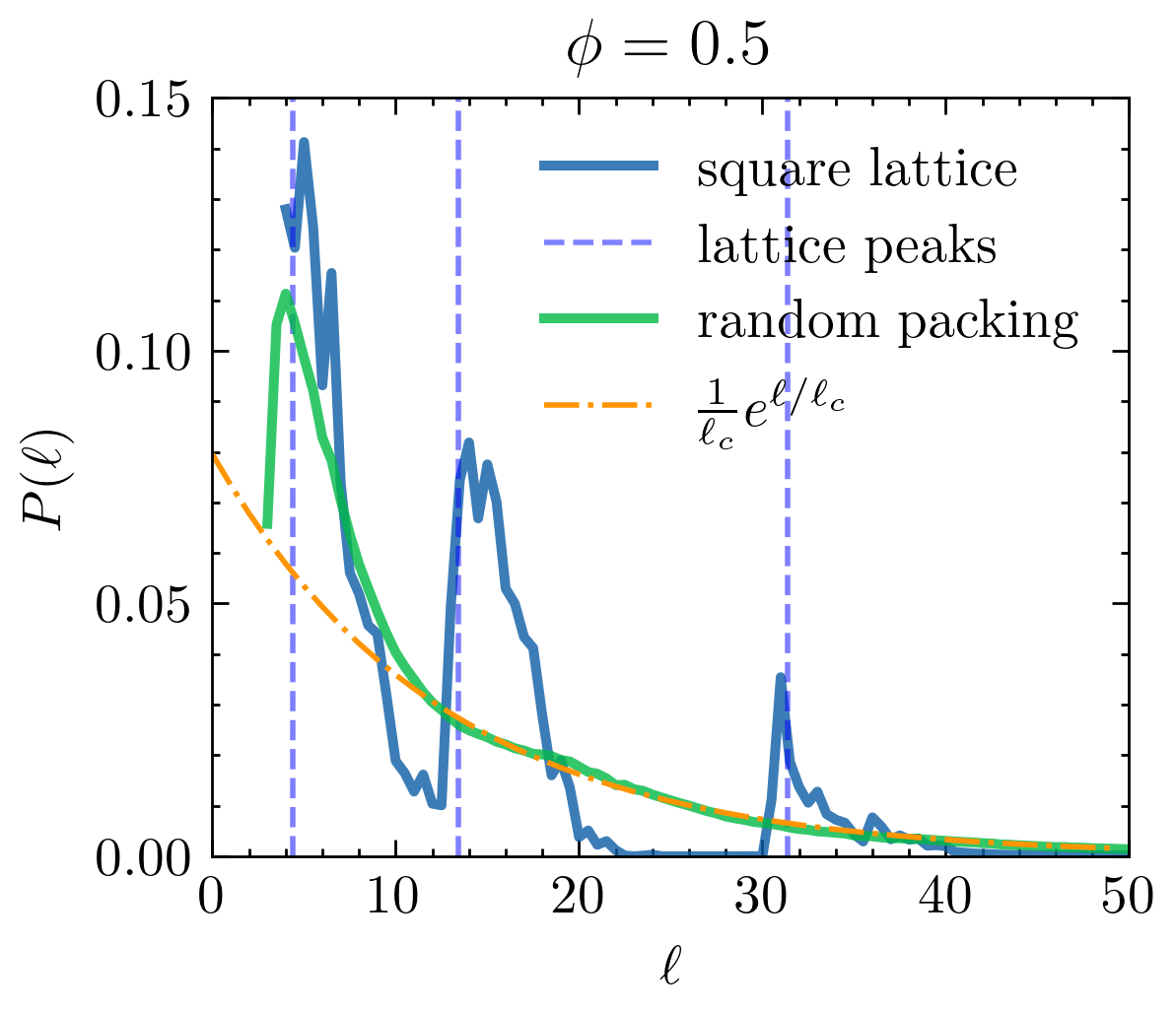}
        \put(2,88){(a)}
    \end{overpic}
    \hfill
     \begin{overpic}[width=0.32\linewidth]{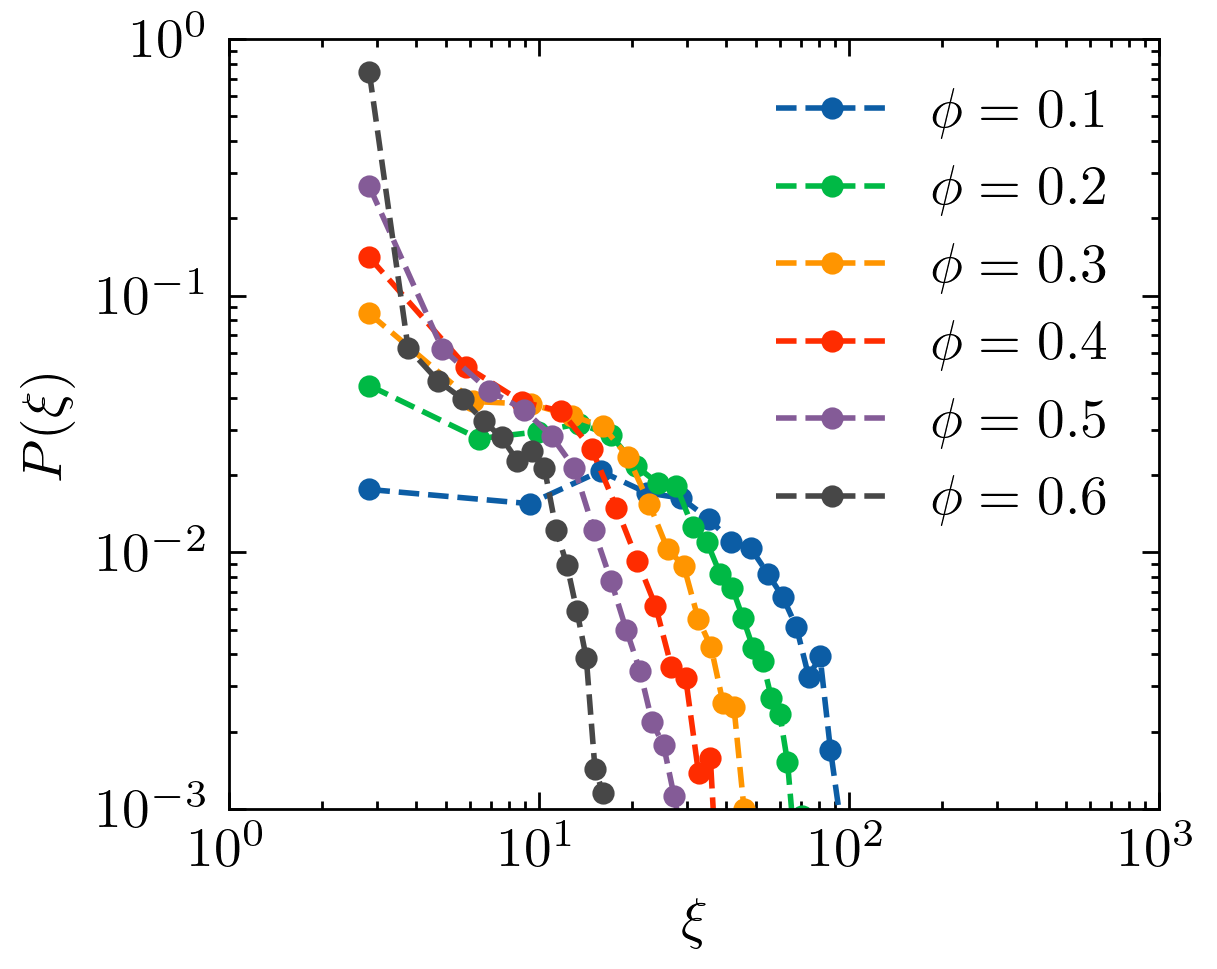}
        \put(2,88){(b)}
    \end{overpic}
      \hfill
     \begin{overpic}[width=0.32\linewidth]{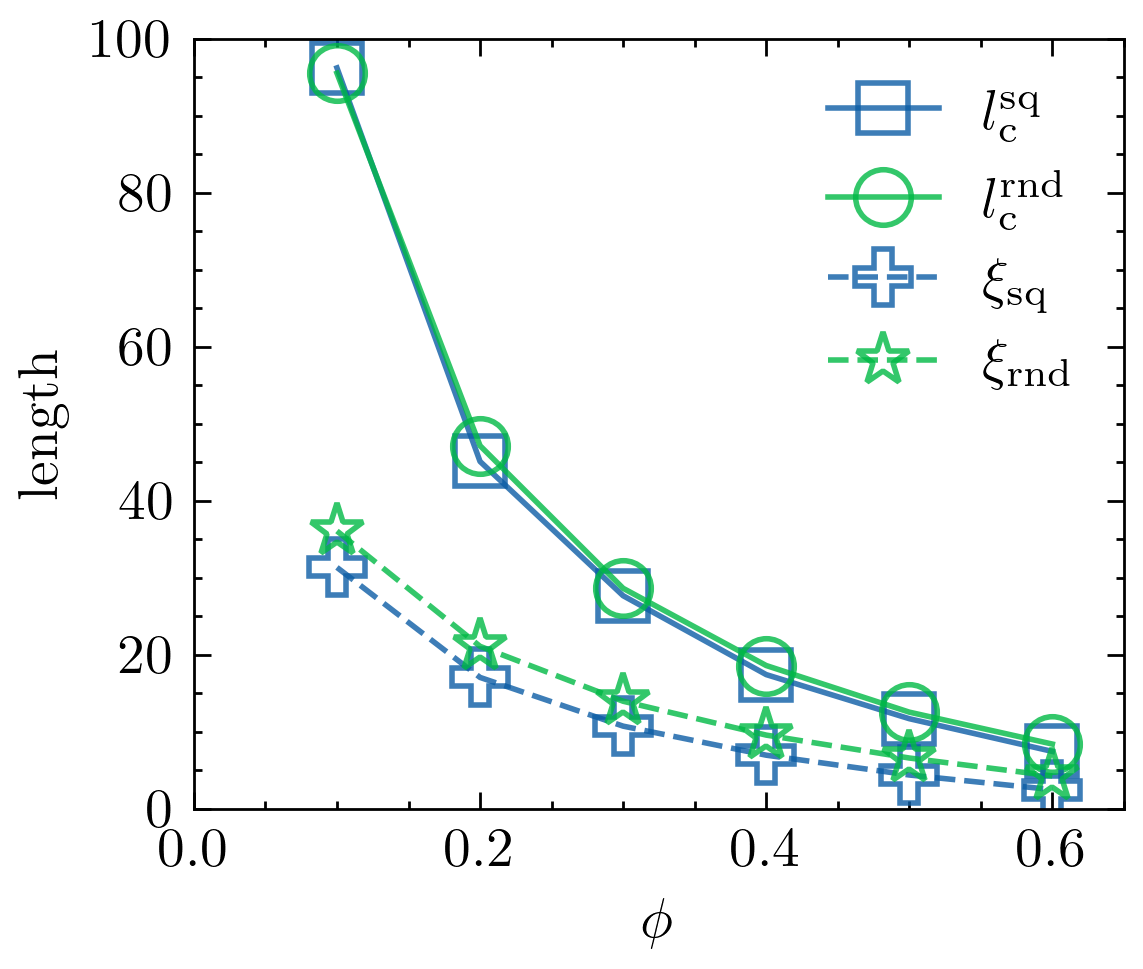}
     \put(2,88){ (c) }
    \end{overpic}
    \caption{Pore-space length scales for ordered and disordered obstacle arrays. (a) Chord-length distribution $P(\ell)$ at \(\phi=0.5\), where the dashed line shows the exponential distribution with a decay length equal to the mean chord length \(\ell_\text{c}\). (b) Probability density of gap widths in random media. (c) Mean chord length and nearest neighbour gap width as functions of obstacle packing fraction \(\phi\).}
    \label{fig:LengthScales}
\end{figure}
Figure~\ref{fig:LengthScales}(a) shows the chord length distribution (CLD) for \(\phi=0.5\) for random and periodic configurations obtained using the \texttt{Python} package \texttt{PoreSpy}~\cite{gostick2019porespy}. 
The CLD for random media follows exponential decay, except for small values where it has a peak around the imposed minimum gap width. The CLD for square lattice has peaks corresponding to successive nearest neighbour distances minus the obstacle diameter.
From the CLDs, we extract the mean chord length $\ell_\text{c}$ as shown in Fig.~\ref{fig:LengthScales}(c). $\ell_\text{c}$ decreases with packing fraction for both types of media. Although the full CLDs differ substantially between random and periodic environments, their mean chord lengths remain similar.

Another relevant geometric length scale is the gap width between neighbouring obstacles, which can influence pore-scale trapping and channelling of active filaments. We define the gap width, $\xi$, as the centre-to-centre distance between neighbouring obstacles minus the obstacle diameter. For square lattices, $\xi$ is simply the lattice spacing minus the obstacle diameter, see Fig.~\ref{fig:schematics}(a). For random arrays, neighbouring obstacles are identified as pairs sharing a Voronoi edge, computed using \texttt{scipy.spatial.Voronoi}; the corresponding gap-width distribution, $P(\xi)$, is then obtained from all such neighbour pairs as presented in Fig.~\ref{fig:LengthScales}(b). The mean gap width obtained from the distribution, $\langle \xi \rangle \equiv \xi_{\text{rnd}}$, is shown in Fig.~\ref{fig:LengthScales}(c). The mean gap width is slightly larger in random arrays than in square lattices. 
Although both length scales $\ell_\text{c}$ and $\xi$ decrease with increasing obstacle packing fraction $\phi$, at low packing fractions the mean chord length is significantly larger than the mean gap size. At the lowest packing fraction, $\phi=0.1$, the mean chord length is larger than the mean gap width, by approximately a factor of $2.5$. This difference decreases with increasing obstacle density, and for $\phi>0.4$ the two length scales become comparable. In the following sections, we examine how these geometric length scales influence the conformation and dynamics of active filaments in porous media.

\subsubsection{Tortuosity of ordered and random media}

Another important geometric characteristic of porous media is the tortuosity, which quantifies how strongly transport pathways through the pore space deviate from straight lines. Tortuosity is a scale-invariant geometric property of the medium and can be defined in several ways. Here, we focus on the bulk-diffusion tortuosity, which measures how the convoluted pore structure increases the effective path length of diffusing particles and thereby reduces the effective diffusivity relative to free space. It is defined as
\(T_{\text{diff}}=\dfrac{D_{\text{free}}}{D_{\text{effective}}}(1-\phi_\text{eff})\), 
where \(D_{\mathrm{free}}\) and \(D_{\mathrm{effective}}\) are the diffusion coefficients in free space and in the porous medium, respectively~\cite{ghanbarian2013tortuosity}. Here, we consider the pore space freely accessible by the centre of a monomer, with effective obstacle radius \(R_\text{eff}=R_\text{o}+\sigma/2=9.2\sigma\) and effective packing fraction \(\phi_\text{eff}=\frac{N_\text{o}\pi R_\text{eff}^2}{L_\text{box}^2}\). We compute \(T_{\mathrm{diff}}\) from pixelised binary images of the porous media using the \texttt{PoreSpy} package~\cite{gostick2019porespy}, which solves a Fickian diffusion problem subject to concentration boundary conditions imposed on opposite sides of the system.

\begin{figure}
    \centering
    \includegraphics[width=0.7\linewidth]{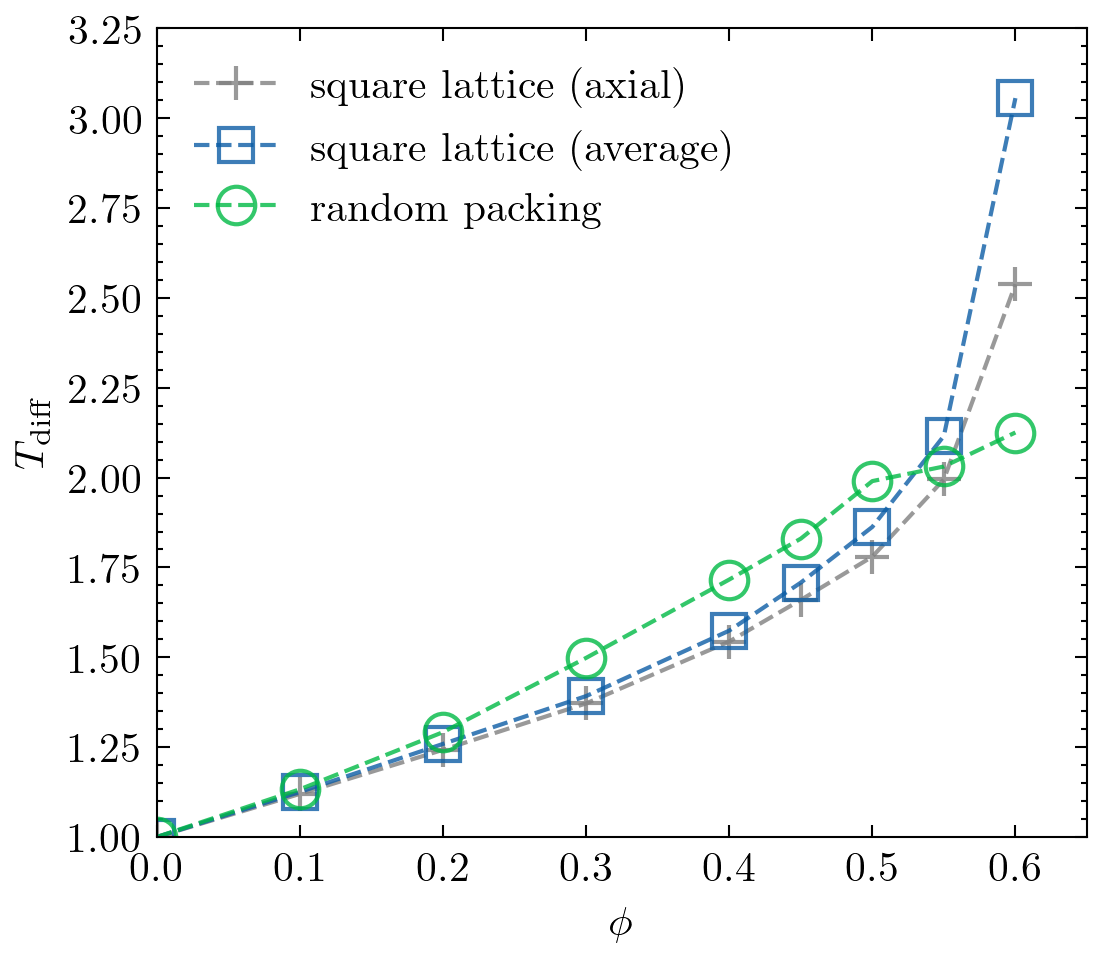}
    \caption{Bulk-diffusion tortuosity \(T_{\text{diff}}\) as a function of obstacle packing fraction \(\phi\). Plus symbols denote \(T_{\text{diff}}\) for square-lattice arrays measured along lattice directions, and squares denote the values averaged over all directions. Circles denote \(T_{\text{diff}}\) for random obstacle arrays.}
    \label{fig:Tor}
\end{figure}

Figure~\ref{fig:Tor} shows \(T_{\mathrm{diff}}\) as a function of obstacle packing fraction \(\phi\) for both random and square-lattice media. For square lattices, we distinguish between the tortuosity averaged over all directions and the tortuosity measured along the primitive lattice axes. These values are similar at low packing fractions, but for \(\phi \gtrsim 0.4\) the axial tortuosity becomes noticeably smaller than the directional average, reflecting the presence of preferential transport channels. As shown later, this anisotropy strongly influences the motion of stiff filaments whose persistence lengths exceed the lattice spacing. Comparing ordered and disordered media, we find that their tortuosities are similar at low packing fractions. At intermediate densities, random media are more tortuous, whereas at high packing fractions (\(\phi \gtrsim 0.5\)) the square lattice exhibits the larger tortuosity.

\section{Long-time diffusion map in ordered and random media}
\label{sec:map}
Having described our simulation method and characterized the porous media, we now turn to the results. We begin by examining how filament flexibility and obstacle packing fraction govern active-filament transport in ordered and disordered environments.
To this end, we compute the mean squared displacement (MSD) of polymer centre of mass:
\begin{equation}
    \label{eq:msd}
    \langle \Delta \mathbf{r}_{\text{cm}}^2 (t)\rangle = \langle |\mathbf{r}_{\text{cm}}(t) - \mathbf{r}_{\text{cm}}(0)|^2 \rangle, 
\end{equation}
where $\langle \cdots \rangle$ includes both ensemble and time averaging. Representative MSD curves are shown in Fig.~\ref{appfig:msd-t-k} in Appendix~\ref{sec:MSD}.
As shown in previous studies~\cite{isele2015self,fazelzadeh2023effects}, the diffusion of active polymers in free space has three regimes. At short times, the motion is dominated by random force and follows normal diffusion. At intermediate times comparable to the mean relaxation time of the end-to-end vector \(\langle\tau\rangle\) (see Sec.~\ref{subsec:dynamics}), active polymers typically display ballistic motion. The dynamics then cross over to diffusive motion again at long times, with an enhanced coefficient of diffusion due to activity. In the presence of obstacles, a similar trend is observed~\cite{fazelzadeh2023active,sinaasappel2025locomotion}, except for flexible (\(\kappa=1\)) polymers at high obstacle packing fraction \(\phi=0.6\), which show subdiffusive behaviour at intermediate timescales, see Fig.~\ref{appfig:msd-t-k}(b).
We quantify active-filament transport using the long-time diffusion coefficient \(D_{l}\) obtained by fitting the centre-of-mass mean squared displacement 
\(\langle \Delta \mathbf{r}_{\text{cm}}^2 \rangle\) in the diffusive regime \(t \gg \langle\tau\rangle\), where \( \langle \Delta \mathbf{r}_{\text{cm}}^2(t)\rangle = 4 D_{l} t\).
\begin{figure}
    \centering
    \begin{overpic}[width=1\linewidth]{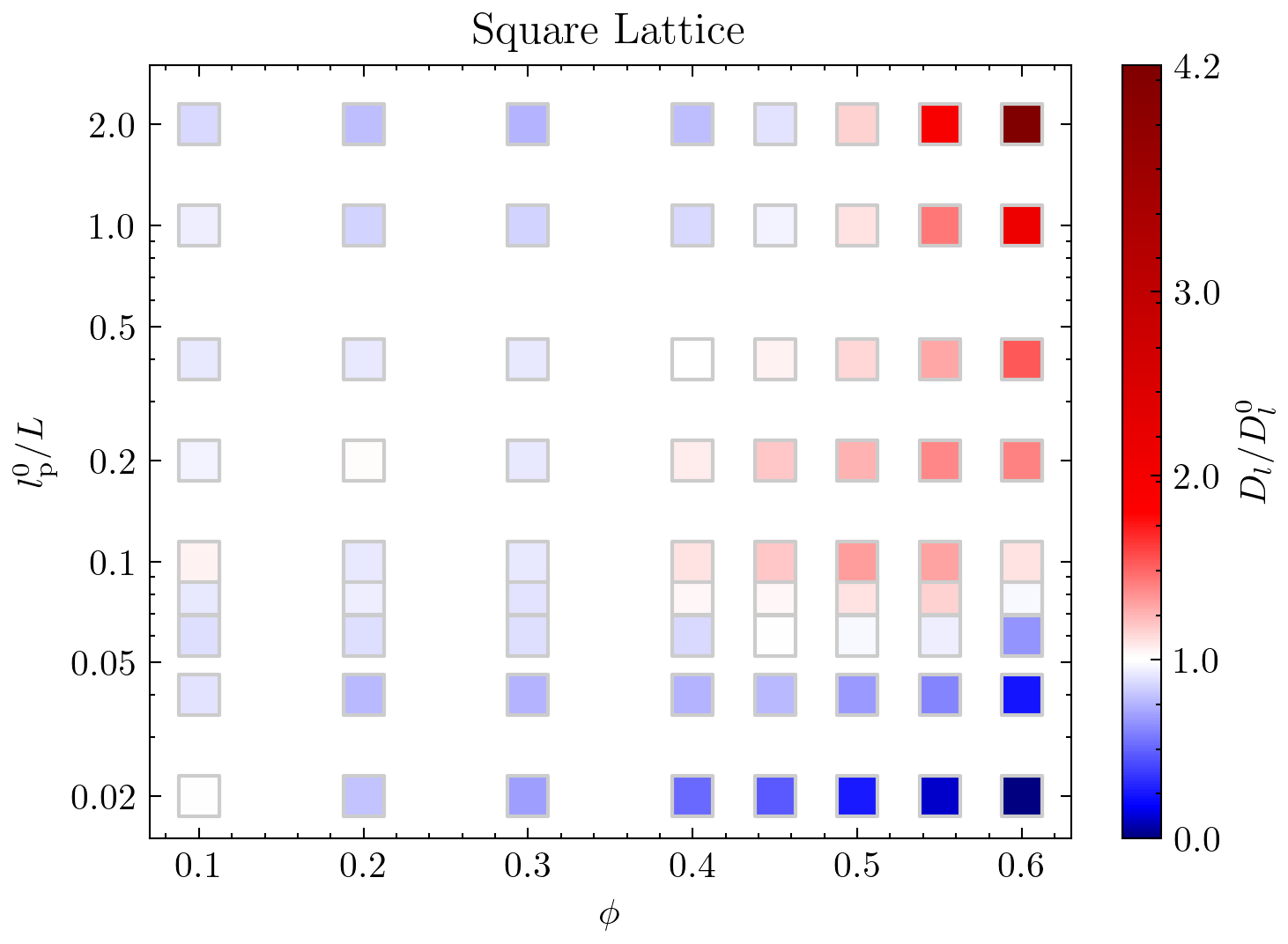}
        \put(2,70){\textbf{(a)}}
    \end{overpic}
    \hfill
    \begin{overpic}[width=1\linewidth]{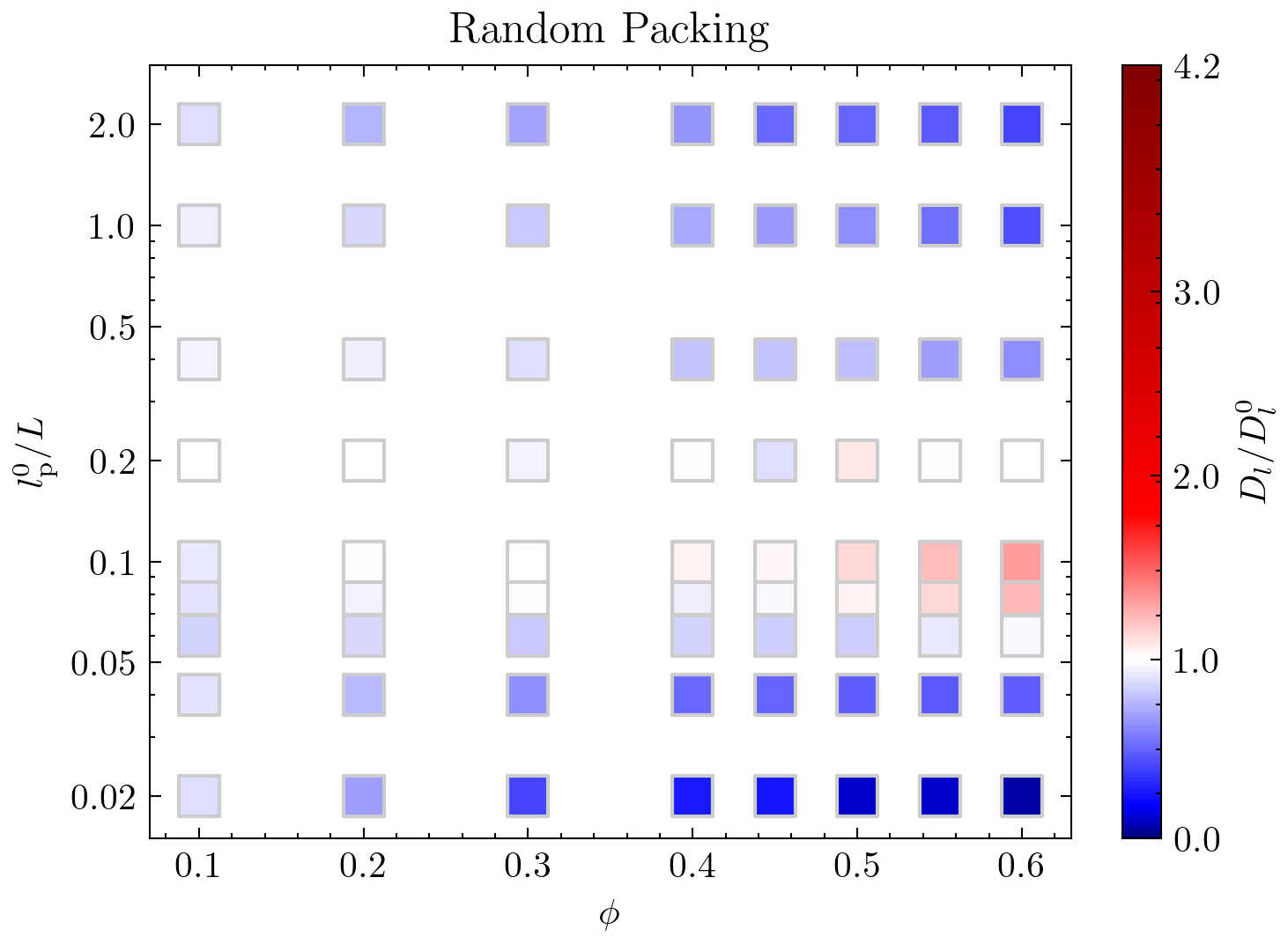}
        \put(2,70){\textbf{(b)}}
    \end{overpic}
    \caption{Colour map of the normalised long-time diffusion coefficient \(D_l(\phi,\kappa)/D_l^{0}(\kappa)\) of polymer centre of mass, where \(D_l^{0}(\kappa)\) denotes the value in free space, for (a) square-lattice and (b) random obstacle arrangements, as a function of obstacle packing fraction \(\phi\) and relative stiffness \(\ell_\text{p}^0/L\). Here, \(\ell_\text{p}^0=2\kappa/D_0\gamma\) is the intrinsic persistence length of passive polymers in free space and \(L=100\) is the polymer contour length.}
    \label{fig:DiffusionPhaseDiagram}
\end{figure}

Figure~\ref{fig:DiffusionPhaseDiagram} presents the normalised long-time diffusion \(D_l(\phi,\ell_{\text{p}}^0)/D_l^{0}(\ell_{\text{p}}^0)\) of active polymers in ordered and disordered obstacle arrays as a function of polymer relative stiffness \(\ell_{\text{p}}^0/L\) (inverse of degree of flexibility) and obstacle packing fraction \(\phi\). Here, $D_l^{0}(\ell_{\text{p}}^0)$ denotes the long-time diffusion in free space. The long-time diffusion landscapes in ordered and disordered media differ markedly, particularly in the strongly confined regime (\(\phi>0.4\)), where the mean chord length and gap size become comparable for less flexible polymers with $\ell_{\text{p}}^0/L>0.1$. These trends highlight the joint role of the underlying geometry of the confining environment and filament flexibility in controlling the diffusive transport of semiflexible active filaments.

\begin{figure}
    \centering
    \begin{overpic}[width=0.485\linewidth]{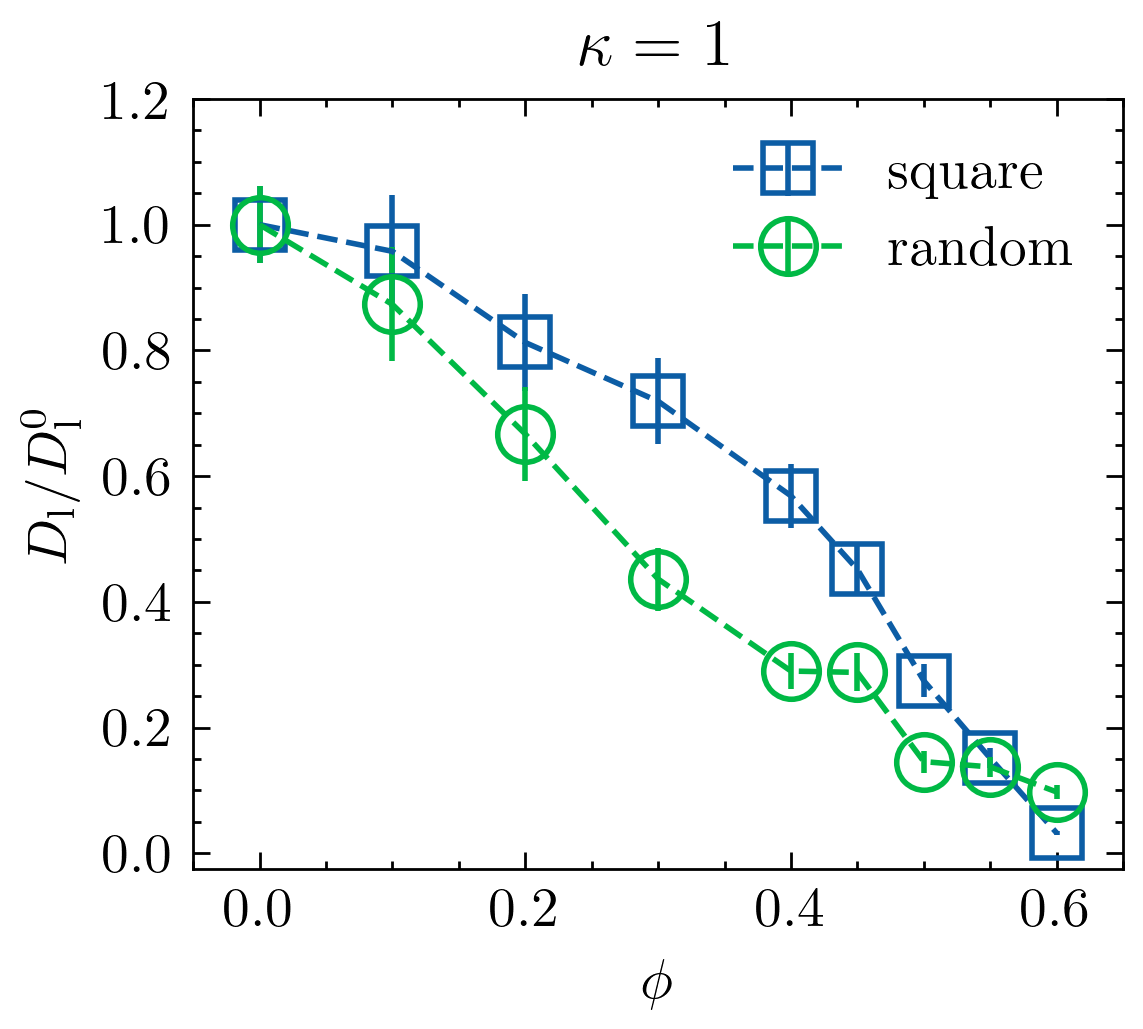}
        \put(2,87){\textbf{(a)}}
    \end{overpic}
    \hfill
    \begin{overpic}[width=0.5\linewidth]{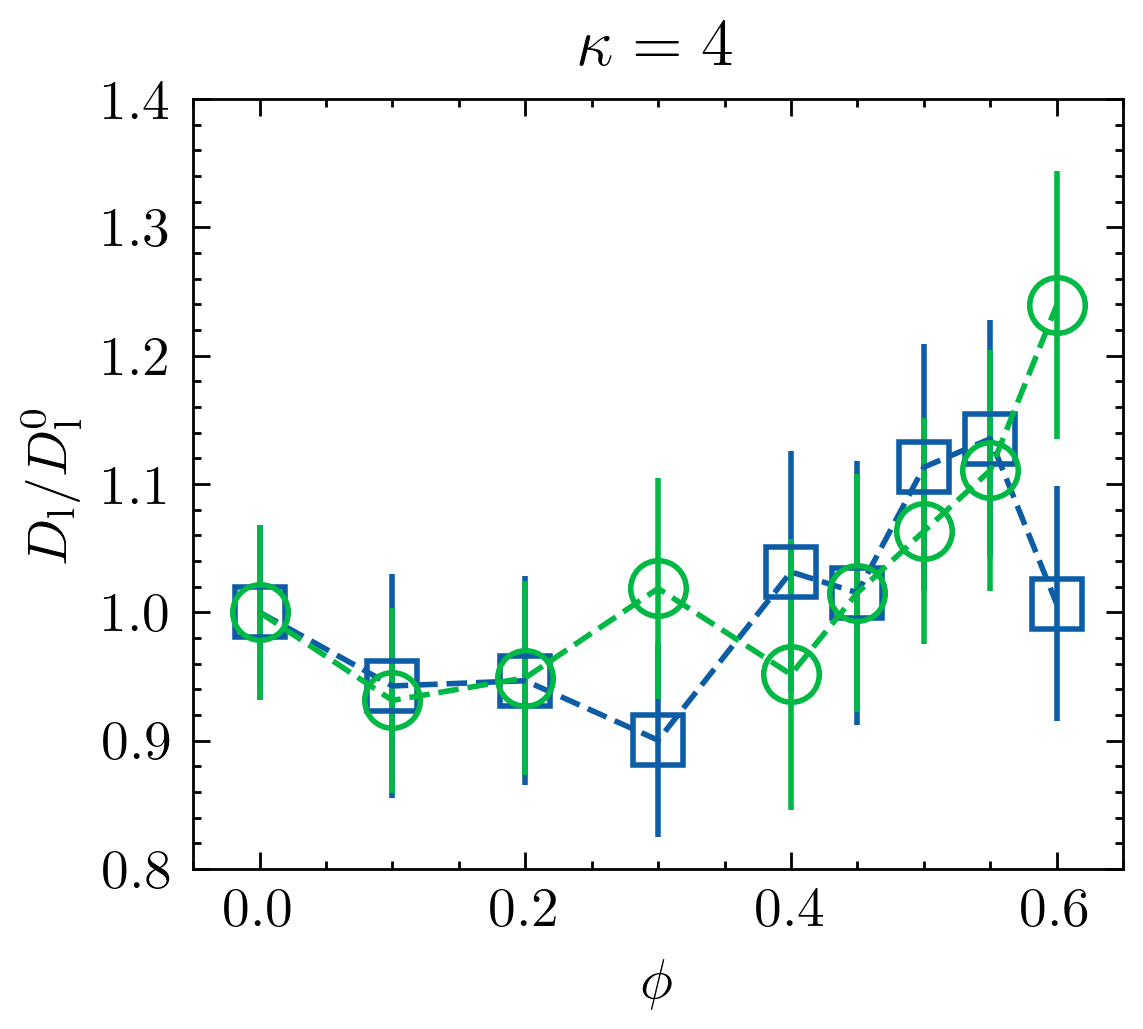}
        \put(2,85){\textbf{(b)}}
    \end{overpic}
    \hfill
    \begin{overpic}[width=0.49\linewidth]{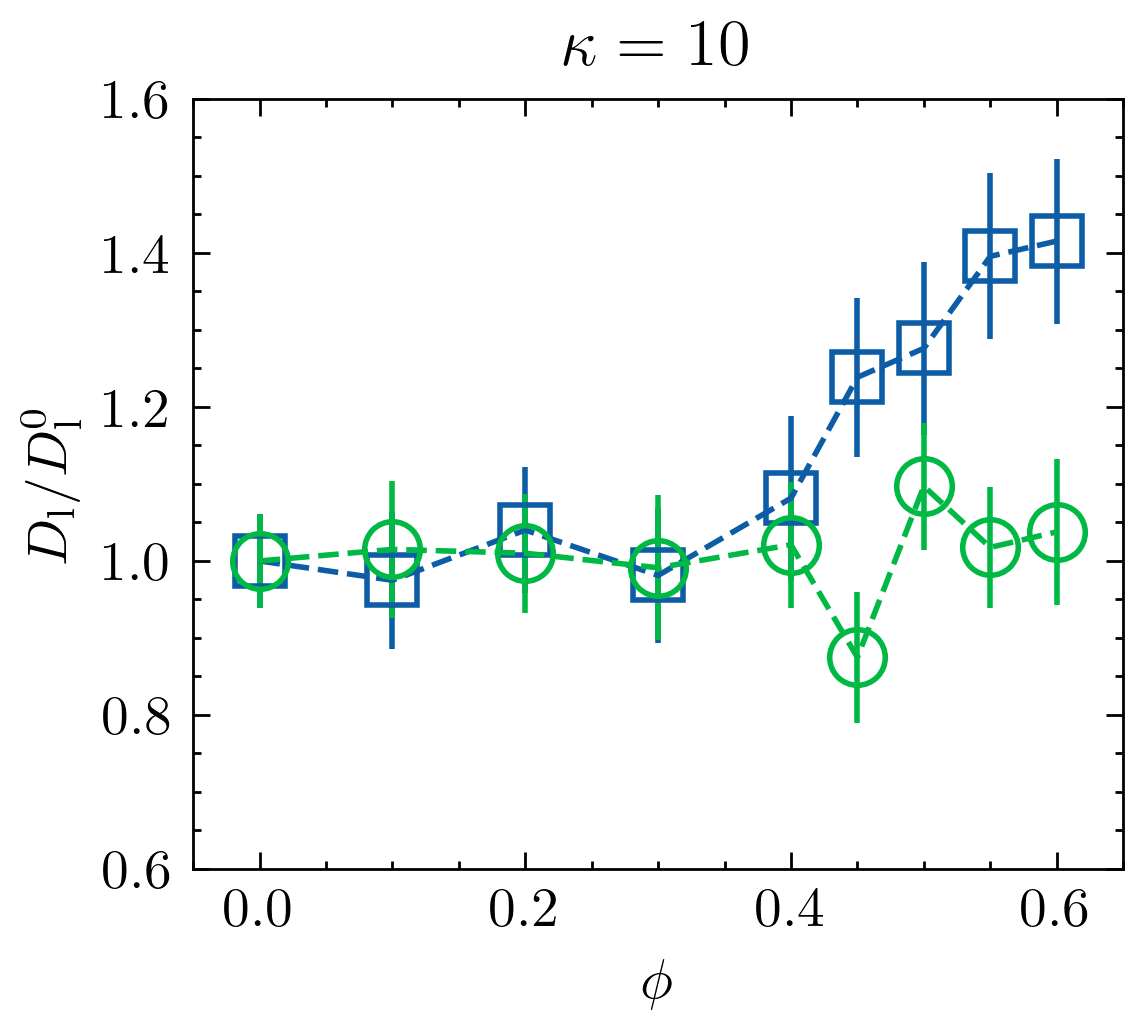}
        \put(2,86.5){\textbf{(c)}}
    \end{overpic}
    \hfill
    \begin{overpic}[width=0.47\linewidth]{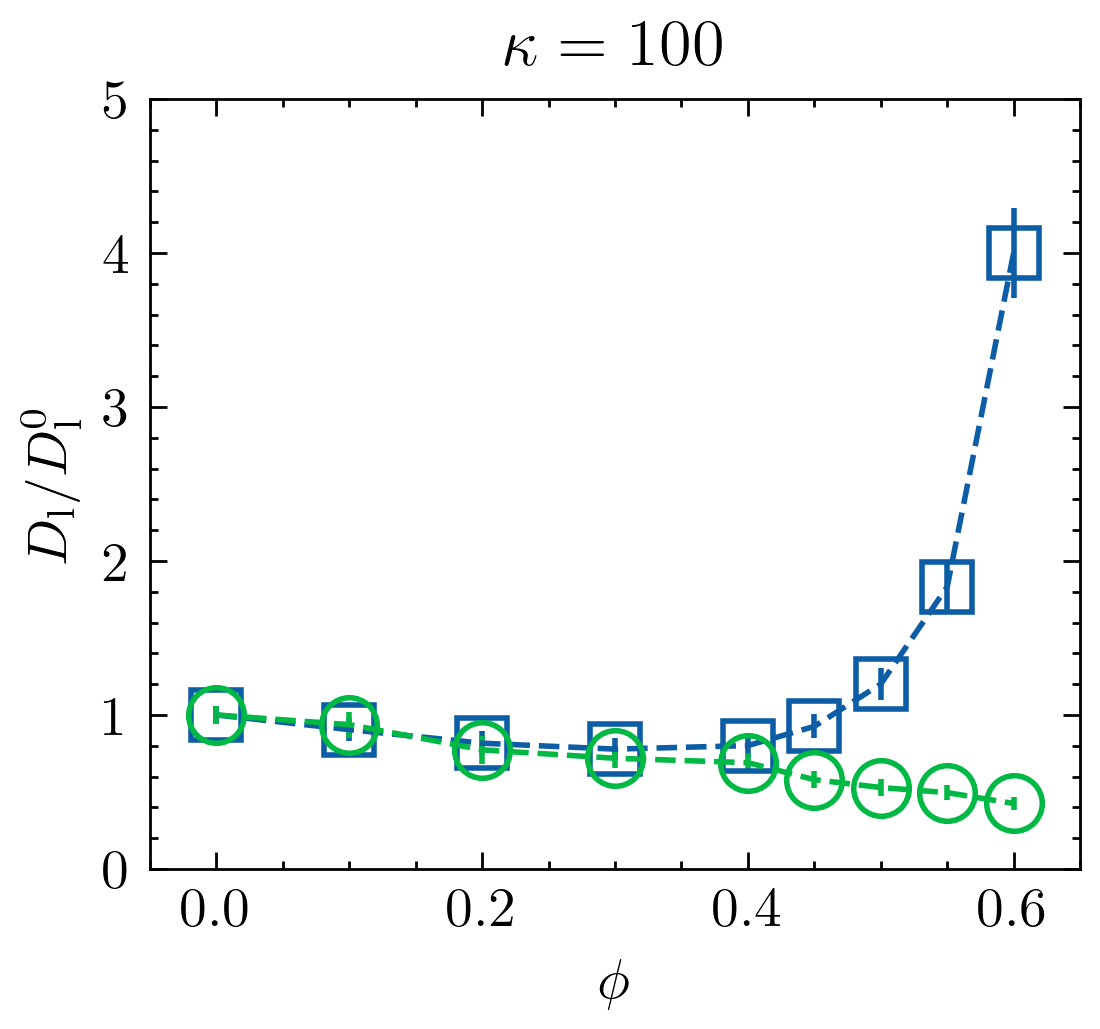}
        \put(2,90){\textbf{(d)}}
    \end{overpic}
    \caption{Normalised long-time diffusion coefficient \(D_l/D_l^{0}\) of polymer centre of mass as a function of obstacle packing fraction \(\phi\). Panels (a)--(d) correspond to polymer bending stiffness values \(\kappa=1,4,10\), and $100$, respectively.}
    \label{fig:dl-phi-k}
\end{figure}

Based on the observed trends for the dependence of long-time diffusion on packing fraction, we can identify three regimes of behaviour:\\
(i) \textbf{Highly flexible regime} ($\ell_{\text{p}}^0/L \ll 1$)\\
For highly flexible polymers with $\ell_{\text{p}}^0/L < 0.08$, increasing obstacle packing fraction slows down diffusive transport in both ordered and disordered media, as shown in Fig.~\ref{fig:dl-phi-k}(a) at \(\kappa=1\) corresponding to $ \ell_{\text{p}}^0/L=0.02$. At intermediate obstacle packing fraction \(\phi\), this slowdown is more pronounced in disordered media. However, at higher packing fractions, in particular at $\phi=0.6$, polymers become more strongly trapped in the square lattice, leading to a stronger suppression of diffusion in ordered media. Interestingly, at the highest packing fraction the tortuosity of the square lattice is higher than the random arrays, consistent with the correlation between slowdown of diffusion and the mean tortuosity. \\
(ii) \textbf{Moderately flexible regime} ($\ell_{\text{p}}^0/L \sim 0.1$) \\
In the regime \(0.08 \leq \ell_{\text{p}}^0/L \le 0.2\), the polymers remain relatively flexible, although the energetic cost of bending is no longer negligible. This regime is characterized by a long-time diffusion that remains comparable to the free-space value at packing fractions below \(\phi_\text{c} \approx 0.4\), followed by a slight enhancement relative to free space under strong confinement for \(\phi \ge 0.4\), where the mean chord length becomes comparable to the mean gap size. 
This behaviour is evident in Fig.~\ref{fig:dl-phi-k}(b)-(c) for \(\kappa=4\) and \(10\).
We note that diffusion data in this intermediate regime display large relative fluctuations.\\
(iii) \textbf{Semiflexible regime }($ \ell_{\text{p}}^0 /L\sim 1$) \\
For $0.2 < \ell_{\text{p}}^0/L \le 2$ crowding affects the long-time diffusion in qualitatively different ways in ordered and disordered media at high obstacle packing fractions \(\phi > 0.3\) as shown in Fig.~\ref{fig:dl-phi-k}(d) for $\kappa=100$ ($\ell_{\text{p}}^0/L=2$). In disordered media, diffusion decreases monotonically with increasing \(\phi\). In contrast, diffusion in the square lattice follows a similar trend only up to \(\phi \approx 0.3\), beyond which it exhibits a pronounced enhancement with increasing confinement. Interestingly, the onset of this deviation coincides with the packing fraction at which the tortuosity of the random medium becomes significantly larger than that of the ordered lattice, see Fig.~\ref{fig:Tor}. As discussed in Ref.~\cite{fazelzadeh2023active}, the enhanced diffusion of semiflexible filaments in ordered media originates from persistent directed motion along the straight narrow channels formed by the square lattice, see the top-right snapshot in Fig.~\ref{fig:snapshot}.

\begin{figure}
    \centering
    \includegraphics[width=0.7\linewidth]{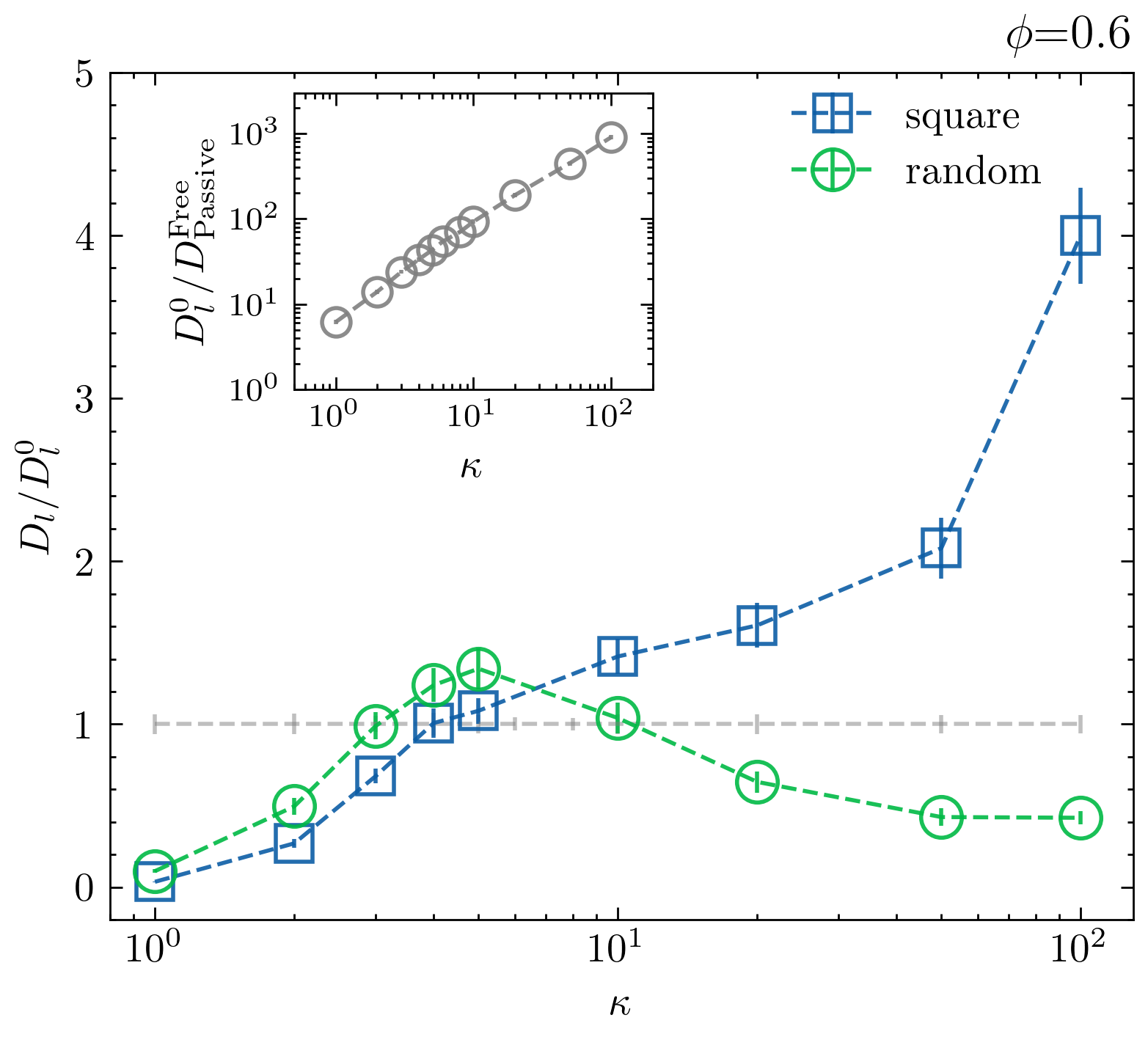}
    \caption{Normalised long-time diffusion coefficient of polymer centre of mass, \(D_l(\kappa,\phi=0.6)/D_l^{0}(\kappa)\), as a function of polymer bending stiffness \(\kappa\) at obstacle packing fraction \(\phi=0.6\). Inset: the corresponding free-space long-time diffusion coefficient \(D_l^0(\kappa)\) normalised by the passive free-space diffusion coefficient \(D^\text{Free}_\text{Passive}=0.01\).}
    \label{fig:dl-k-phi0.6}
\end{figure}

In summary, the long-time diffusion maps indicate that ordered and disordered obstacle environments give rise to qualitatively similar transport behaviour for flexible active polymers, but 
exhibit contrasting trends for semiflexible polymers at high packing fractions \(\phi > 0.3\). This behaviour is illustrated in Fig.~\ref{fig:dl-k-phi0.6}, which shows the long-time diffusion at \(\phi = 0.6\), normalised by its free-space value $D^0_l$ at the same active force \(f^a = 0.1\), as a function of bending stiffness. At high packing fractions, the transport of highly and moderately flexible active filaments depends only weakly on the obstacle arrangement. In this regime, the long-time diffusion is slightly larger in disordered media and exhibits a modest enhancement relative to free space around \(\kappa=5\). In contrast, for semiflexible polymers (\(\kappa \gtrsim 10\)), the geometry of the porous medium strongly affects transport. While dense ordered media enhance long-time diffusion, disordered media suppress it, although less strongly compared to highly flexible filaments. As we will discuss in detail in the following sections, these distinct behaviours in different flexibility regimes arise from fundamentally different transport mechanisms. Flexible filaments primarily move through a sequence of trapping and hopping events, whereas semiflexible filaments exhibit channel-guided motion.

\section{Effects of porous media on conformation and orientational dynamics}
\label{sec:results}
\subsection{Conformational properties}
\label{subsec:conformation}
\begin{figure}
    \begin{parbox}[c]{0.05\linewidth}
        \centering
        \rotatebox{90}{~~~~~~~~square}
    \end{parbox}
    \hfill
    \begin{overpic}[width=0.3\linewidth]{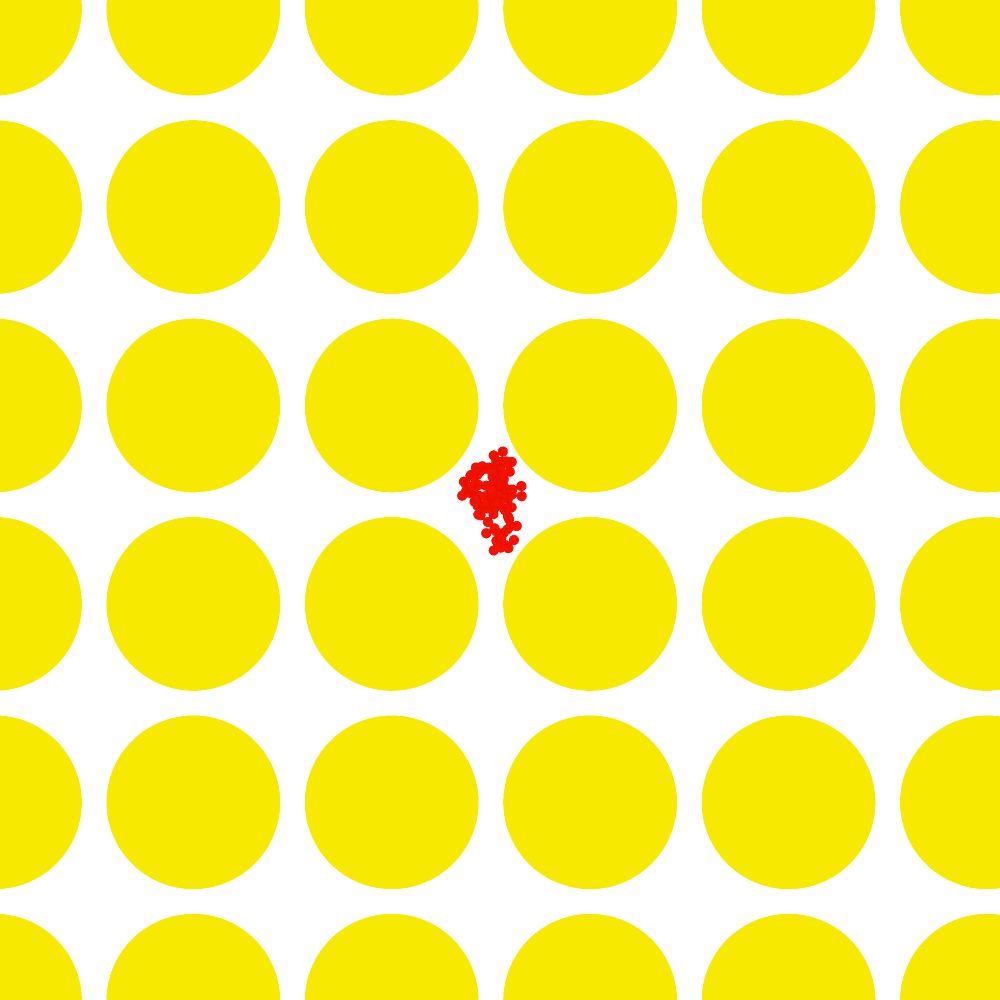}
        \put(35,105){\textbf{\(\kappa=1\)}}
    \end{overpic}
    \hfill
    \begin{overpic}[width=0.3\linewidth]{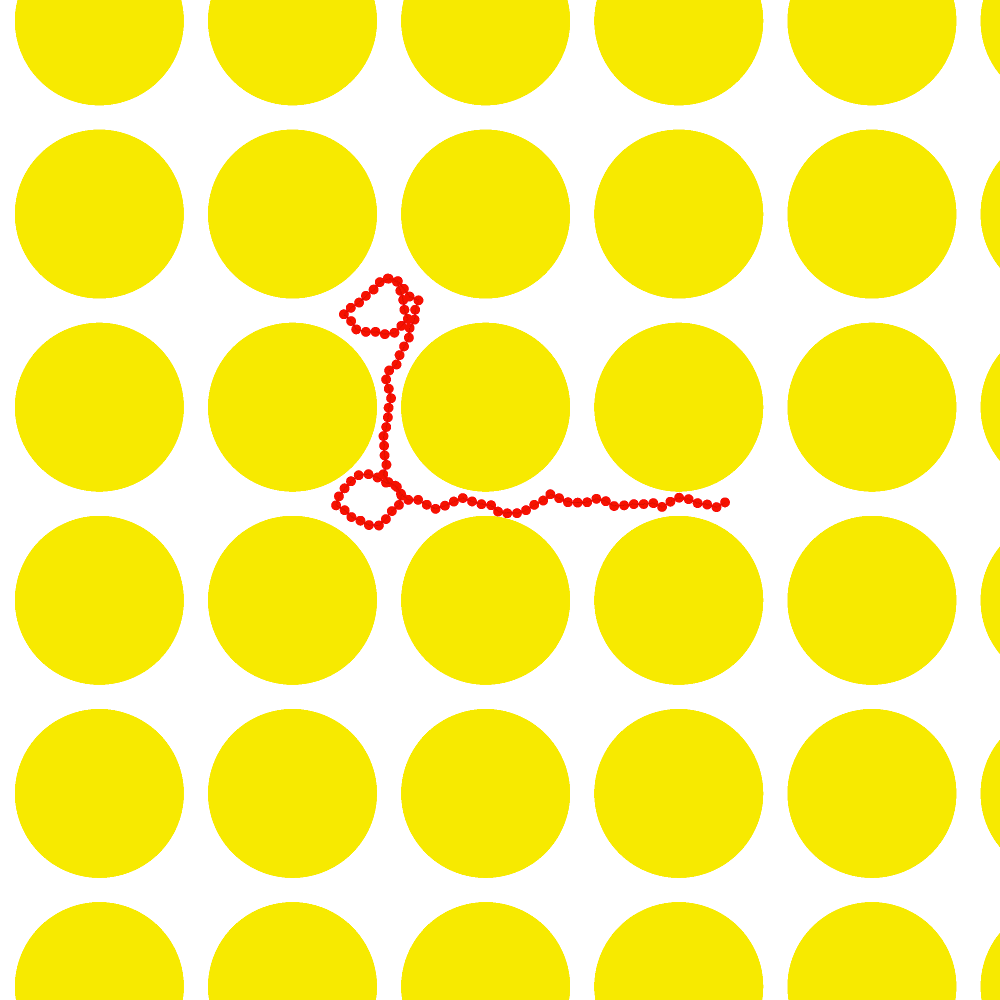}
        \put(35,105){\textbf{\(\kappa=5\)}}
    \end{overpic}
    \hfill
    \begin{overpic}[width=0.3\linewidth]{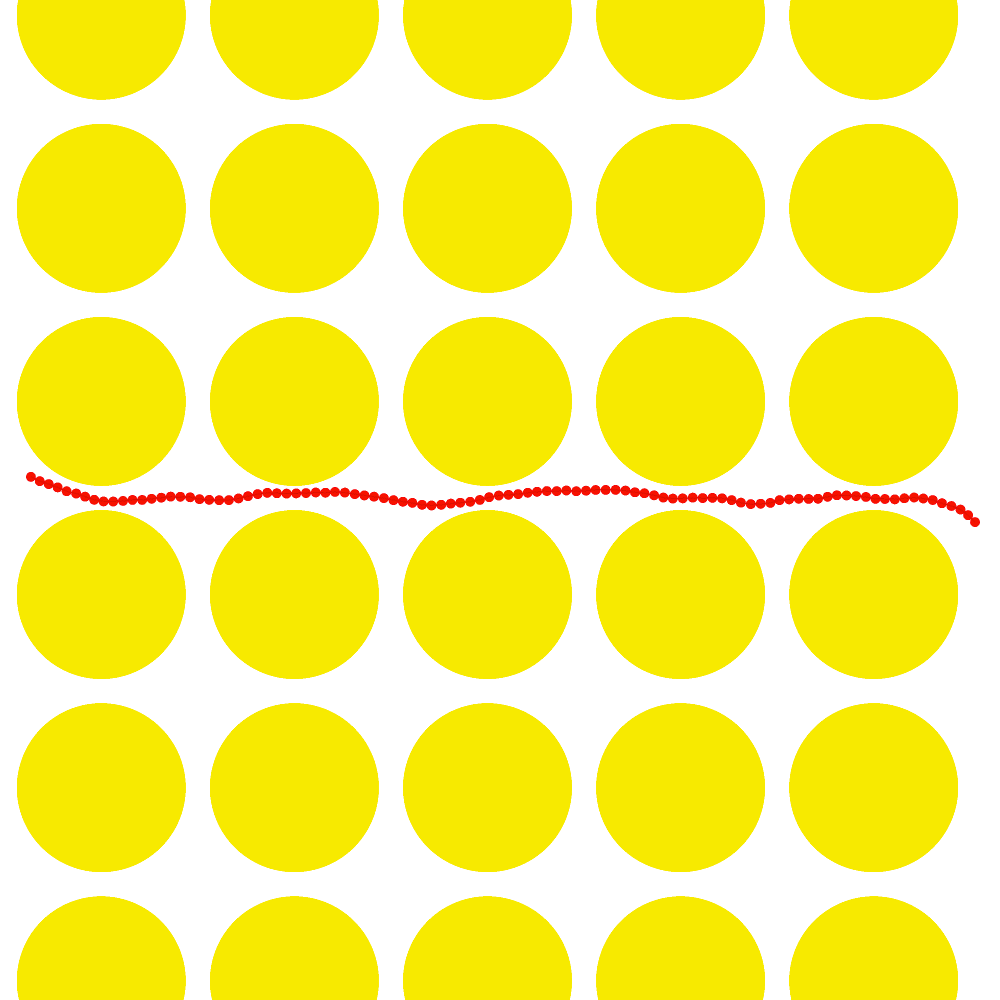}
        \put(30,105){\textbf{\(\kappa=100\)}}
    \end{overpic}\\[2mm]
    \hfill
    \begin{parbox}[c]{0.05\linewidth}
        \centering
        \rotatebox{90}{~~~~~~~random}
    \end{parbox}
    \hfill
    \begin{overpic}[width=0.3\linewidth]{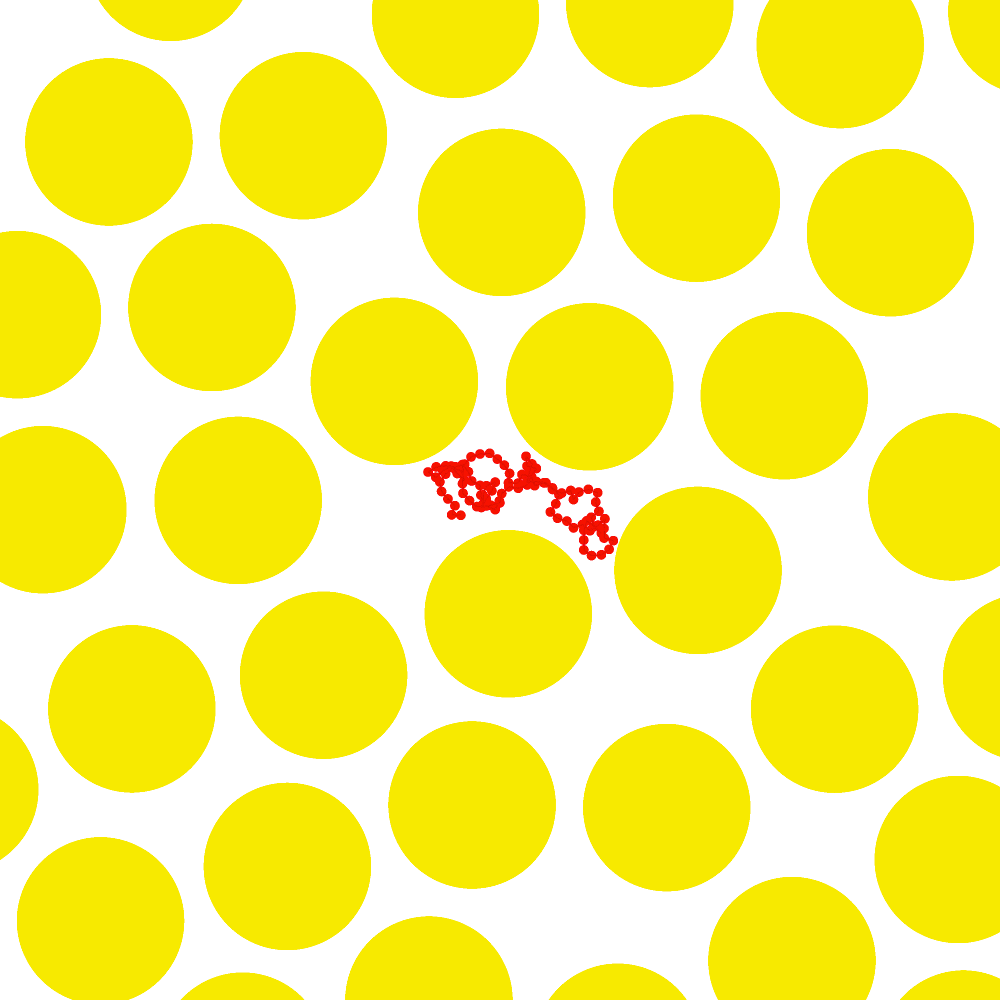}
    \end{overpic}
    \hfill
    \begin{overpic}[width=0.3\linewidth]{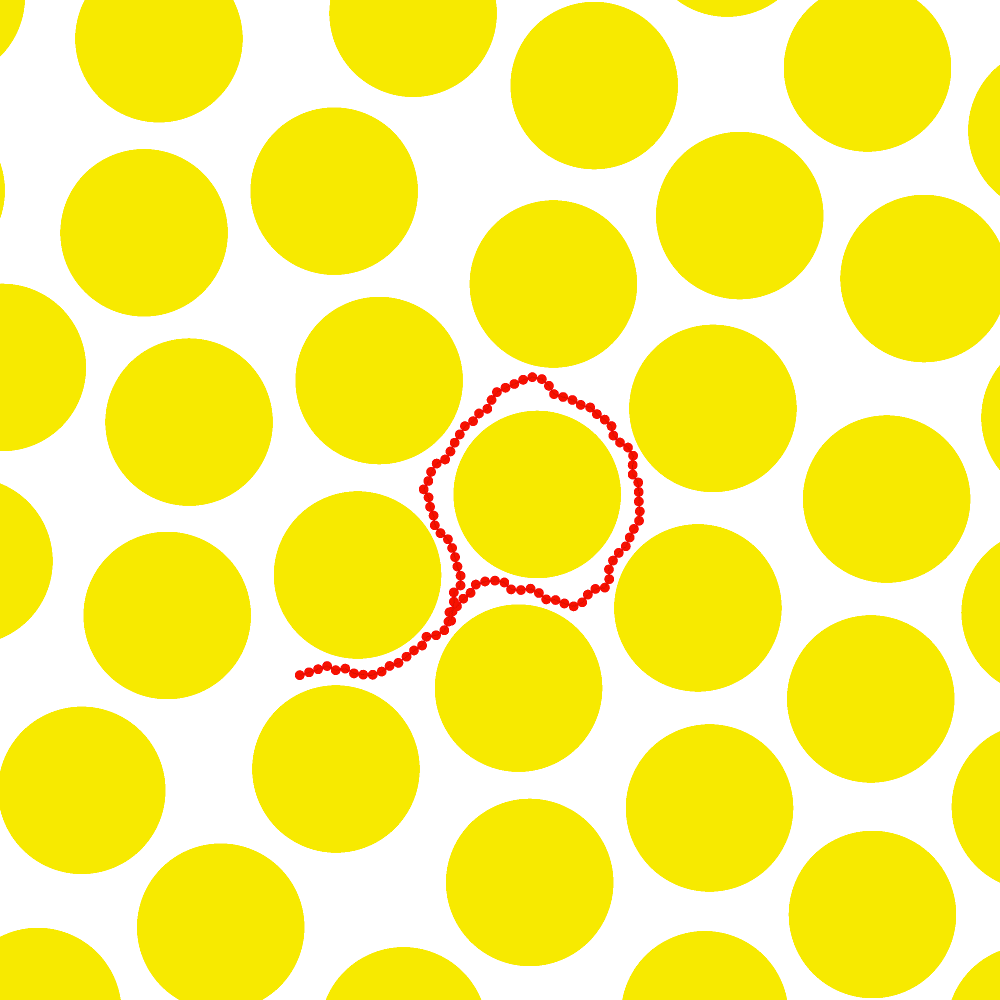}
    \end{overpic}
    \hfill
    \begin{overpic}[width=0.3\linewidth]{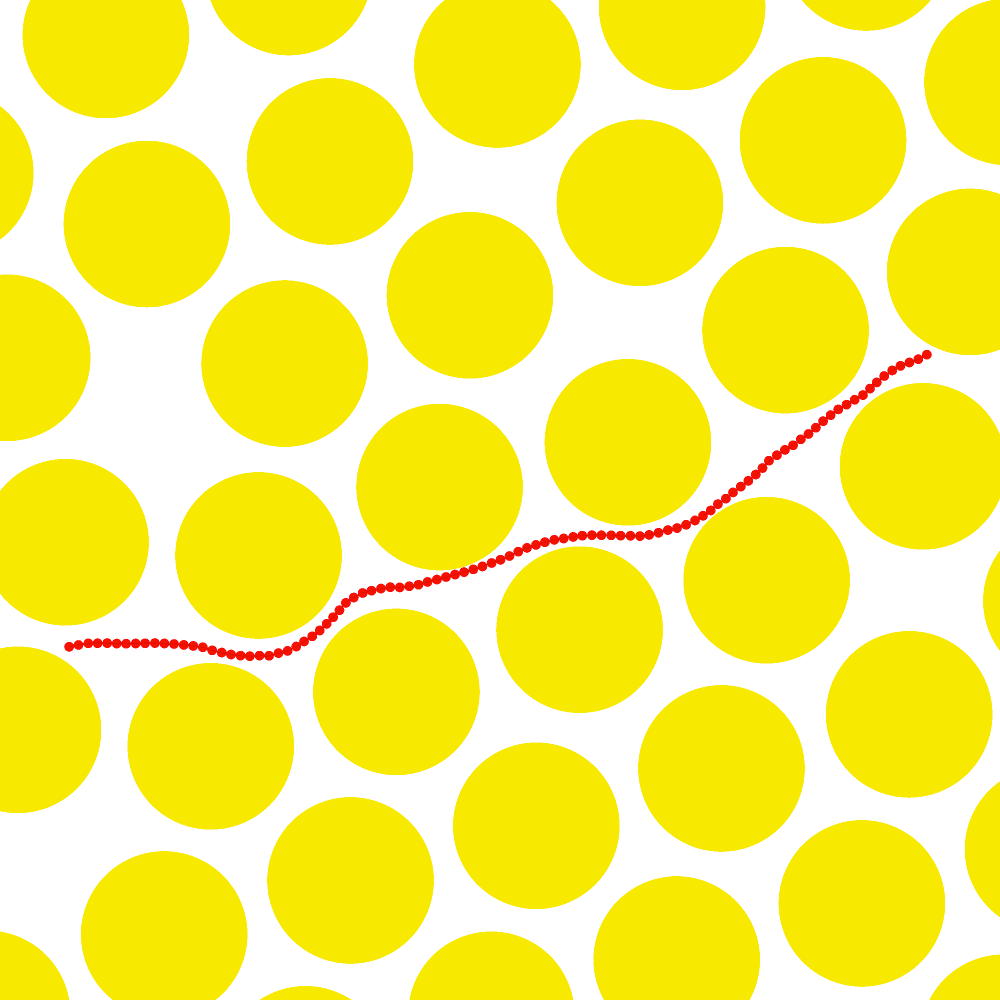}
    \end{overpic}
    \caption{Representative simulation snapshots of 2D phantom active polymers with different bending stiffness values \(\kappa\) in ordered (top) and disordered (bottom) obstacle arrays at \(\phi=0.6\). Corresponding movies are provided in the Supplemental Material.} 
    \label{fig:snapshot}
\end{figure}

In this section, we investigate how the obstacle configuration influences the conformational properties of active filaments. We begin with a visual inspection of active filaments in ordered and disordered obstacle arrays under strong confinement at $\phi=0.6$, where the most pronounced changes in diffusive transport are observed. Figure~\ref{fig:snapshot} shows simulation snapshots of active polymers with bending stiffness \(\kappa = 1, 5,\) and \(100\) in both ordered and disordered obstacle arrays at \(\phi = 0.6\).

From these snapshots, we make the following qualitative observations: (i) highly flexible polymers (\(\kappa = 1\)) adopt collapsed conformations and remain primarily trapped within pore spaces in both ordered and disordered media with occasional hopping events in which polymers transiently extend to pass through the narrow channels; (ii) moderately flexible polymers (\(\kappa = 5\)) whose persistence length is comparable to the obstacle size bend around multiple obstacles to conform to the tight curvilinear channels of the porous environment 
and (iii) semiflexible polymers (\(\kappa = 100\)) in the square lattice maintain extended conformations and preferentially align with straight channels, whereas in disordered media they undergo stronger bending to adapt to tortuous, curvilinear pathways.

To quantify the effect of porous media on conformations, we examine the distribution of end-to-end distance, namely the magnitude of the end-to-end vector \(\mathbf{R}_{\text{e}}(t)\) defined as the vector going from the tail to the head bead at time \(t\):
\begin{equation}
    \label{eq:Re}
    \mathbf{R_{\text{e}}}(t) = \mathbf{r}_N(t)-\mathbf{r}_1(t).
\end{equation}

\begin{figure} 
    \centering
    \includegraphics[width=0.99\linewidth]{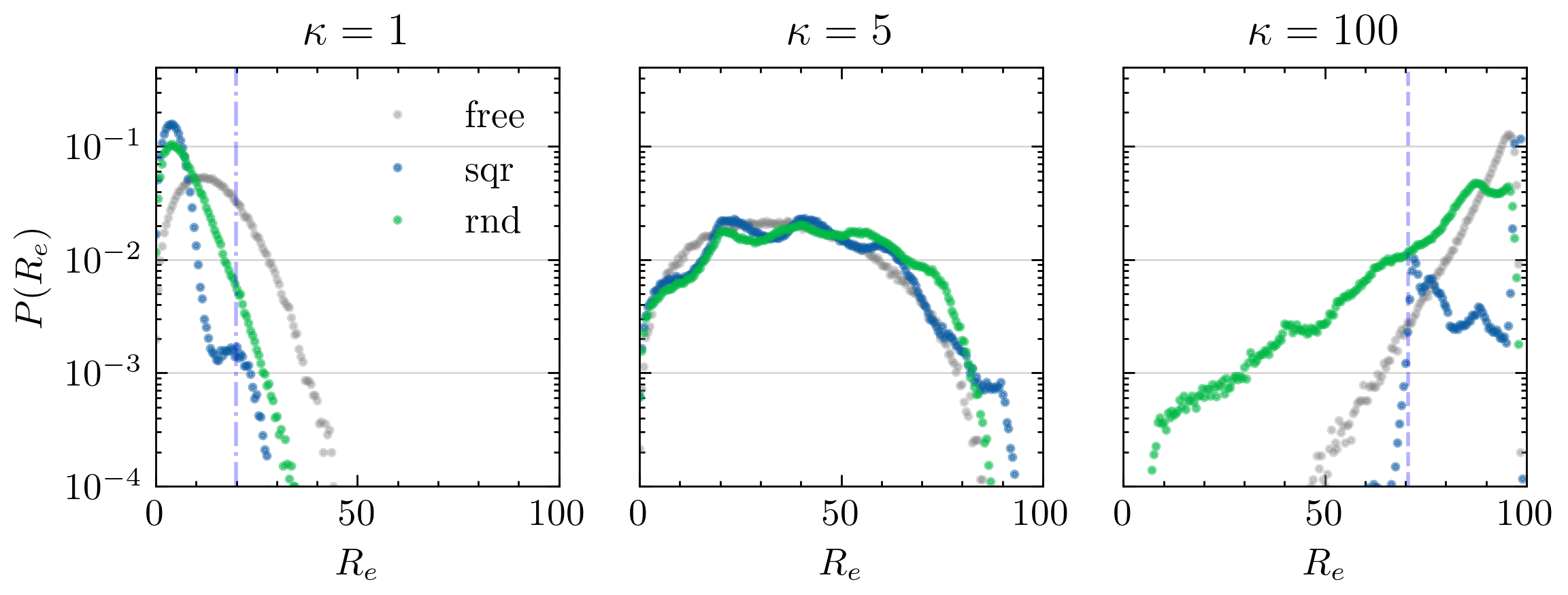}
    \caption{Probability density function of end-to-end distance \(R_{\text{e}}\) for \(\kappa=1,5,100\) at \(\phi=0.6\). The vertical blue line in the left panel indicates the lattice spacing, and the vertical blue line in the right panel marks the minimum \(R_{\text{e}}\) value corresponding to a \(90^\circ\) turning motion of the semiflexible polymer in square lattice channels.}
    \label{fig:PRe}
\end{figure}

Figure~\ref{fig:PRe} shows the probability density of the end-to-end distance at \(\phi = 0.6\) for three bending stiffness values, \(\kappa = 1, 5,\) and \(100\). For flexible polymers with \(\kappa = 1\), the distributions of \(R_{\text{e}}\) in both ordered and disordered media are shifted toward smaller values and become narrower relative to freely moving polymers. Notably, for a square lattice, a secondary peak appears due to the fixed hopping length between lattice sites. For the moderate flexibility regime with \(\kappa = 5\), the distributions are similar to those of active polymers in free space, with additional secondary peaks and a slight shift toward larger values. These secondary features stem from the geometry of the pore space. They correspond to characteristic conformations whose end-to-end distances coincide with peaks of the two-point correlation function \(S_2(r)\), which measures the probability that two points separated by a distance $r$ both lie within the pore space.
For semiflexible polymers with \(\kappa = 100\), the distribution of \(R_{\text{e}}\) is substantially broader in random media than in square-lattice arrays, reflecting enhanced bending induced by the tortuous pore structure.  In square lattices, the distribution is dominated by a sharp peak near \(R_{\text{e}} = L=100\), corresponding to filaments that remain nearly straight and aligned with the axial channels of the lattice. A smaller secondary peak appears near \(\frac{\sqrt{2}}{2}L\), consistent with channel-switching conformations in which the filament forms an approximately \(90^\circ\) bend and distributes itself nearly equally between two perpendicular channels.

\begin{figure}
    \centering
    \begin{overpic}[width=0.48\linewidth]{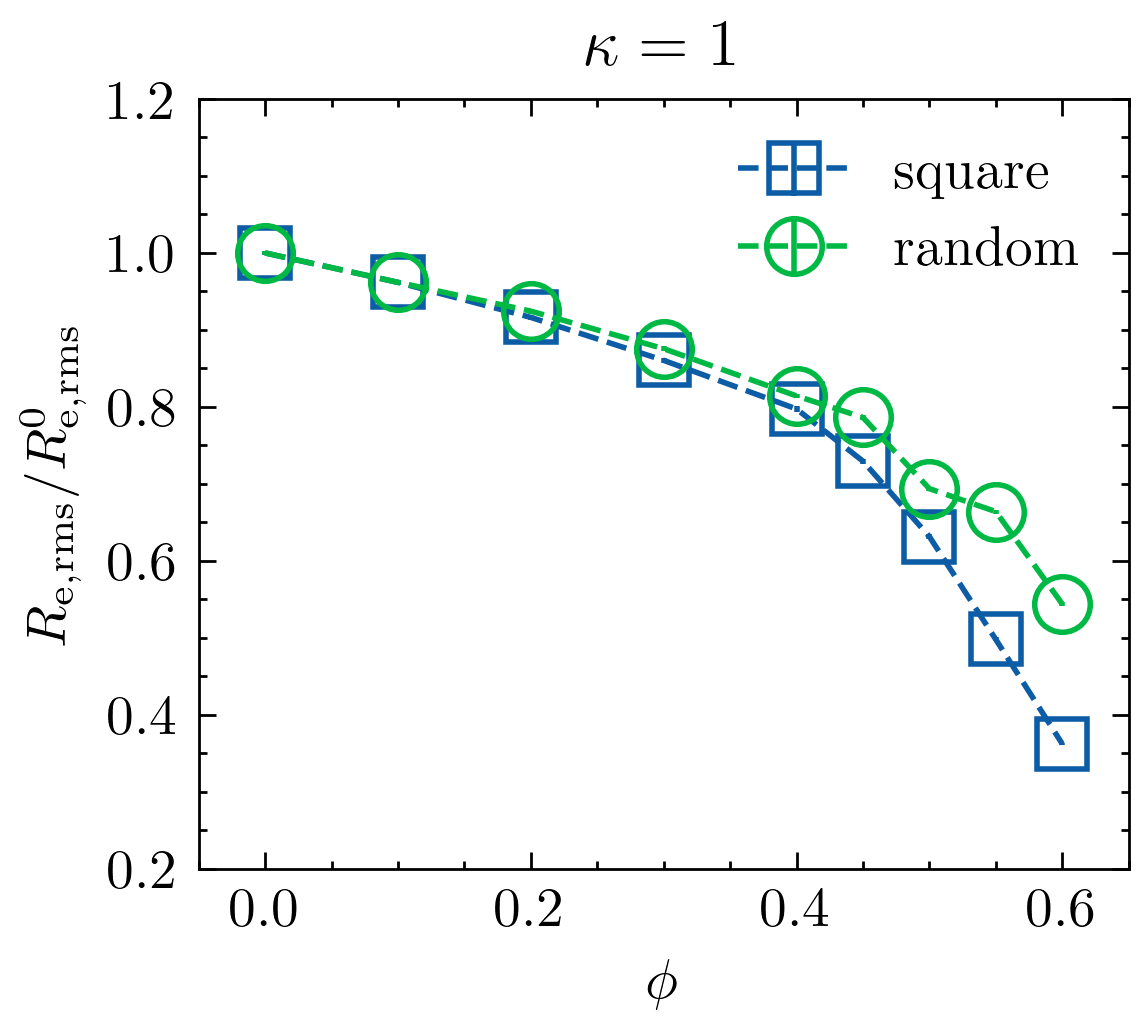}
        \put(2,86.5){\textbf{(a)}}
    \end{overpic}
    \hfill
    \begin{overpic}[width=0.49\linewidth]{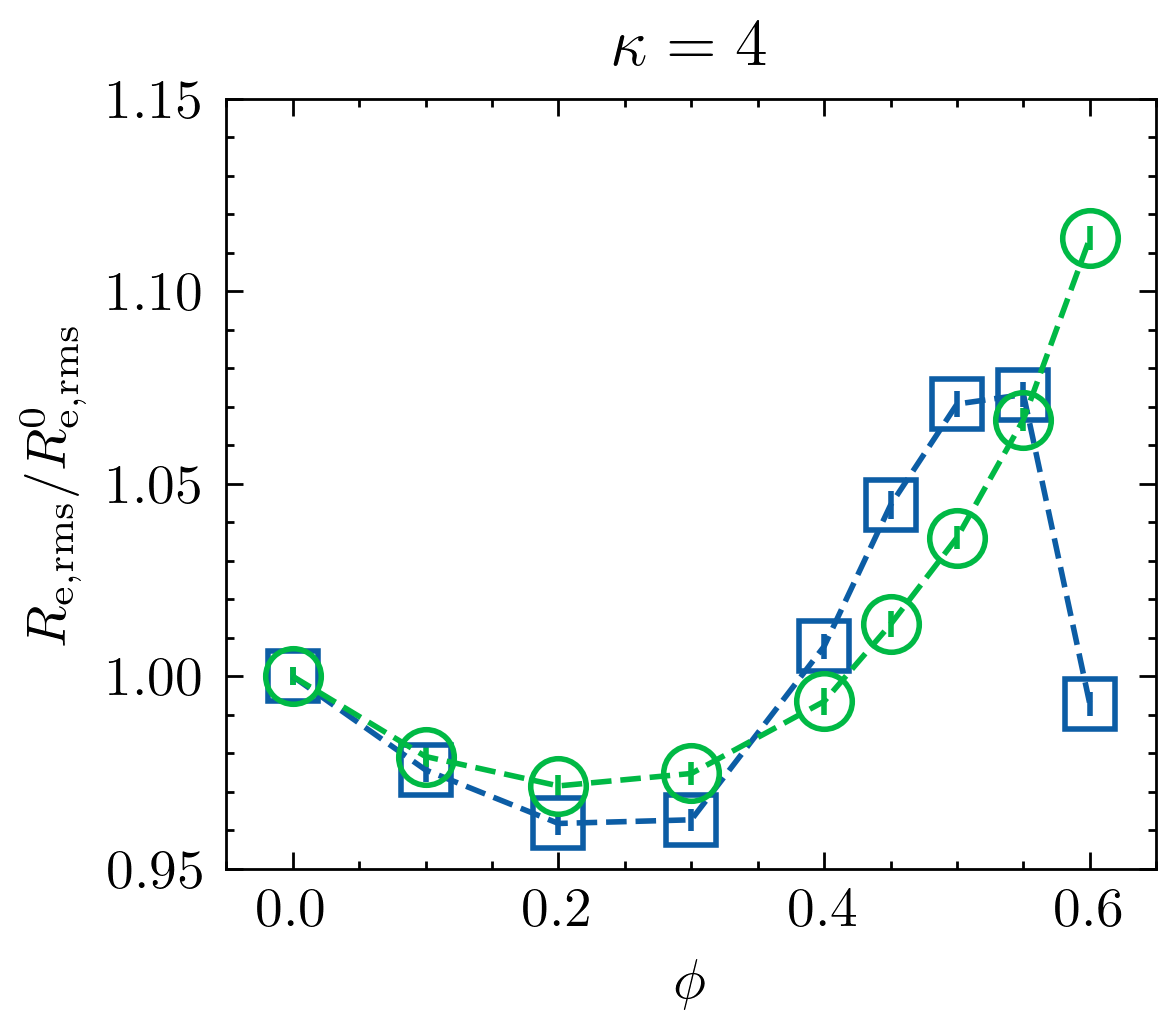}
        \put(2,85){\textbf{(b)}}
    \end{overpic}
    \hfill
    \begin{overpic}[width=0.49\linewidth]{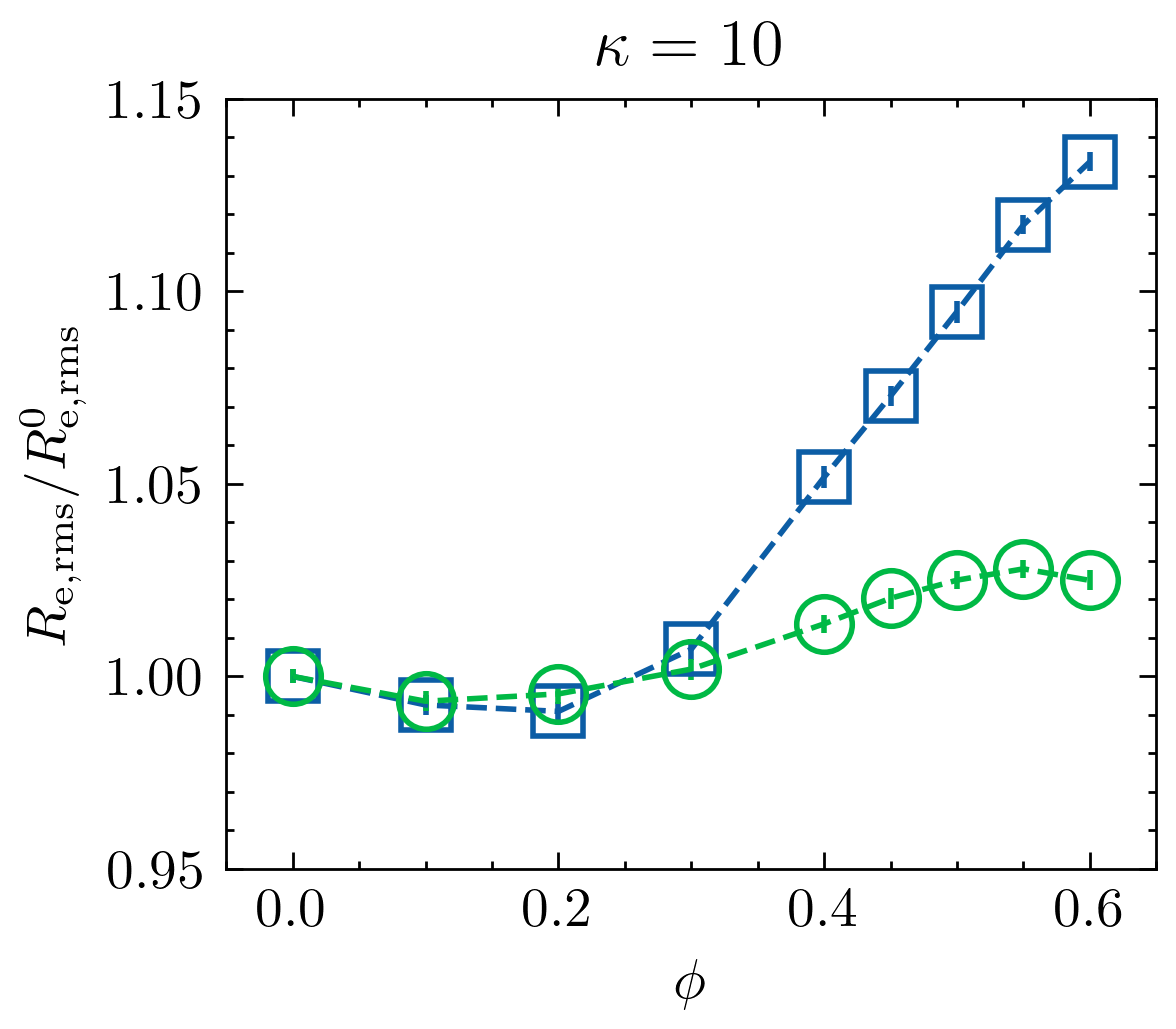}
        \put(2,85){\textbf{(c)}}
    \end{overpic}
    \hfill
    \begin{overpic}[width=0.49\linewidth]{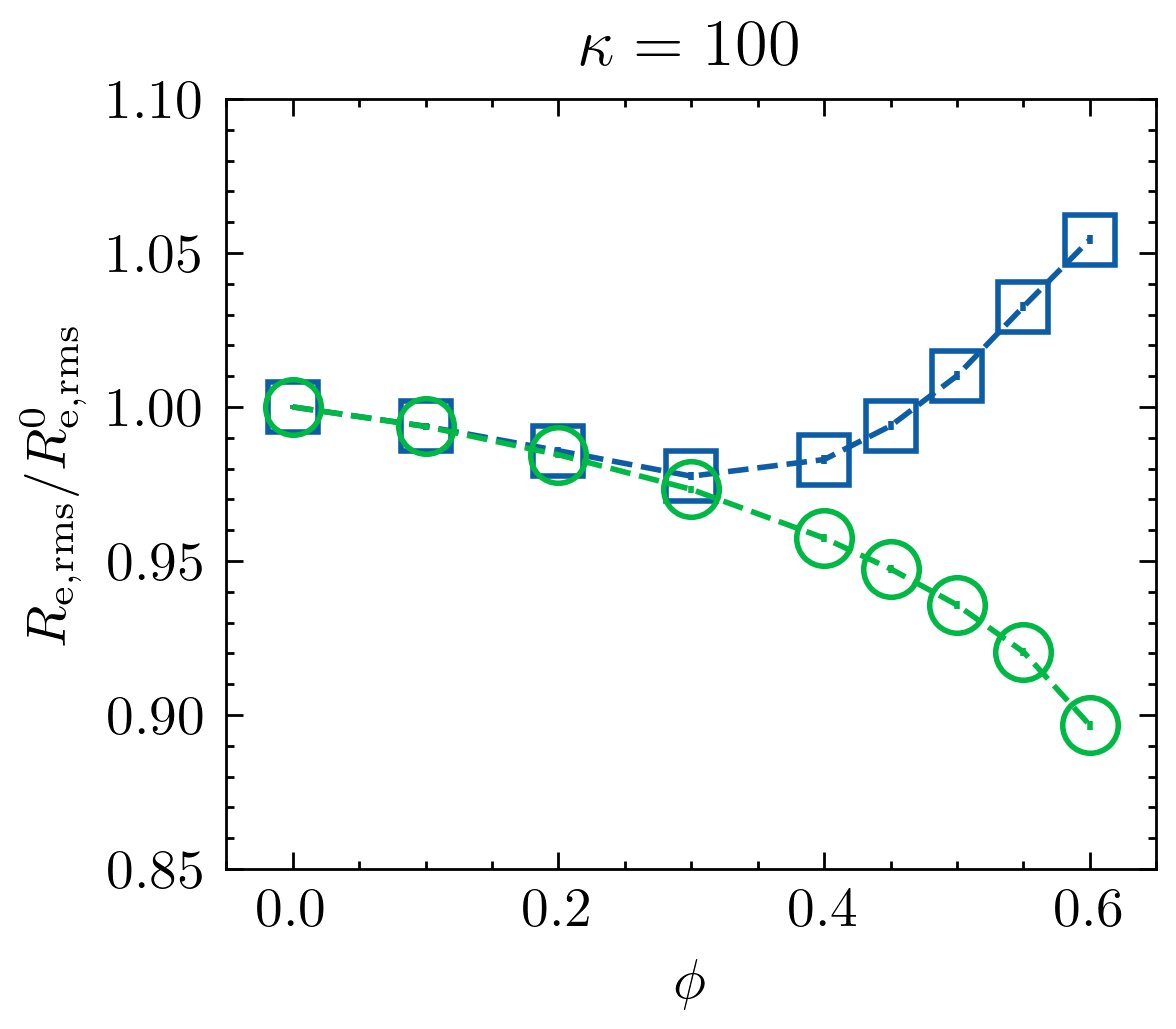}
        \put(2,85){\textbf{(d)}}
    \end{overpic}
    \caption{Normalised root-mean-square end-to-end distance, \(R_{\text{e,rms}}(\phi,\kappa)/R_{\text{e,rms}}^{0}(\kappa)\), as a function of obstacle packing fraction \(\phi\). Panels (a)--(d) correspond to polymer bending stiffness \(\kappa=1,4,10,100\), respectively. Here, \(R_{\text{e,rms}}^{0}(\kappa)\) denotes the root-mean-square end-to-end distance of active polymers in free space. The corresponding values are \(16.0,35.4,54.8,91.3\) for \(\kappa=1,4,10,100\) respectively.}
    \label{fig:re-phi-k}
\end{figure}

Having examined the effects of porous media on the distribution of end-to-end distance, next we look into the dependence of the root-mean-square end-to-end distance, 
\(R_{\text{e,rms}}=\sqrt{\langle R_{\text{e}}^2\rangle}\), on the obstacle packing fraction $\phi$. Figure~\ref{fig:re-phi-k} shows 
\(R_{\text{e,rms}}\) of active polymers normalised by their free-space values \(R_{\text{e,rms}}^0\) as a function of \(\phi\) in different flexibility regimes in both ordered and disordered media. 
For flexible polymers with \(\kappa=1\), shown in Fig.~\ref{fig:re-phi-k}(a), \(R_{\text{e,rms}}\) decreases markedly with increasing \(\phi\) 
indicating progressive chain compaction under tighter spatial confinement.  
For $\phi \gtrsim 0.4$, when the free-space radius of gyration exceeds the average pore size, the decrease is especially strong in ordered media, where the pore size is fixed by the lattice periodicity.
For polymers with moderate flexibility (\(\kappa=4\) and $10$), \(R_{\text{e,rms}}\), presented in Fig.~\ref{fig:re-phi-k}(b) and (c), exhibits a slight decrease with \(\phi\) up to \(\phi \approx 0.2\). At higher packing fractions, strong confinement leads to more extended conformations and therefore an increase in the mean end-to-end distance. An exception occurs for \(\kappa=4\) in the square lattice, where for \(\phi>0.5\) the polymers shrink again due to confinement within the narrow periodic channels. For semiflexible polymers (\(\kappa=100\)), shown in Fig.~\ref{fig:re-phi-k}(d), the trends differ qualitatively between environments. In disordered media, \(R_{\text{e,rms}}\) decreases monotonically with increasing \(\phi\). In contrast, in the square lattice it initially follows a similar decreasing trend at low packing fractions, but transitions to a stretched conformation for \(\phi>0.3\). We note that for all active polymers with $\kappa \ge 4$, the changes in mean end-to-end distance are at most \(15\%\), nevertheless they are visible and affect the polymer transport as will become clear from the analytical theory presented in Sec.~\ref{sec:theory}.

Additional simulations in random media with different obstacle radii show that the intermediate-\(\kappa\) maximum in \(R_{\text{e,rms}}/R_{\text{e,rms}}^{0}\) shifts with obstacle size, consistent with geometric matching between polymer flexibility and pore-space length scales (Appendix~\ref{sec:midk}, Fig.~\ref{appfig:k-r-relation}).

\subsection{Reorientational dynamics}
\label{subsec:dynamics}
To characterise the orientational dynamics of active polymers, we compute the normalised time autocorrelation function (TACF) of the end-to-end vector:
\begin{equation}
    \label{eq:tacf}
    \hat{C}_{\text{e}}(t) = \frac{\langle \mathbf{R}_{\text{e}}(0) \cdot \mathbf{R}_{\text{e}}(t) \rangle}{\langle R_{\text{e}}^2 \rangle}.
\end{equation}
This quantity measures the persistence of the active polymer orientation, which defines the direction of total active force and probes the timescale over which the filament loses memory of its initial direction due to conformational changes, and interactions with the confining environment.

\begin{figure}
    \centering
    \begin{overpic}[width=0.49\linewidth]{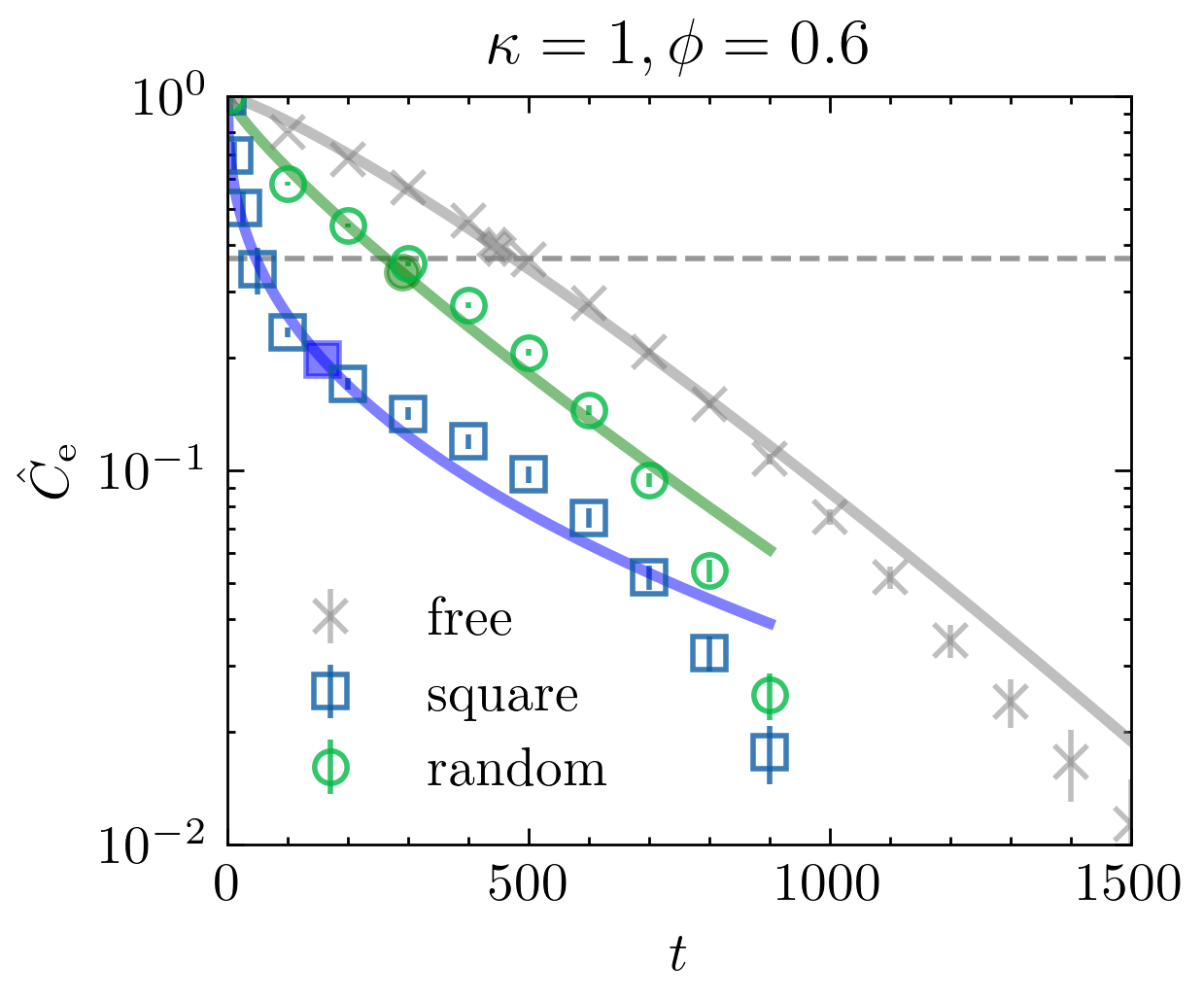}
        \put(2,82){\textbf{(a)}}
    \end{overpic}
    \hfill
    \begin{overpic}[width=0.49\linewidth]{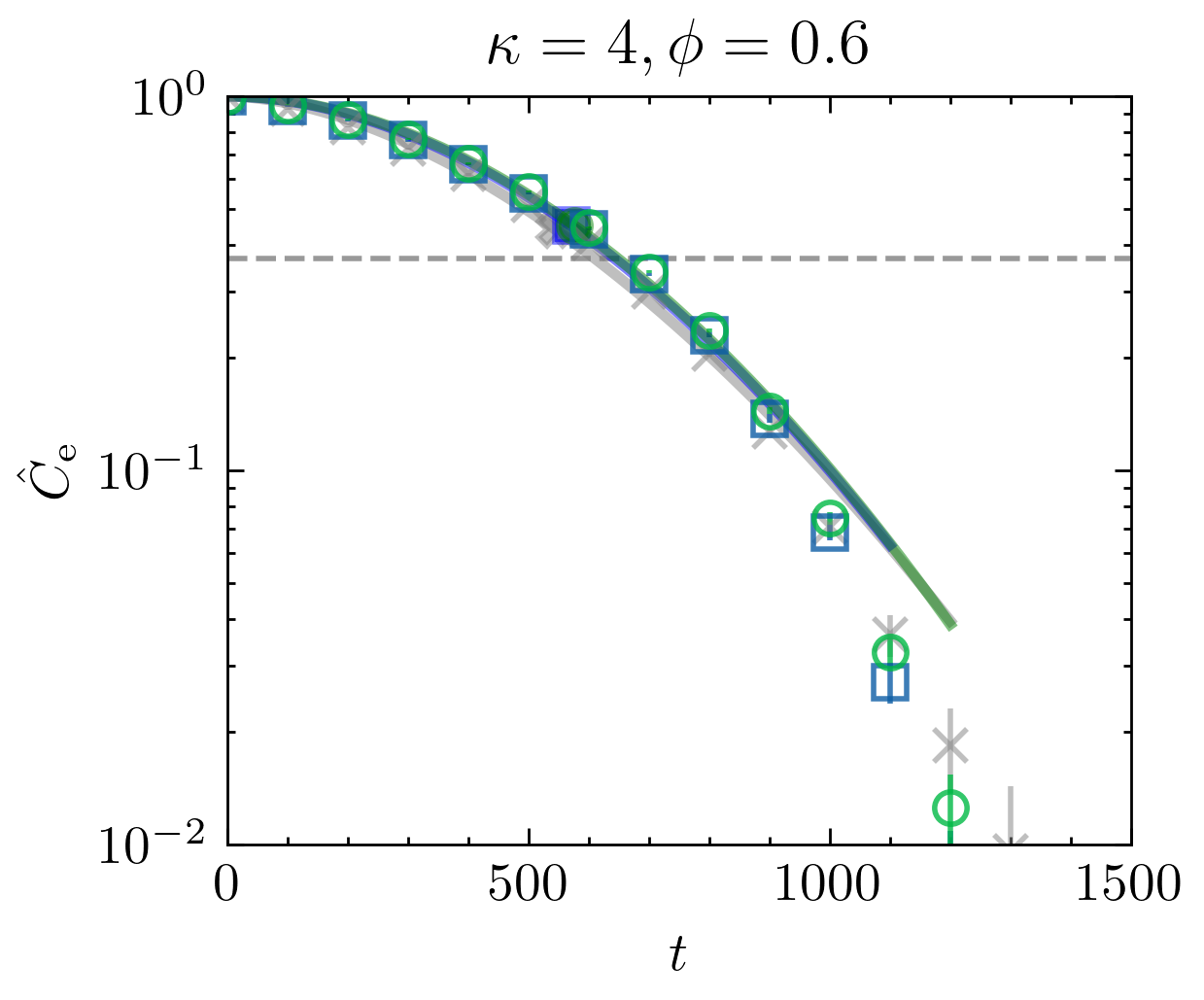}
        \put(2,82){\textbf{(b)}}
    \end{overpic}
    \hfill
    \begin{overpic}[width=0.49\linewidth]{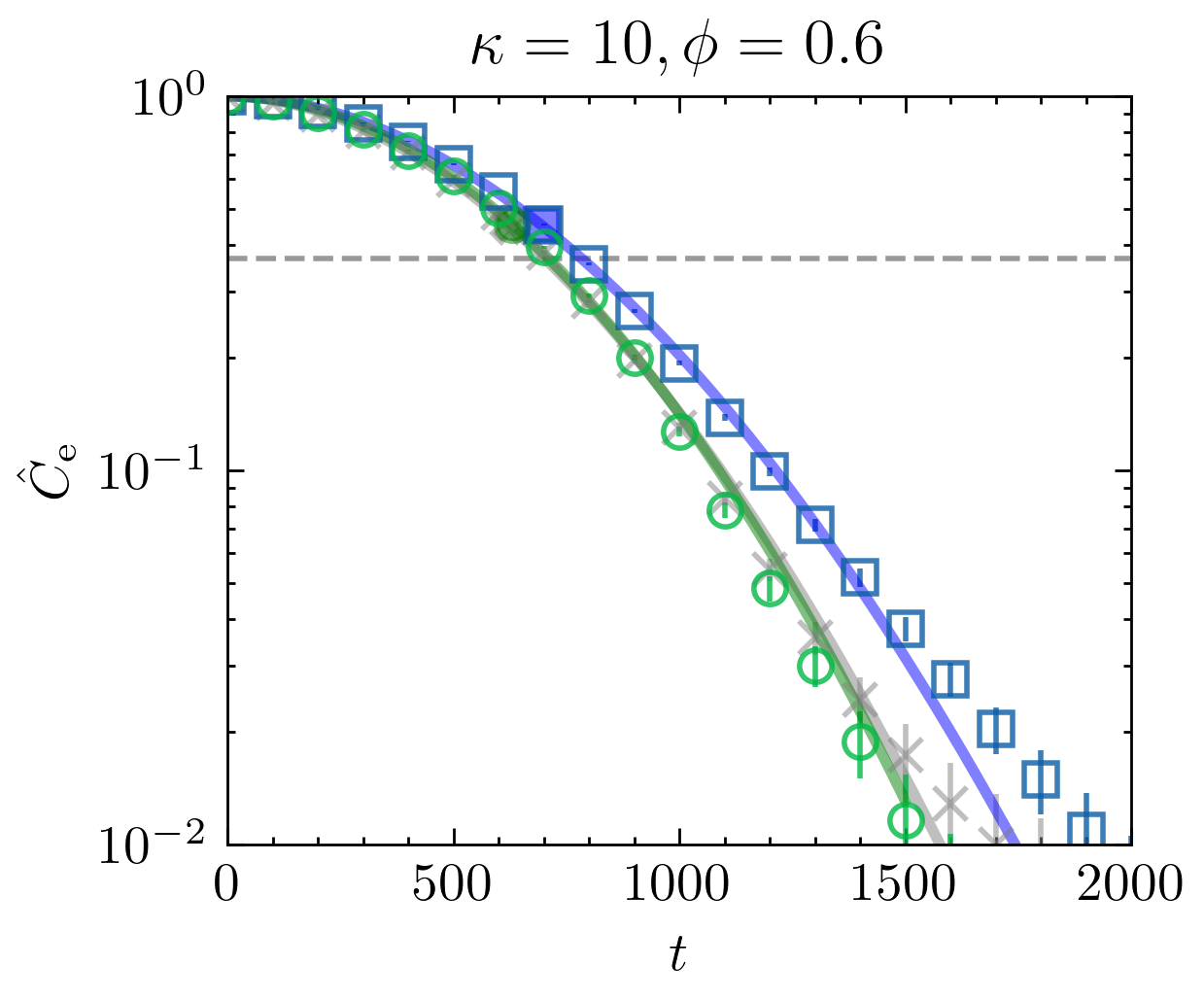}
        \put(2,82){\textbf{(c)}}
    \end{overpic}
    \hfill
    \begin{overpic}[width=0.49\linewidth]{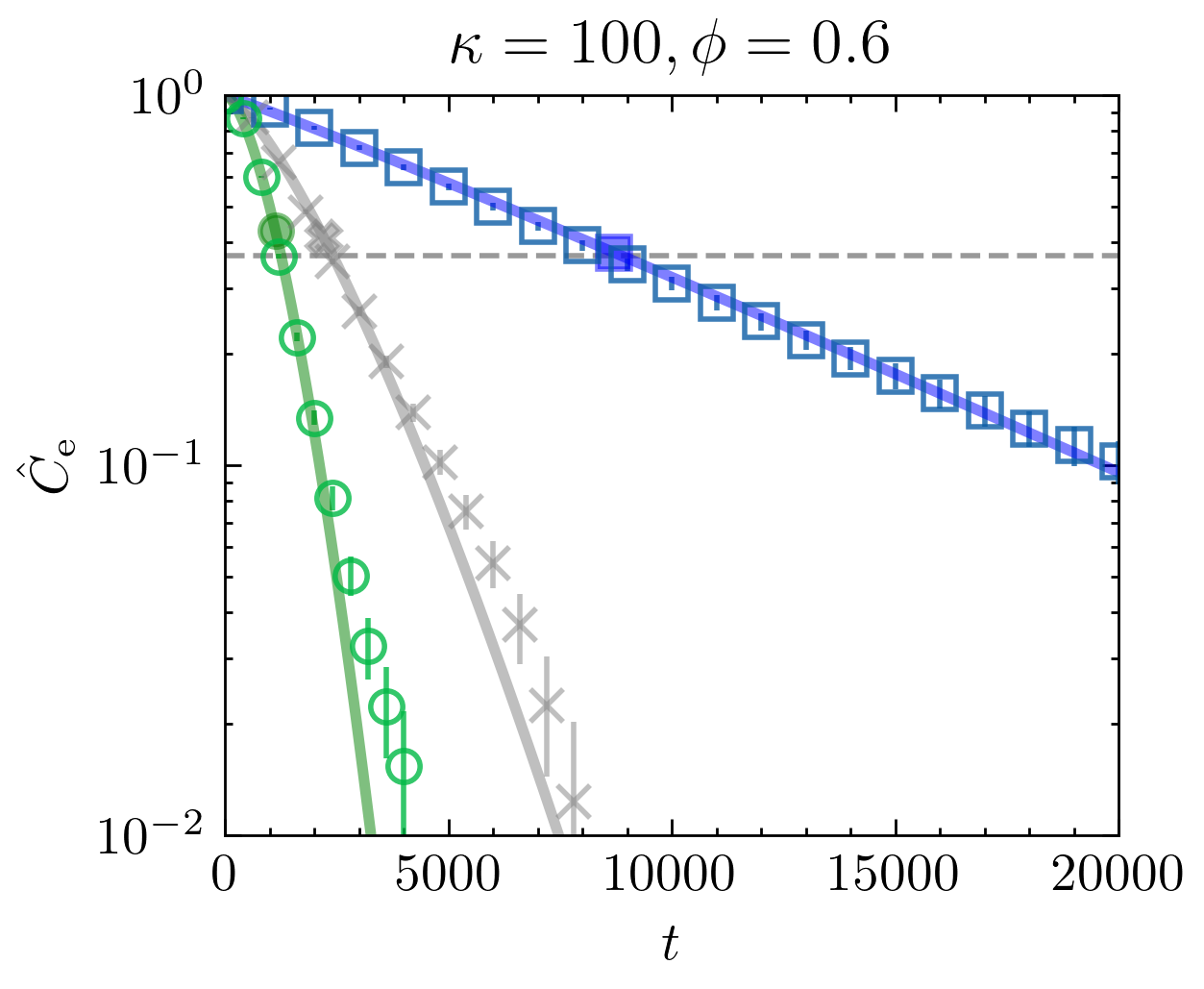}
        \put(2,82){\textbf{(d)}}
    \end{overpic}
    \caption{Normalised time autocorrelation function of the polymer end-to-end vector defined by Eq.~\eqref{eq:tacf} at \(\phi=0.6\). Panels (a)--(d) correspond to polymer bending stiffness values \(\kappa=1,4,10,100\), respectively. Solid lines represent fits with compressed or stretched exponential, and solid symbols indicate the mean relaxation time \(\langle\tau\rangle\).}
    \label{fig:Ce-t-k}
\end{figure}

Figure~\ref{fig:Ce-t-k} shows \(\hat{C}_{\text{e}}(t)\) at \(\phi=0.6\) for different flexibility regimes corresponding to bending rigidity values $\kappa=1,4,10$, and $100$ in both ordered and disordered media. For all flexibility regimes and obstacle configurations, the decay of the TACF is reasonably well described by a stretched- or compressed-exponential of the form
\begin{equation}
     \hat{C}_{\text{e}}(t) \approx \exp\!\left[-\left(\frac{t}{\tau_k}\right)^\beta\right],
\end{equation}
as indicated by the solid lines in Fig.~\ref{fig:Ce-t-k}. Here, \(\tau_k\) denotes a characteristic relaxation timescale, and \(0<\beta<1\) corresponds to a stretched exponential, whereas \(1<\beta<2\) corresponds to a compressed exponential. For \(\beta=1\), the expression reduces to a simple exponential decay associated with a single relaxation timescale, whereas \(\beta \neq 1\) quantifies deviations from single-timescale orientational relaxation. The stretched-exponential case \(\beta<1\) reflects the presence of a broad spectrum of slow relaxation modes and long-lived orientational memory. In porous environments, such behaviour typically arises from heterogeneous confinement, trapping events, and the coexistence of multiple filament relaxation modes. In contrast, a compressed-exponential decay with \(\beta>1\) implies a coherent or superdiffusive relaxation faster than exponential. Such behaviour is commonly associated with driven, out-of-equilibrium systems. 
 
\begin{figure}
    \centering
    \begin{overpic}[width=0.49\linewidth]{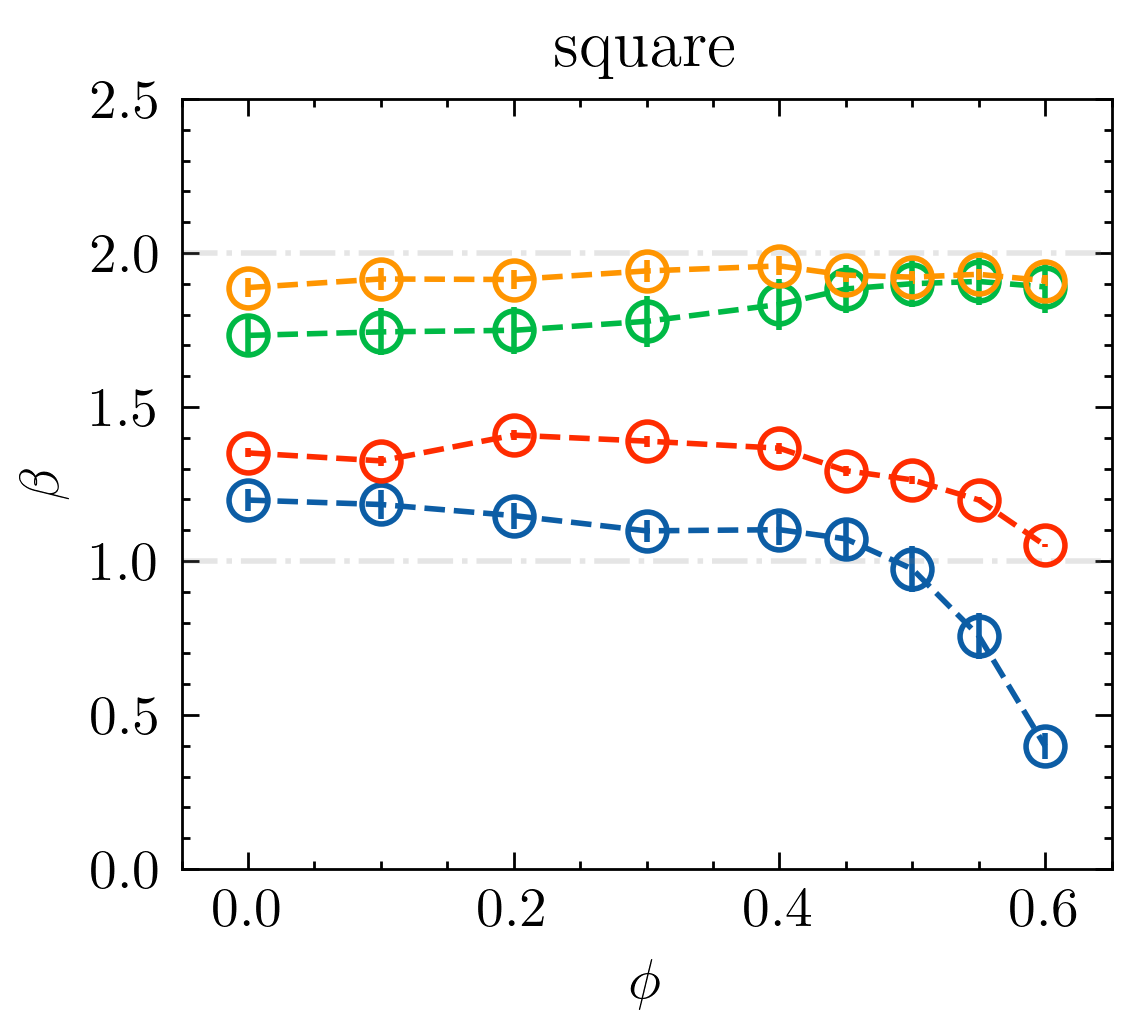}
        \put(2,87){\textbf{(a)}}
    \end{overpic}
    \hfill
    \begin{overpic}[width=0.49\linewidth]{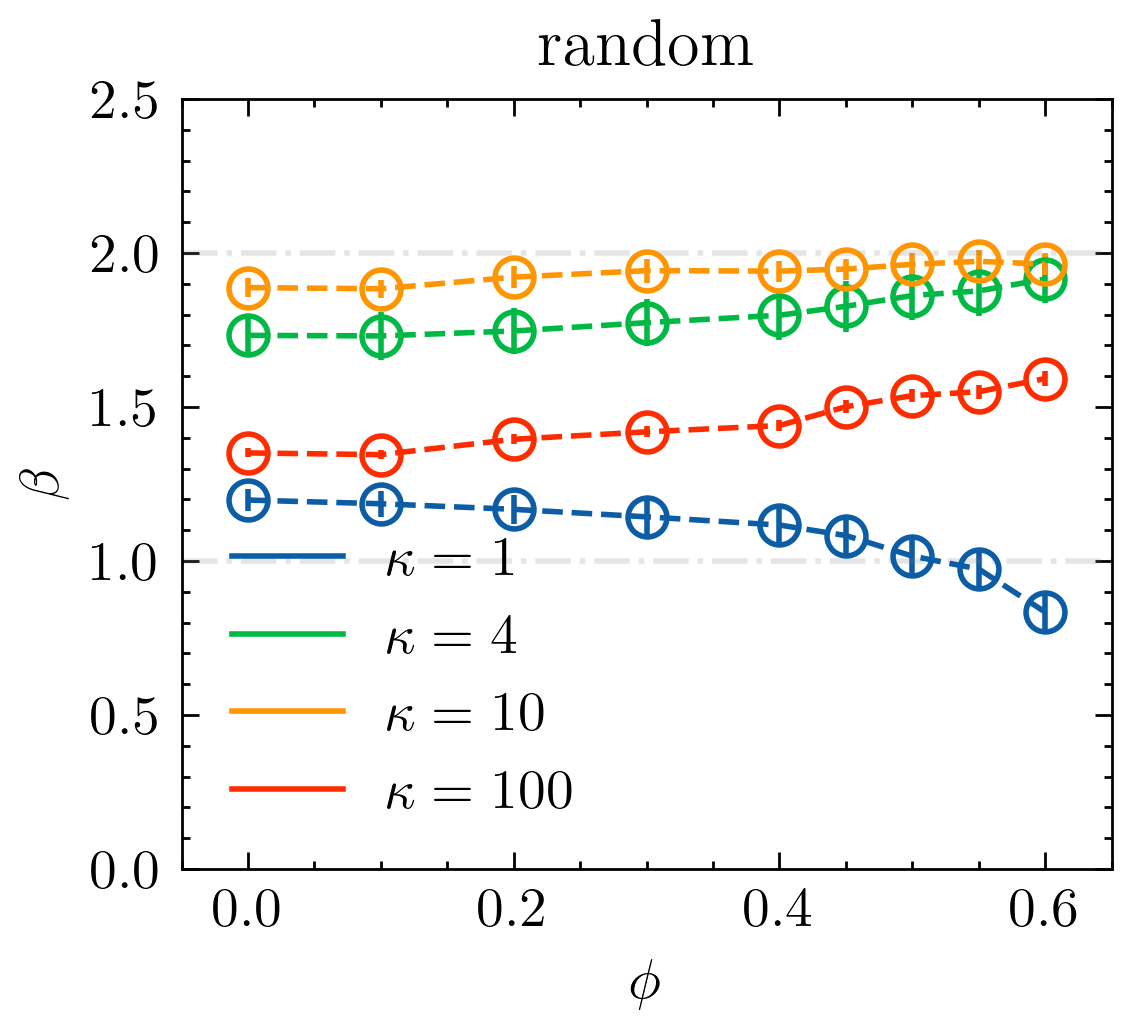}
        \put(2,87){\textbf{(b)}}
    \end{overpic}
    \begin{overpic}[width=0.49\linewidth]{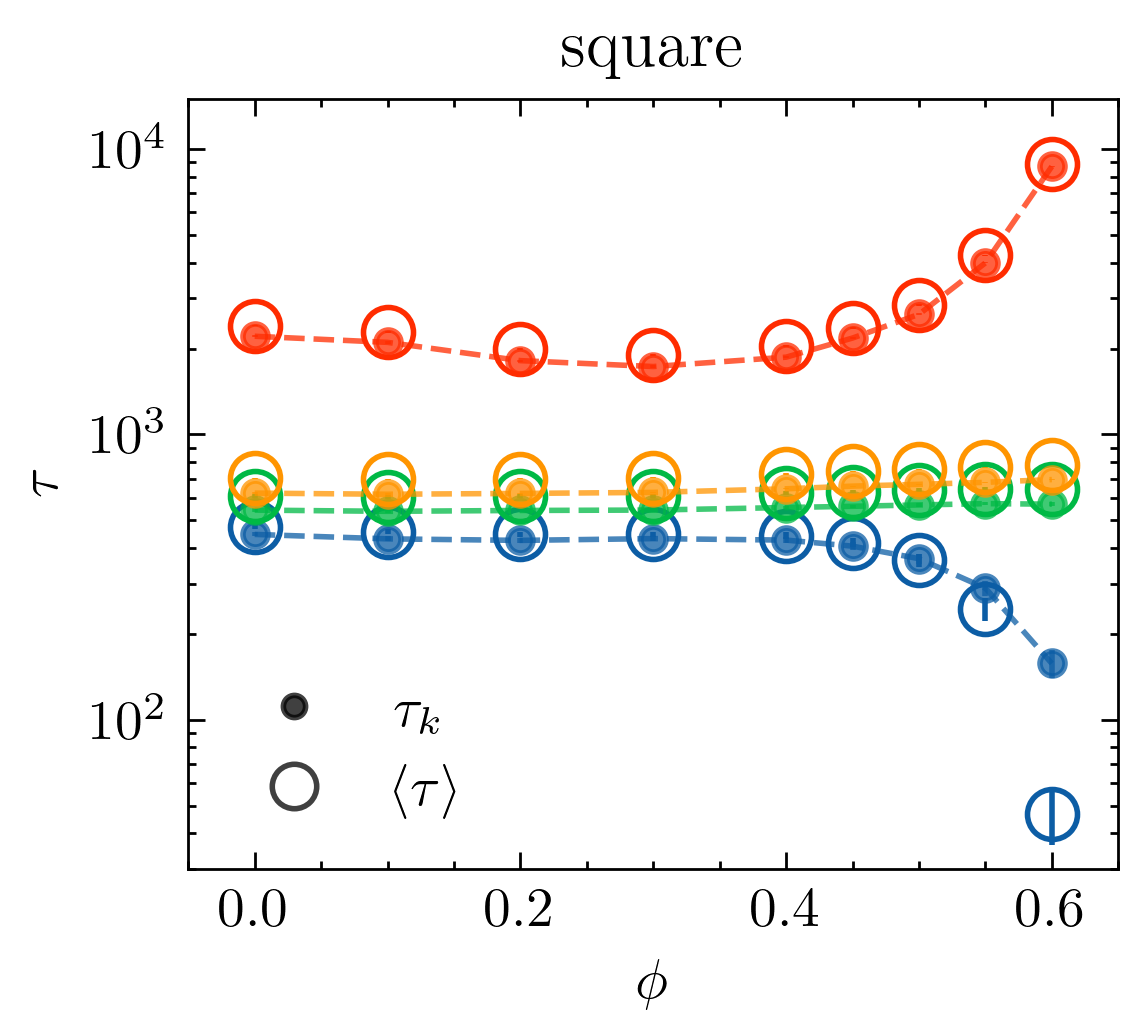}
     \hfill
        \put(2,87){\textbf{(c)}}
    \end{overpic}
    \hfill
    \begin{overpic}[width=0.49\linewidth]{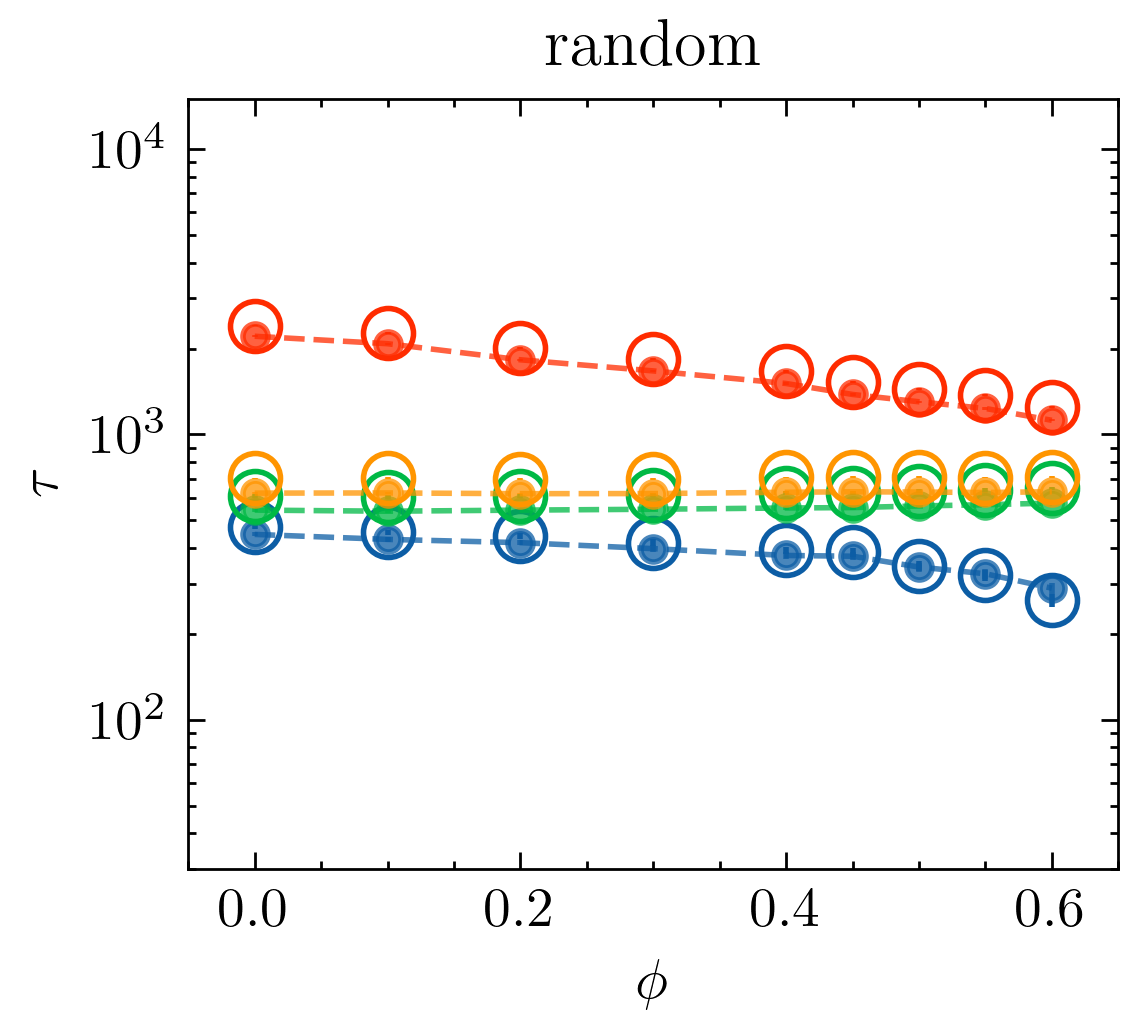}
        \put(2,87){\textbf{(d)}}
    \end{overpic}
    \caption{Parameters of the compressed-/stretched-exponential fit,
\(\hat{C}_{\mathrm e}(t)=\exp[-(t/\tau_k)^\beta]\), to the normalised end-to-end vector TACF, with fits restricted to \(\hat{C}_{\mathrm e}(t)>0.01\). Panels (a) and (b) show the exponent \(\beta\) against obstacle packing fraction \(\phi\) for square-lattice and random packings, respectively, for different bending stiffness values $\kappa=1$, $4$, $10$, and $100$. Panels (c) and (d) show the corresponding fitted relaxation time \(\tau_k\) and mean relaxation time \(\langle \tau \rangle=\tau_k\Gamma(1/\beta)/\beta\), for square-lattice and random packings as a function of \(\phi\).}
    \label{fig:taubeta-phi-k}
\end{figure}

The $\beta$ values extracted from fits as a function of obstacle packing fraction \(\phi\) are shown in Fig.~\ref{fig:taubeta-phi-k}(a)--(b) in both ordered and disordered media for different bending stiffness values $\kappa$. In free space, $\beta>1$ and varies non-monotonically with bending rigidity, reaching its maximum at $\kappa=10$. Such compressed-exponential relaxation is consistent with persistent, coherent motion that produces faster-than-exponential loss of orientational memory.

The dependence of $\beta$ on obstacle density varies with filament flexibility and medium structure. For highly flexible polymers ($\kappa=1$), $\beta$ decreases with $\phi$ in both ordered and disordered media, although more weakly in the latter, and falls below unity at high densities. This stretched-exponential relaxation reflects increasingly heterogeneous trapping-and-hopping dynamics under tight confinement. For moderately flexible polymers ($\kappa=4$ and $10$), $\beta$ instead increases in both geometries, approaching $\beta\simeq2$, indicative of confinement-enhanced, nearly ballistic reorientation at intermediate times. For semiflexible polymers ($\kappa=100$), $\beta$ decreases toward unity in square lattices, corresponding to nearly exponential orientational relaxation, but increases with $\phi$ in random media, indicating more coherent orientational dynamics under confinement in curvilinear channels.

Figures~\ref{fig:taubeta-phi-k}(c) and (d) show the relaxation timescale $\tau_k$ as a function of obstacle packing fraction $\phi$ in ordered and disordered media. Its dependence on $\phi$ varies strongly with filament flexibility: $\tau_k$ decreases for flexible polymers, remains nearly constant for moderately flexible polymers ($\kappa=4$ and $10$), and exhibits opposite trends for semiflexible polymers, increasing in square lattices but decreasing in random media.

To quantify the orientational persistence time, we define it as the mean relaxation time obtained from integral of TACF,
\begin{equation}
\label{eq:tau}
    \langle \tau \rangle \equiv \int_0^\infty \exp\!\left[-\left(\frac{t}{\tau_k}\right)^\beta\right]dt = \frac{\tau_k}{\beta} \Gamma\!\left(\frac{1}{\beta}\right).
\end{equation} 
The resulting values of \(\langle\tau\rangle\) are indicated by open symbols in Figs.~\ref{fig:taubeta-phi-k}(c) and (d).
In general, $\tau_k$ and $\langle \tau \rangle$ are of comparable magnitude, except for the case $\kappa = 1$ and $\phi = 0.6$ on a square lattice, where $\langle \tau \rangle$ is significantly smaller than $\tau_k$ and exhibits large uncertainty.

\begin{figure}
    \centering
    \begin{overpic}[width=0.48\linewidth]{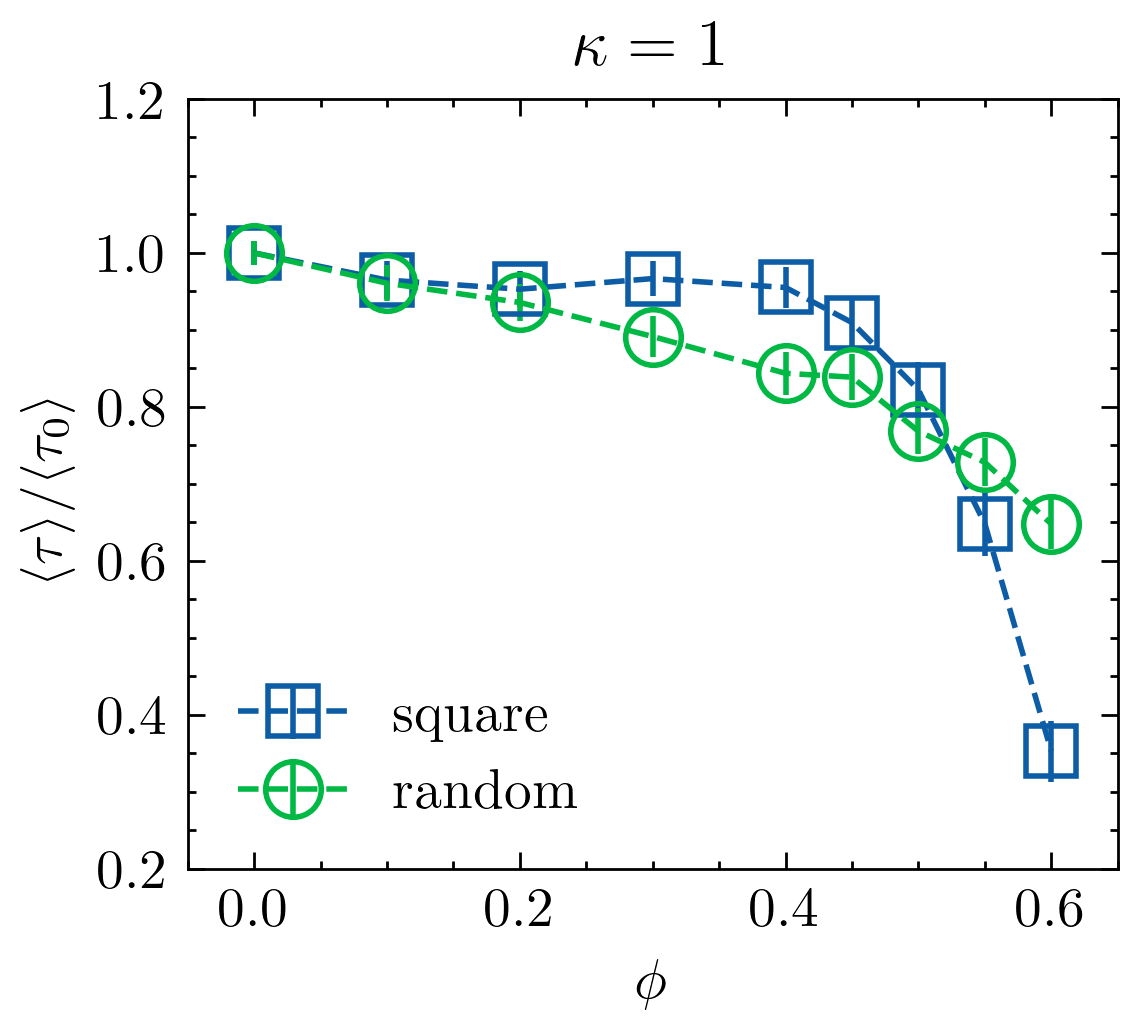}
        \put(2,87){\textbf{(a)}}
    \end{overpic}
    \hfill
    \begin{overpic}[width=0.49\linewidth]{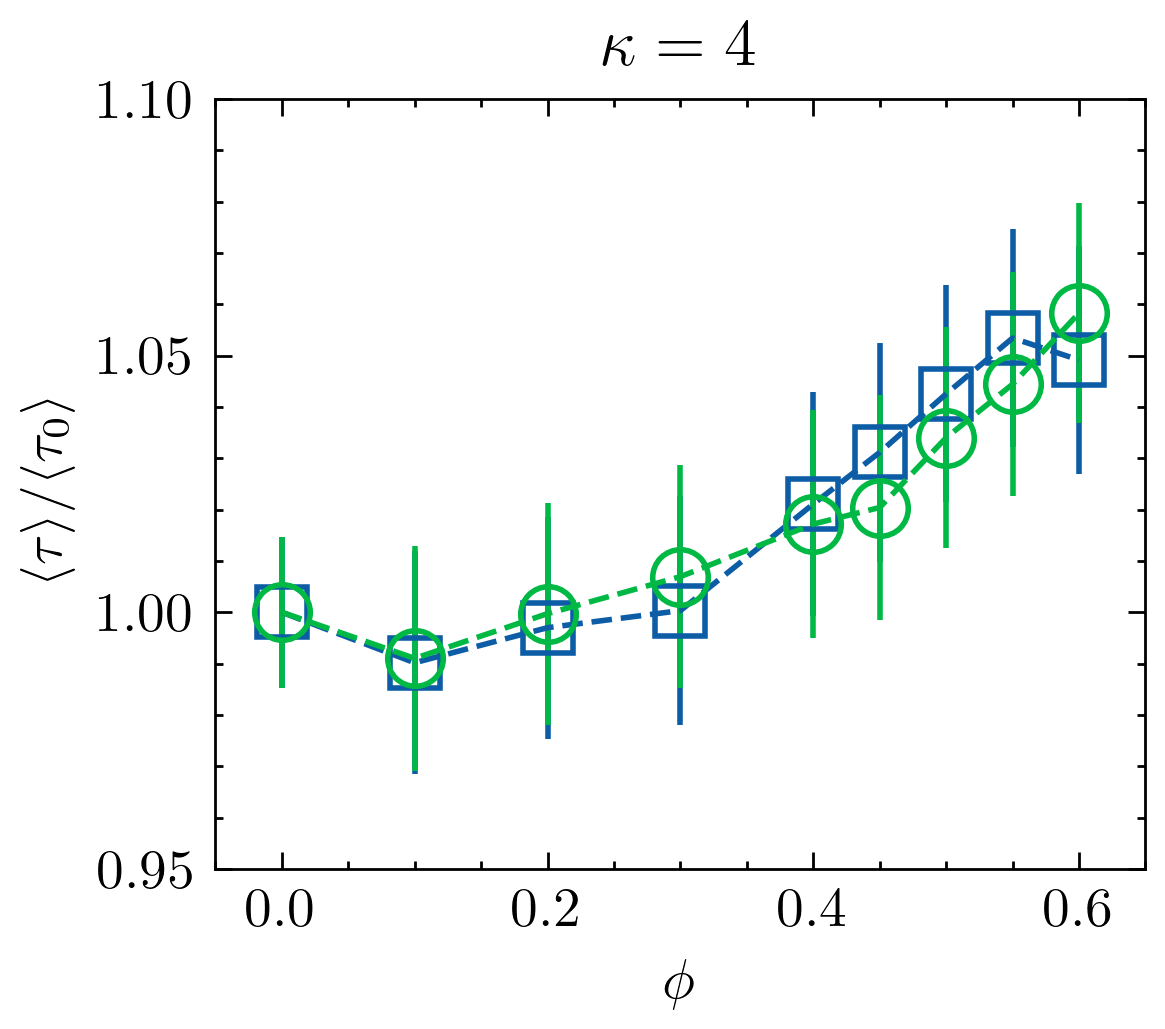}
        \put(2,85){\textbf{(b)}}
    \end{overpic}
    \hfill
    \begin{overpic}[width=0.49\linewidth]{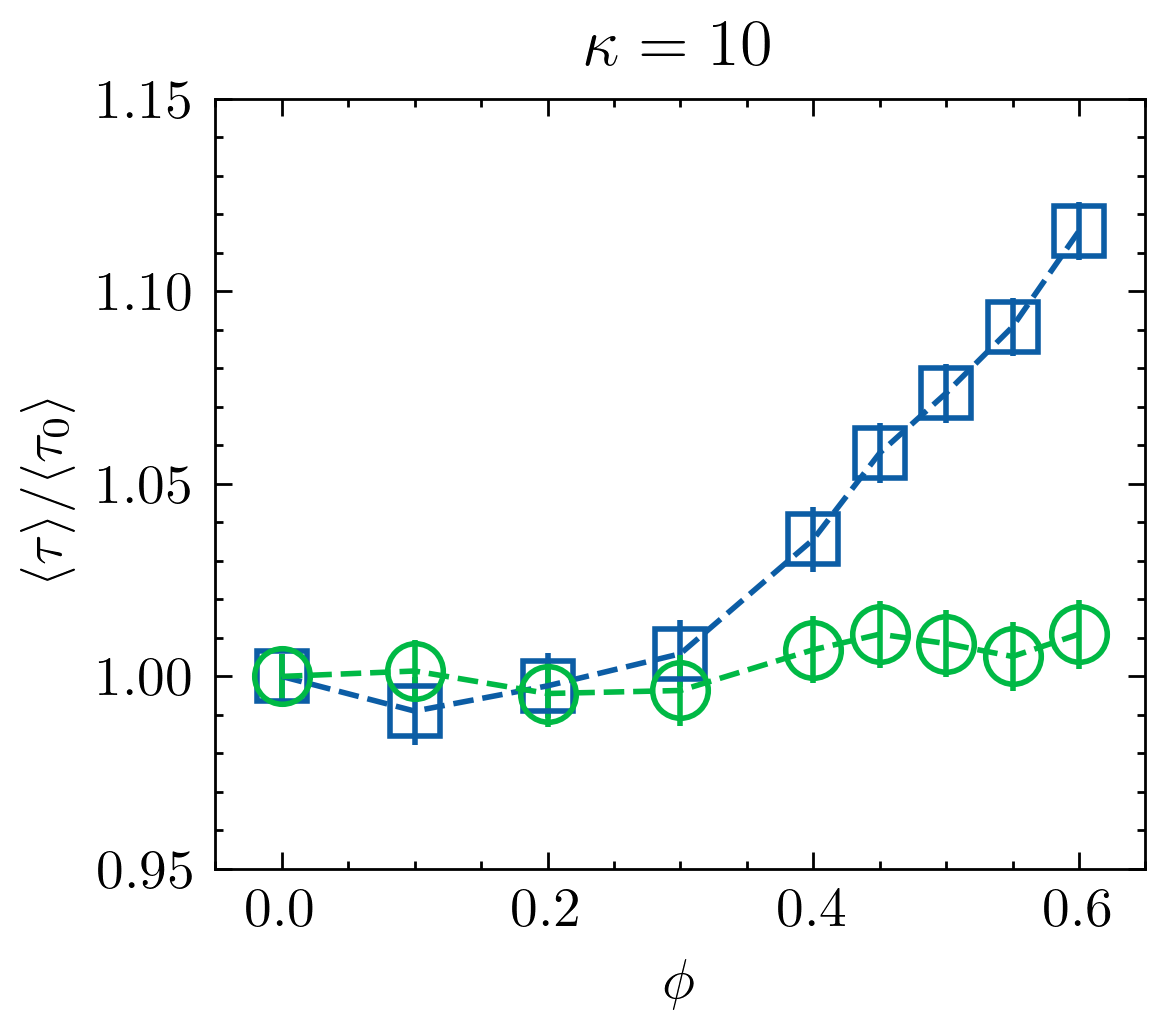}
        \put(2,85){\textbf{(c)}}
    \end{overpic}
    \hfill
    \begin{overpic}[width=0.46\linewidth]{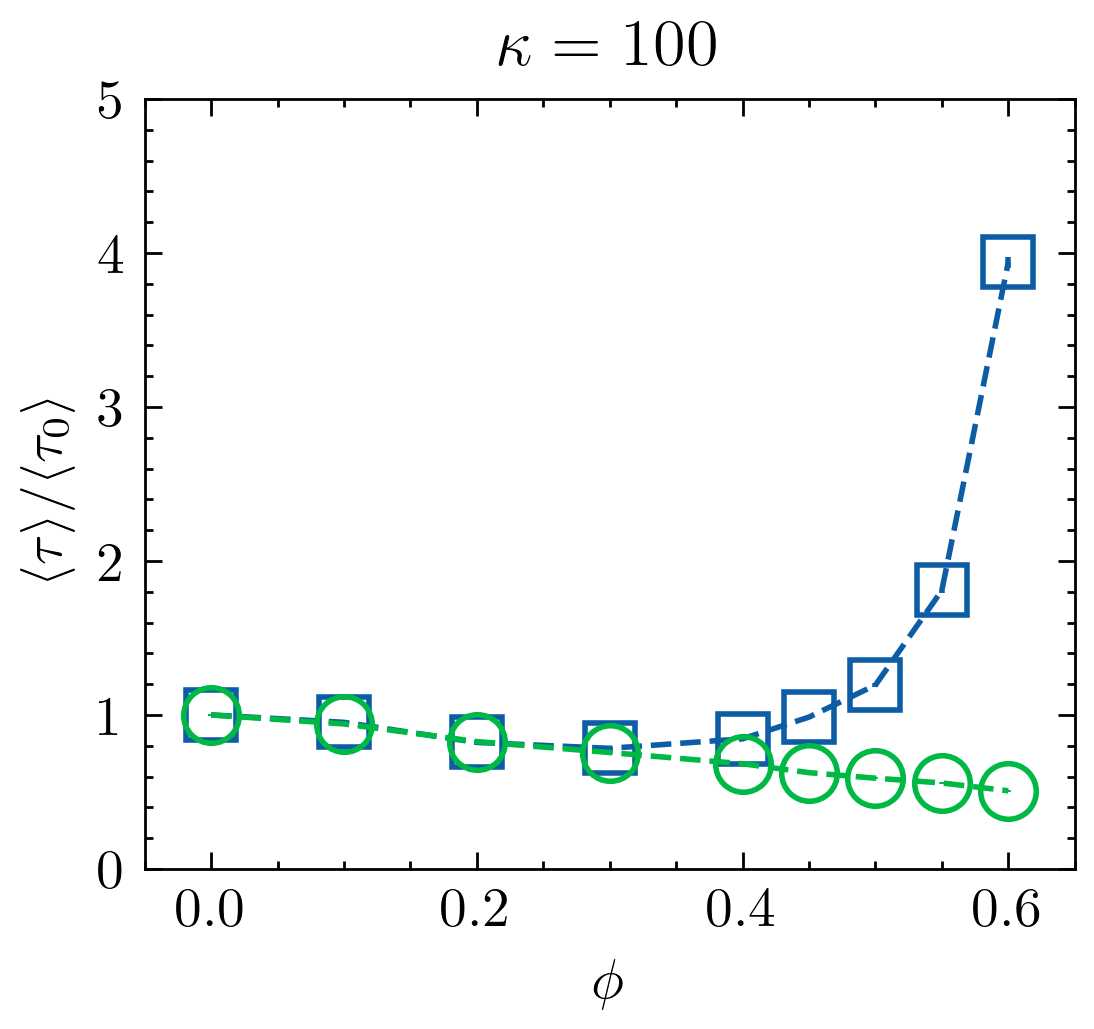}
        \put(2,91){\textbf{(d)}}
    \end{overpic}
    \caption{Normalised mean relaxation time of the polymer end-to-end vector, \(\langle\tau\rangle/\langle\tau_0\rangle\), as a function of obstacle packing fraction \(\phi\), with \(\langle\tau_0\rangle(\kappa)\) being the value in free space. Panels (a)--(d) correspond to polymer bending stiffness \(\kappa=1,4,10,100\), respectively.}
    \label{fig:tau-phi-k}
\end{figure}

Figure~\ref{fig:tau-phi-k} shows the mean relaxation time, normalised by its free-space value, as a function of obstacle packing fraction \(\phi\) for active polymers with different flexibility regimes. For highly flexible polymers (\(\kappa=1\)), the mean relaxation time \(\langle\tau\rangle\) decreases with increasing \(\phi\) in both ordered and disordered media, where for the ordered media with high density of obstacles the decrease of \(\langle\tau\rangle\) is more marked. This indicates accelerated orientational decorrelation under confinement. In contrast, moderately flexible polymers (\(\kappa=4\)) exhibit a weak increase in \(\langle\tau\rangle\) with \(\phi\) with little sensitivity to the underlying obstacle arrangement. As the bending rigidity increases to \(\kappa=10\), confinement induces a more pronounced growth of \(\langle\tau\rangle\) in the square lattice. The strongest contrast emerges for semiflexible polymers (\(\kappa=100\)), where the two media produce qualitatively different trends. In random media, \(\langle\tau\rangle\) decreases monotonically with \(\phi\), whereas in the square lattice it follows a similar trend only up to \(\phi\approx0.3\), beyond which it increases with obstacle density. This divergence highlights the increasingly important role of porous medium geometry in governing the orientational dynamics of semiflexible active filaments.

Having established how porous confinement influences both the conformational properties and orientational dynamics of active filaments, we now examine how these effects combine to determine their long-time diffusive transport. To this end, we derive an analytical estimate in the following section that relates the observed diffusion behaviour to the mean polymer conformation and mean reorientational relaxation.

\section{Analytical theory for long-time diffusion and comparison to simulation results}
\label{sec:theory}
In this section, extending the approach developed in Ref.~\cite{fazelzadeh2023active}, we derive an analytical expression for the long-time diffusion coefficient \(D_l\) in terms of two key polymer properties: the mean end-to-end distance squared \(\langle R_{\text{e}}^2 \rangle\) and the mean relaxation time \(\langle\tau\rangle\).

The long-time diffusion coefficient can be expressed through the Green--Kubo relation as the time integral of the velocity autocorrelation function of polymer centre of mass, \(D_l = \frac{1}{2}\int_{t=0}^\infty \langle \mathbf{v}_\text{cm}(t)\cdot\mathbf{v}_\text{cm}(0) \rangle dt\).

The centre-of-mass velocity $\mathbf{v}_{\text{cm}}=\frac{1}{N}\sum_i \mathbf{v}_i$ follows directly from summing the equations of motion over all beads. The resulting force balance contains contributions from the total active force \(\mathbf{F}^\text{a}=\sum_{i=1}^N \mathbf{f}_i^a\), total random force \(\mathbf{F}^\text{r}=\sum_{i=1}^N \mathbf{f}_i^r\), and obstacle-induced forces \(\mathbf{F}^\text{o}\),
\begin{equation}
\label{eq:v-f}
    \gamma\mathbf{v}_{\text{cm}}(t)=\frac{1}{N}(\mathbf{F}^\text{a}(t)+\mathbf{F}^\text{r}(t)+\mathbf{F}^\text{o}(t)).
\end{equation} 
Using the relation $\mathbf{F}^\text{a}=\frac{f^a}{\ell_0} \, \mathbf{R}_\text{e}$, the velocity autocorrelation function can be decomposed into active, obstacle-induced, and noise contributions,
\begin{equation}
\label{eq:v-expand}
\langle \mathbf{v}_\text{cm}(t)\cdot\mathbf{v}_\text{cm}(0) \rangle =
\begin{aligned}[t]
 & \left(\frac{f^{\text{a}}}{N\gamma \ell_0}\right)^2 
   \langle \mathbf{R}_\text{e}(t)\cdot\mathbf{R}_\text{e}(0)\rangle \\
 &\hspace{-7em}+ \frac{f^{\text{a}}}{N^2\gamma^2 \ell_0} 
   \Big[\langle \mathbf{R}_{\text{e}}(t)\cdot\mathbf{F}^\text{o}(0)\rangle 
        + \langle \mathbf{F}^\text{o}(t)\cdot \mathbf{R}_{\text{e}}(0)\rangle\Big] \\
 &\hspace{-7em}+ \frac{1}{N^2\gamma^2} 
   \Big[\langle \mathbf{F}^\text{o}(t)\cdot \mathbf{F}^\text{o}(0)\rangle
        + \langle \mathbf{F}^\text{r}(t)\cdot \mathbf{F}^\text{r}(0)\rangle \Big],
\end{aligned}
\end{equation}
where terms containing a single random force vanish upon averaging due to its zero mean property, and are therefore omitted. 

Furthermore, under the dilute-collision approximation, the cross-correlation terms in the second line are expected to be small, except for highly flexible polymers at large $\phi$, where obstacle interactions become significant. Hence, they are neglected. The remaining force-correlation terms in the third line account for passive transport in the porous medium. In free space, the random force contribution reduces to the passive diffusion coefficient ($D_{\text{Passive}}^{\text{Free}}=\frac{D_0}{N}$). Assuming that the number of collisions is not significantly modified in the low activity limit, obstacle interactions primarily renormalise this passive contribution. Therefore, the combined effects of the two terms in the last line can be described by an effective diffusion coefficient $D_l^{\text{Passive}}(\phi,\kappa) <\frac{D_0}{N}$, which decreases with increasing confinement.

The dominant contribution therefore results from activity given by the first line, which couples diffusion directly to the conformational and orientational dynamics of the filament.
We make two assumptions: i) fluctuations of the end-to-end distance are small, and 
ii) the end-to-end vector time autocorrelation can be approximated as stretched- or compressed-exponential $\langle \mathbf{R}_\text{e}(t)\cdot \mathbf{R}_\text{e}(0)\rangle
\simeq \langle R_\text{e}^2\rangle e^{-(t/\tau_k)^\beta}.$ 
This approximation is expected to break down for highly flexible polymers $\kappa<2$ under strong confinement (\(\phi>0.3\)), where large conformational fluctuations become important due to trapping-and-hopping events.

Integrating the autocorrelation function yields
\begin{equation}
\label{eq:tacfre-int}
    \int_{t=0}^{\infty}\langle \mathbf{R}_\text{e}(t)\cdot\mathbf{R}_\text{e}(0)\rangle dt \approx \int_{t=0}^{\infty} \langle R_\text{e}^2\rangle e^{-(t/\tau_k)^\beta}dt=\langle R_\text{e}^2\rangle \langle\tau\rangle,
\end{equation}
where $\langle \tau \rangle$ is the mean orientational relaxation time defined by Eq.~\eqref{eq:tau}. Substituting this result into the Green--Kubo expression leads to the analytical estimate:
\begin{equation}
\label{eq:analyticalformula}
    D_l\approx D_l^{\text{Passive}}(\phi,\kappa) + \frac{(f^{\text{a}})^2\langle R_{\text{e}}^2\rangle \langle\tau\rangle}{2 N^2 \ell_0^2 \gamma^2}.
\end{equation} 

Equation~\eqref{eq:analyticalformula} provides a direct connection between long-time transport and two key polymer properties: its typical extension, quantified by $\langle R_\text{e}^2\rangle$, and its mean orientational decay time, characterized by $\langle\tau\rangle$. We argue that the contribution from $D_l^{\text{Passive}}$ is negligible except for $\kappa = 1$ at high packing fractions. This follows from the estimate $D_{\text{Passive}}^{\text{Free}} = \frac{D_0}{N} = 0.01$, with the expectation that the presence of obstacles further suppresses diffusion.

\begin{figure}
    \raggedright
    \begin{overpic}[width=0.99\linewidth]{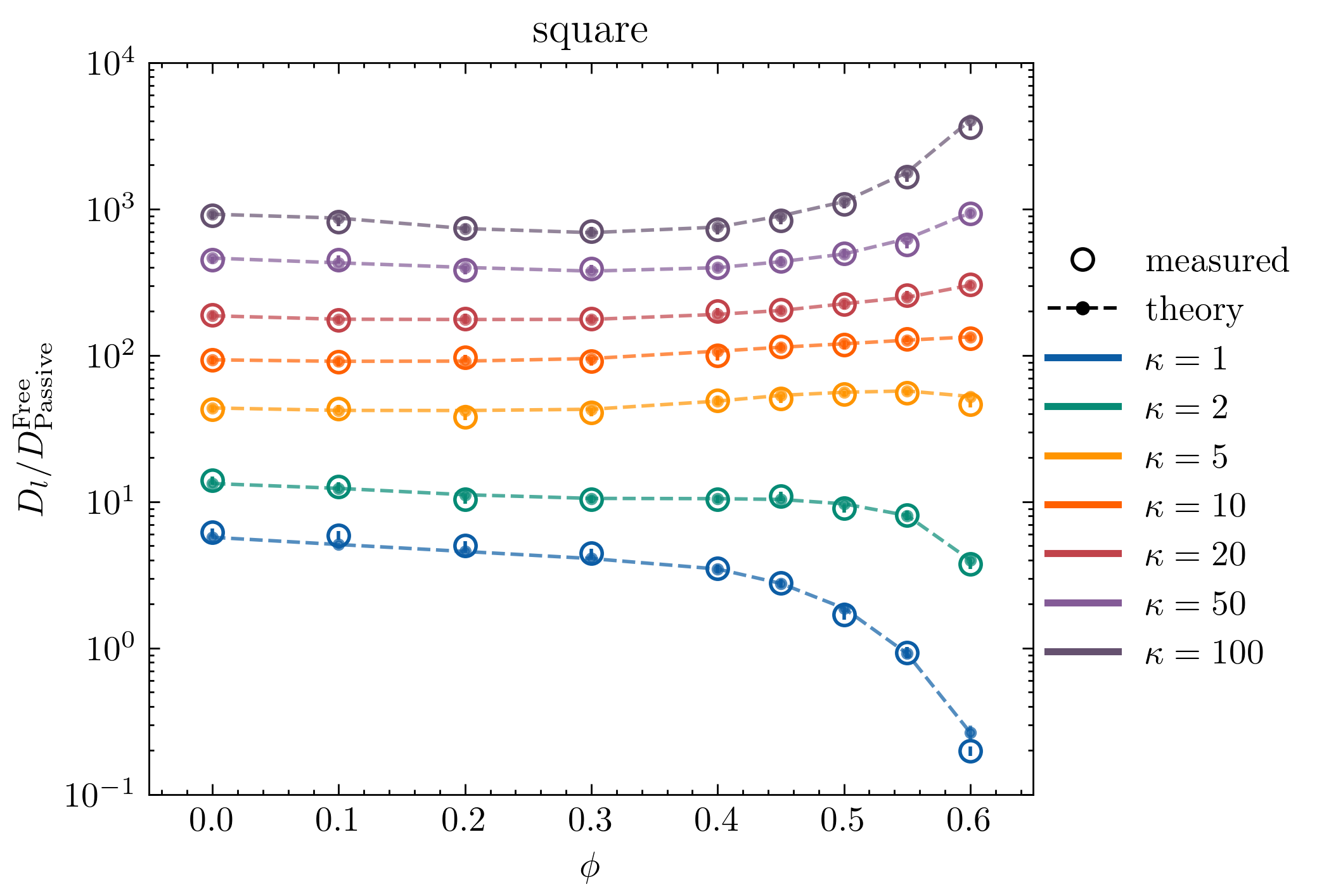}
        \put(2,67){\textbf{(a)}}
    \end{overpic}
    \hfill
    \begin{overpic}[width=0.78\linewidth]{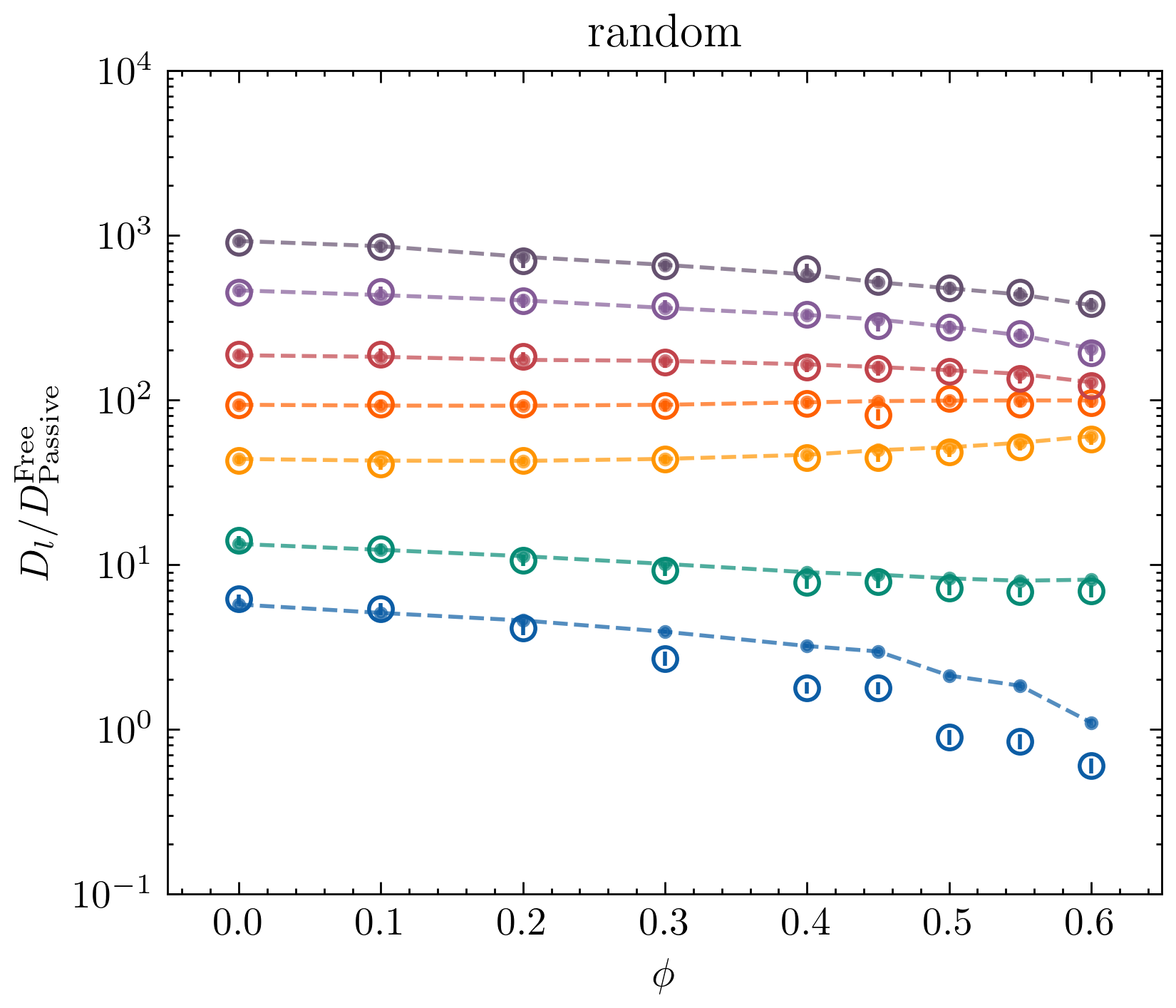}
        \put(2,85){\textbf{(b)}}
    \end{overpic}
    \caption{Normalised long-time diffusion coefficient of polymer centre of mass, \(D_l/D^\text{Free}_\text{Passive}\), as a function of obstacle packing fraction \(\phi\) in (a) square-lattice and (b) random packing. Here, \(D^\text{Free}_\text{Passive}=0.01\) is the diffusion coefficient of passive polymers in free space. Empty symbols show simulation values obtained from \(\langle \Delta\mathbf{r}^2_{\text{cm}}\rangle=4D_lt\) at long times, and filled symbols show values calculated from the analytical estimate based on \(\langle R_{\text{e}}^2\rangle \langle\tau\rangle\), where the mean relaxation time \(\langle\tau\rangle\) is determined from compressed-/stretched-exponential fits.}
    \label{fig:dl-phi-k-theory}
\end{figure}

To assess the validity of the analytical estimate, Fig.~\ref{fig:dl-phi-k-theory} compares the active contribution $D_l^{\text{a}} (\phi,\kappa)=\dfrac{f^{\text{a}2}\langle R_{\text{e}}^2\rangle \langle\tau\rangle}{2 N^2 \ell_0^2 \gamma^2}$ with the diffusion coefficients measured in simulations across different flexibility regimes. The analytical estimate is in good agreement with the simulation results for moderately flexible and semiflexible polymers, capturing both the magnitude and trends of the diffusion coefficient. In contrast, it systematically overestimates diffusion for highly flexible polymers $\kappa=1$, particularly in disordered media, where the assumptions break down because strong conformational fluctuations and enhanced polymer--obstacle interactions become increasingly important. 

\section{Discussion and conclusion}
\label{sec:discussion-conclusion}

\begin{figure}
    \centering
    \includegraphics[width=0.7\linewidth]{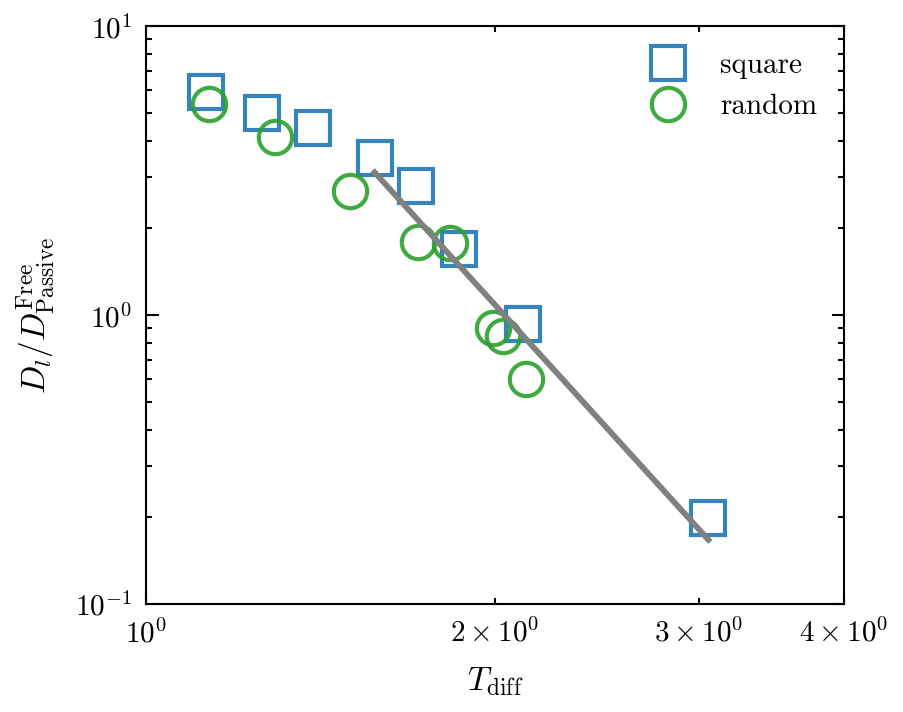}
    \caption{Normalised long-time diffusion coefficient \(D_l/D^\text{Free}_\text{Passive}\) as a function of bulk-diffusion tortuosity \(T_\text{diff}\) for highly flexible polymers (\(\kappa=1\)) in ordered and random obstacle arrays. Here, \(D^\text{Free}_\text{Passive}=0.01\) is the diffusion coefficient of passive polymers in free space. For high packing fractions \(\phi\ge0.4\), where trapping-and-hopping events become the dominant transport mechanism, \(D_l\sim\frac{1}{T_\text{diff}^{\alpha}}\) with \(\alpha=4.4\).}
    \label{fig:Dl_tor}
\end{figure}

The transport of active filaments in complex environments is governed by a subtle interplay between their deformability, their motility mode, and the structure of the surrounding medium. To elucidate the interplay between flexibility and motility in active filaments, we have combined active matter and polymer physics to investigate their impact on the conformational and dynamical behaviour of active filaments in two-dimensional ordered and disordered porous media. We have performed large-scale Brownian dynamics simulations of tangentially driven active filaments over a wide range of degree of flexibility and obstacle packing fractions combined with analytical calculations for the centre of mass dynamics. Based on thorough analysis of conformational and dynamical properties of active polymers, we identified three distinct transport mechanisms depending on filament flexibility. 

In the highly flexible regime, $\ell_{\mathrm{p}}^0/L\ll1$, polymers adopt compact, coil-like conformations with a characteristic size $R_{\mathrm{g}}\sim6$, which decreases as obstacle density increases and the available pore space becomes more restricted. Transport occurs through trapping-and-hopping dynamics: filaments remain temporarily confined within pores and occasionally stretch through narrow gaps into neighbouring pores. Consequently, the long-time diffusivity $D_l$ decreases monotonically with obstacle packing fraction $\phi$ in both ordered and disordered media. This suppression is consistent with the simultaneous decrease in mean chord length and gap width shown in Fig.~\ref{fig:LengthScales}, which restrict motion to increasingly narrow and tortuous pathways. When the long-time diffusion is plotted against the pore-network tortuosity $T_{\mathrm{diff}}$, the diffusivities in ordered and disordered media collapse onto a single master curve, as shown in Fig.~\ref{fig:Dl_tor}. This collapse indicates that transport of highly flexible active polymers is primarily controlled by pore-space tortuosity. For $\phi\ge0.4$, where the polymer size becomes comparable to the mean gap width, we find a power-law relation $D_l \sim T_{\mathrm{diff}}^{-\alpha}$, with $\alpha \approx 4.4$, revealing a strong sensitivity to pathway geometry. Thus, media with greater tortuosity suppress diffusion more strongly, even at the same obstacle packing fraction. This trend parallels classical porous-media transport, for which the effective diffusivity generally decreases with tortuosity as $D_{\mathrm{eff}}\sim D_0 T_{\mathrm{diff}}^{-\alpha}$~\cite{ghanbarian2013tortuosity}, suggesting that tortuosity provides a common geometric descriptor for passive tracers and highly flexible active polymers.

In the moderately flexible regime, \(\ell_\text{p}^0/L \sim 0.1\), filament extension becomes important. Although these filaments remain highly deformable, they retain orientational correlations over distances comparable to their persistence length while undergoing frequent reorientations and large conformational fluctuations. Under confinement, they exploit the local pore geometry by adopting curved conformations and wrapping around obstacles, thereby maintaining connectivity across neighbouring pores. Unlike highly flexible filaments, whose transport is governed by trapping-and-hopping between isolated pores, moderately flexible filaments can span multiple pores, reducing confinement-induced trapping and facilitating transport through the interconnected pore network. The resulting increase in mean end-to-end distance and orientational persistence time leads to a modest enhancement of long-time diffusion in dense disordered media. The obstacle size is an important length scale in this regime, as our additional simulation results show that the bending rigidity corresponding to maximal diffusivity in dense random media depends on the obstacle radius, see Fig.~\ref{appfig:k-r-relation}.

In the semiflexible regime, $\ell_\text{p}^0/L \sim 1$, transport is governed by the interplay between polymer persistent motion over length scales comparable to the polymer contour length and pore geometry. Rather than being strongly compactified, filaments retain extended conformations and navigate the porous medium through persistent motion along interconnected curvilinear channels. Consequently, the architecture of the medium plays a decisive role in controlling transport. In weakly confined disordered media and square lattices, where occasional encounters with obstacles disrupt orientational persistence and enhance reorientation, diffusion is suppressed. Under tight confinement, repeated collisions with obstacles disrupt orientational persistence and suppress diffusion strongly at high packing fractions. By contrast, dense ordered lattices act as geometrical guides that align filaments with straight pore channels, reducing reorientation events and extending the persistence of directed motion. This confinement-induced rectification leads to a pronounced enhancement of long-time diffusion, highlighting a transition from tortuosity-controlled transport in flexible filaments to persistence-controlled transport in semiflexible ones.

To rationalise the observed transport behaviour across a broad range of filament flexibilities and porous geometries, we derived an analytical relation linking the long-time diffusivity to two key polymer properties: the mean squared end-to-end distance, $\langle R_\text{e}^2\rangle$, and the mean conformational relaxation time, $\langle\tau\rangle$. Specifically, we find that
$D_l \propto \langle R_\text{e}^2\rangle\langle\tau\rangle$, which captures how confinement simultaneously modifies polymer conformations and relaxation dynamics. Although the theory neglects direct collisions between filament segments and obstacles, it quantitatively reproduces the simulation results across a wide range of flexibility degrees and pore geometries. The only notable deviation occurs for highly flexible filaments in dense disordered media, where frequent collisions with obstacles and large polymer conformational fluctuations due to hopping events become important and invalidate our theoretical assumptions.

In summary, our results show that active transport through porous media emerges from the interplay between filament flexibility and pore geometry. The system is governed by competing geometric length scales---including the obstacle size, nearest-neighbour gap, and mean chord length---and filament length scales, such as the contour length, persistence length, radius of gyration, and end-to-end distance. The ratio of contour to persistence length sets the flexibility regime and thereby determines which length-scale competition controls transport.

For highly flexible filaments, transport is primarily tortuosity-controlled: the ratio of the gyration radius to the mean gap size determines the onset of strong confinement, where the long-time diffusivity decreases approximately as a power of the inverse tortuosity. For moderately flexible filaments in disordered media, the obstacle size appears to set the flexibility at which confinement-induced diffusion enhancement is maximal, rather than the mean gap or nearest-neighbour distance. For semiflexible filaments, whose end-to-end distance fluctuates weakly, transport is governed by the reorientation of the end-to-end vector. The ratio of persistence length to mean chord length then controls the ability to sustain motion along straight or curved channels. Because motion preferentially follows directions of lower tortuosity, ordered square arrays promote persistent axial transport through their straight channels. Thus, no single geometric length scale governs all regimes; instead, flexibility selects the relevant competition between filament and pore-scale lengths.

These findings establish a unified physical framework for understanding the migration of elongated active agents ranging from biological systems like bacteria and active biopolymers to robotic worms in crowded biological and synthetic environments, in terms of filament flexibility degree and porous medium geometry. Our results show that the same porous environment can either hinder or promote active transport depending on the filament persistence length. Thus, our work suggests new routes for controlling active transport through the design of porous architectures and tuning flexibility degree of active filament.

The present work focuses on porous media with fully connected pore spaces and monodisperse circular obstacles. Extending this framework to systematically explore the effects of obstacle size, shape, and pore-space topology represents a natural next step. Particularly intriguing are media containing dead-end pores, bottlenecks, or finite maximum chord lengths~\cite{kurzthaler2021geometric}, where the latter has been found to govern optimal spreading of stiff active polymers in porous media. 
Beyond static porous environments, an important challenge will be to understand how active filaments navigate dynamically evolving three-dimensional media, such as granular materials, polymer solutions, and biopolymer networks. Addressing these questions will help establish general principles governing the transport of deformable active matter in the complex and heterogeneous environments encountered in nature.

\appendix

\section{Mean squared displacement}
\label{sec:MSD}
\begin{figure}[h]
    \centering
    \begin{overpic}[width=0.49\linewidth]{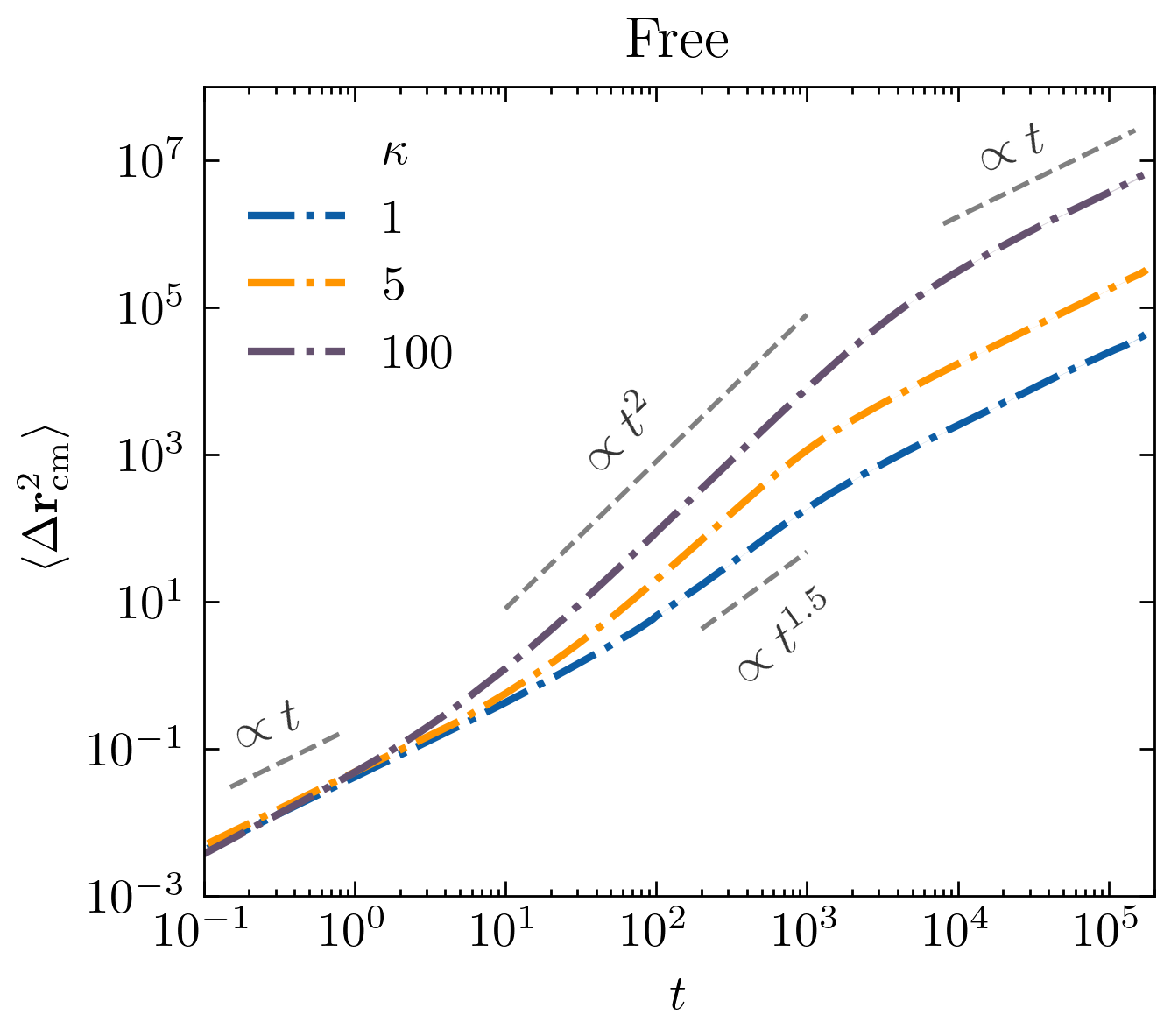}
        \put(2,85){\textbf{(a)}}
    \end{overpic}
    \hfill
    \begin{overpic}[width=0.49\linewidth]{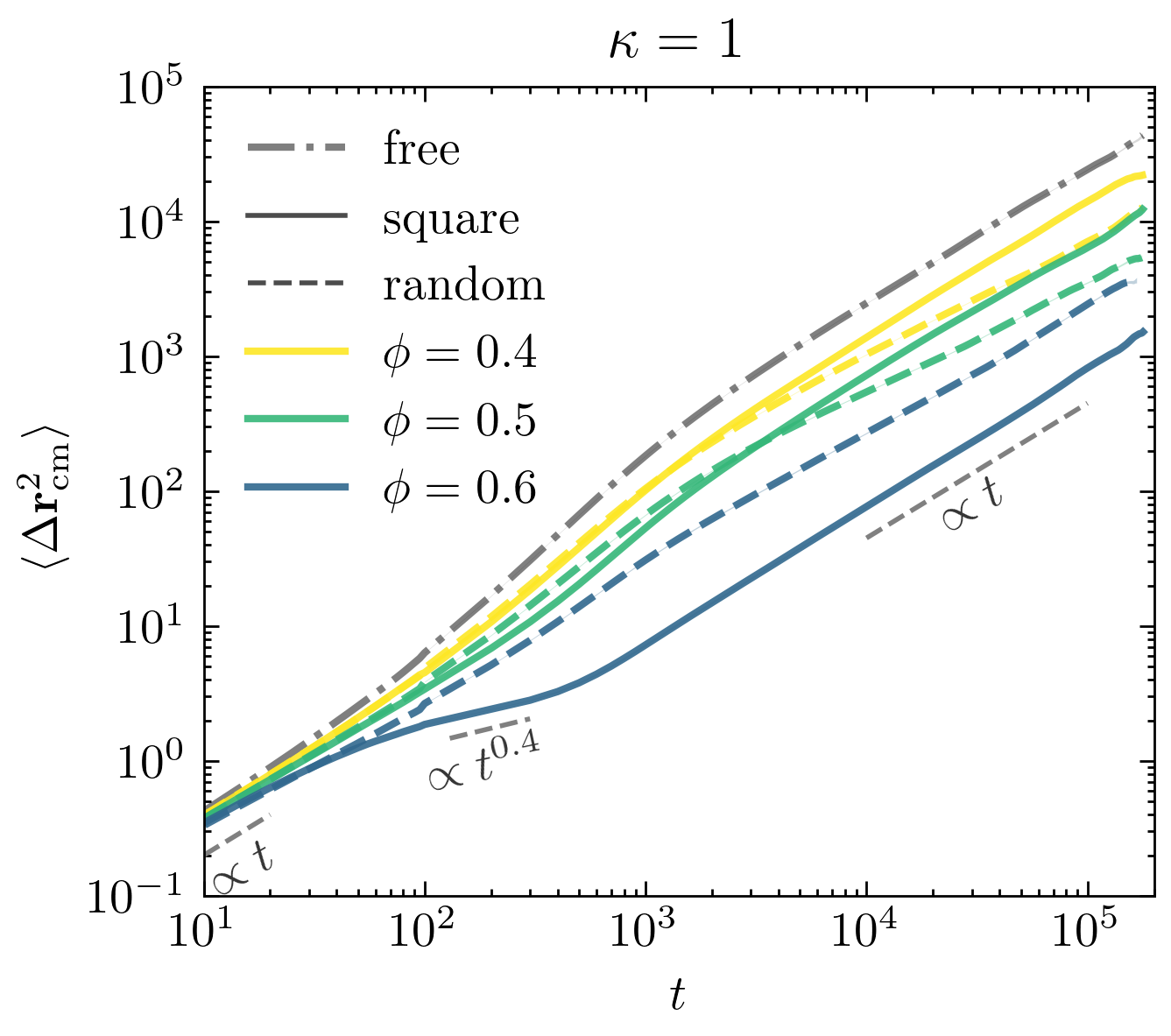}
        \put(2,85){\textbf{(b)}}
    \end{overpic}
    \hfill
    \begin{overpic}[width=0.49\linewidth]{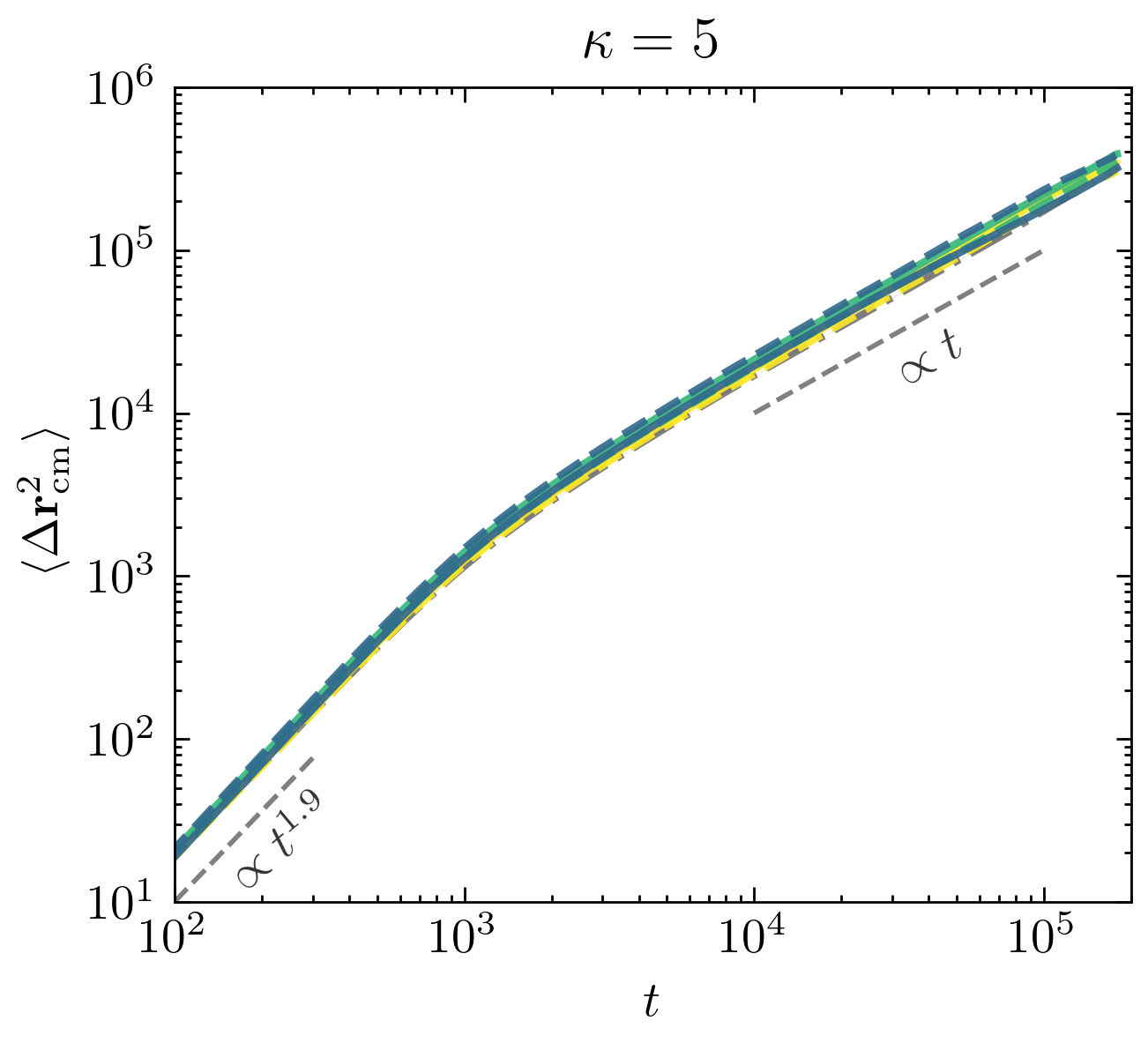}
        \put(2,85){\textbf{(c)}}
    \end{overpic}
    \hfill
    \begin{overpic}[width=0.49\linewidth]{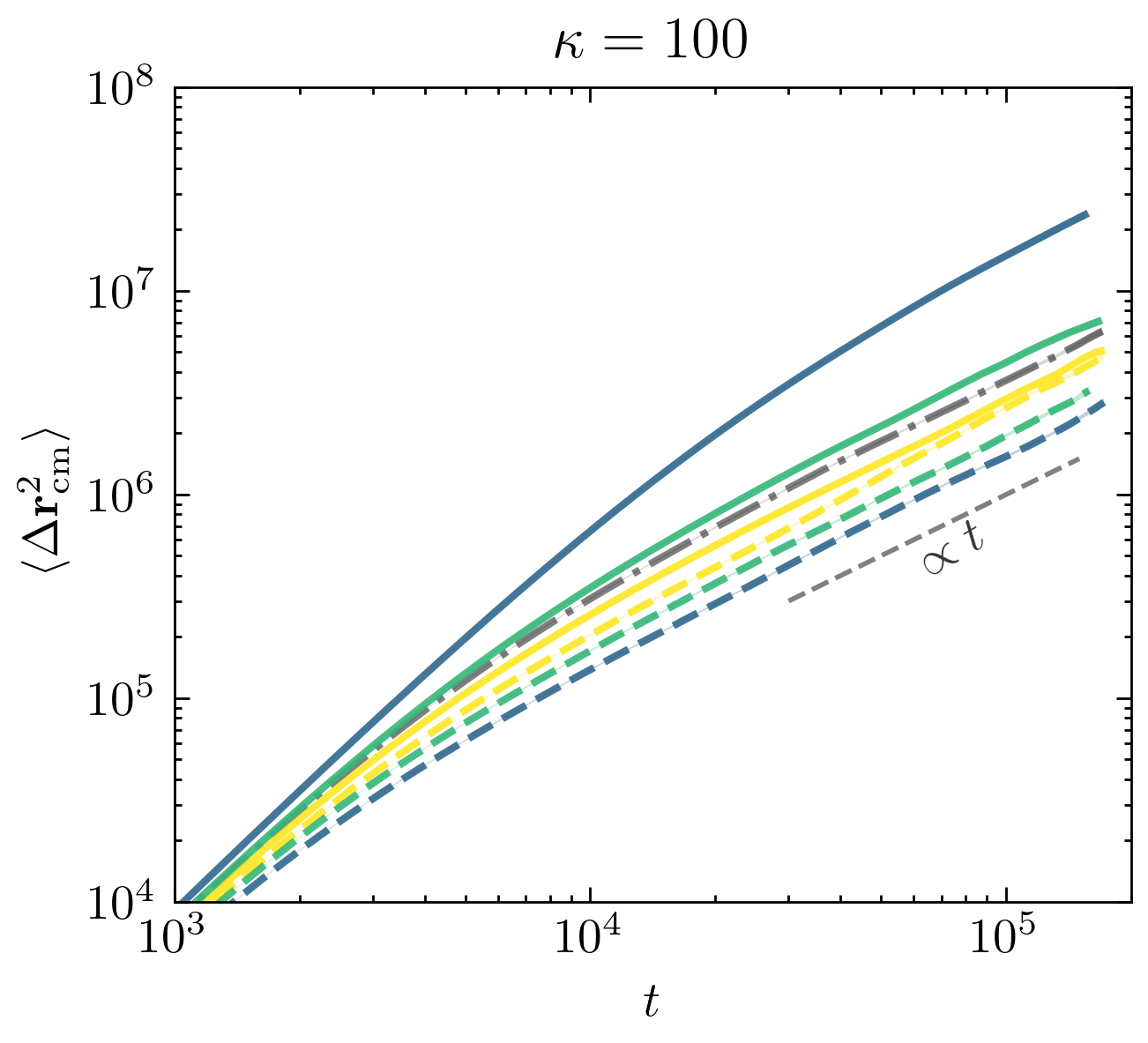}
        \put(2,85){\textbf{(d)}}
    \end{overpic}
    \caption{Mean squared displacement of active polymer centre of mass, \(\langle \Delta\mathbf{r}_{\text{cm}}^2\rangle\), as a function of time. Panel (a) corresponds to polymers in free space. Panels (b)-–(d) correspond to polymers with bending stiffness \(\kappa=1,5,100\), respectively; medium types and packing fractions in panels (b)--(d) are indicated by the legend in panel (b).}
    \label{appfig:msd-t-k}
\end{figure}

To complement the long-time diffusion analysis in the main text, we examine the full time dependence of the mean squared displacement (MSD) of the polymer centre of mass $\langle \Delta \mathbf{r}_{\text{cm}}^2 \rangle$, where the average is taken over trajectories and time origins. Figure~\ref{appfig:msd-t-k} shows MSD curves for representative bending stiffnesses in free space and in ordered and disordered obstacle arrays.

In free space, Fig.~\ref{appfig:msd-t-k}(a), three distinct regimes are observed. At short and long times, $\langle \Delta \mathbf{r}_{\text{cm}}^2 \rangle$ exhibits normal diffusion, while at intermediate times a superdiffusive regime emerges. The scaling exponent is $\alpha \approx 1.5$ for flexible active polymers and approaches ballistic behaviour ($\alpha = 2$) for semiflexible chains. The long-time diffusion coefficient increases with bending stiffness $\kappa$, while the crossover to the asymptotic diffusive regime shifts to later times as \(\kappa\) increases. 

Crowding modifies these regimes in a stiffness-dependent manner. For highly flexible polymers ($\kappa = 1$), Fig.~\ref{appfig:msd-t-k}(b), increasing obstacle packing fraction \(\phi\) suppresses diffusion at intermediate and long times. In the square lattice at $\phi = 0.6$, a pronounced subdiffusive regime appears at intermediate times, consistent with trapping within individual pore spaces and intermittent hopping between neighbouring pores. In random media at intermediate-to-high packing fractions ($\phi = 0.4$ and $0.5$), a weak subdiffusive regime is observed before the onset of long-time diffusion. 

For the moderately flexible case (\(\kappa=5\)), Fig.~\ref{appfig:msd-t-k}(c), the $\langle \Delta \mathbf{r}_{\text{cm}}^2 \rangle$ curves are comparatively insensitive to obstacle arrangement and packing fraction. The intermediate-time superdiffusive regime and the subsequent crossover to long-time diffusion remain close to their free-space behaviour. This indicates that polymers in this stiffness range can deform around obstacles without becoming strongly trapped or strongly channelled by the medium. This regime is comparable to the bending stiffness estimated for biological worms~\cite{sinaasappel2025locomotion}.

For semiflexible polymers (\(\kappa=100\)), Fig.~\ref{appfig:msd-t-k}(d), the response depends strongly on the geometry of the obstacle array. At moderate packing fractions, increasing \(\phi\) reduces the long-time diffusivity in both ordered and disordered media. At higher packing fractions, however, the two environments diverge. In the square lattice, aligned channels promote persistent directed motion and enhance the long-time MSD, although the crossover to the normal diffusion regime is delayed. In random media, the tortuous pore space instead increases bending and reorientation, leading to a further reduction of the long-time diffusivity.

\section{Effect of obstacle size}
\label{sec:midk}

The main text shows that polymers with moderate bending stiffness display enhanced transport in dense disordered media. To test whether this regime is associated with a characteristic geometric length scale of the obstacle array, we compare active polymers in disordered arrays with different obstacle radii. 

\begin{figure}
    \centering
    \begin{overpic}[width=0.485\linewidth]{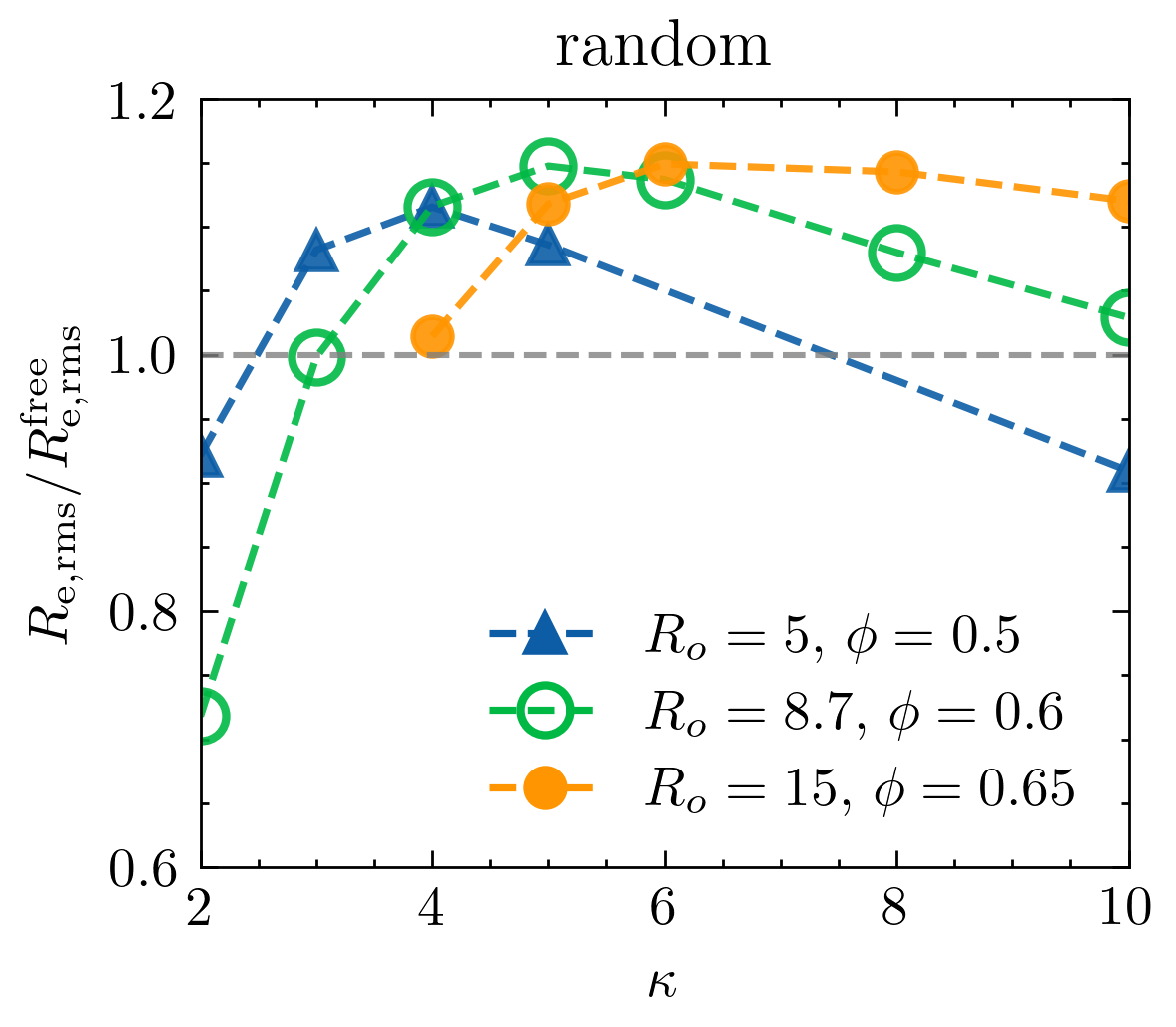}
        \put(2,86){\textbf{(a)}}
    \end{overpic}
    \hfill
    \begin{overpic}[width=0.495\linewidth]{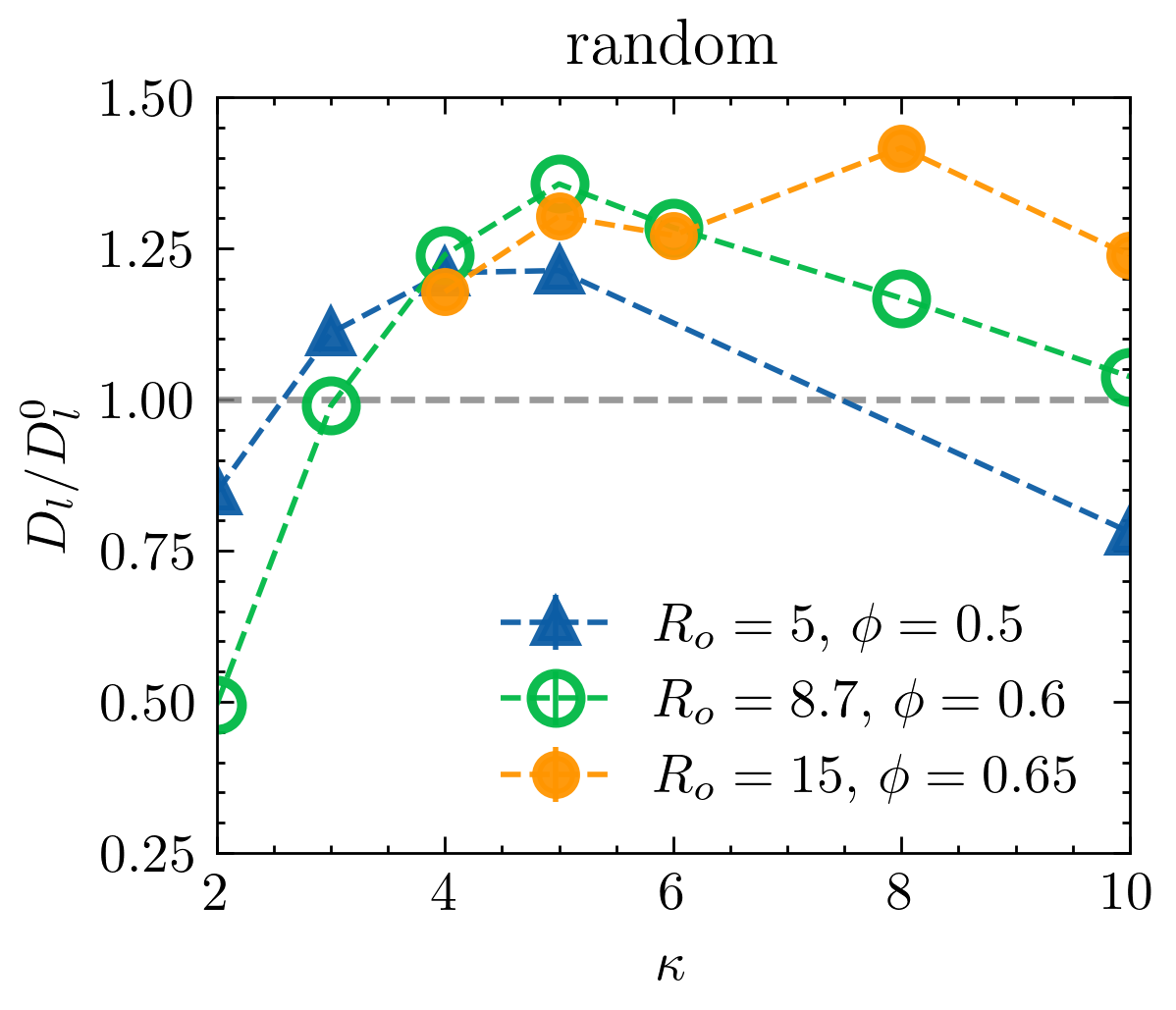}
        \put(2,84){\textbf{(b)}}
    \end{overpic}
    \caption{Properties of active polymers in disordered arrays with different obstacle radii (\(R_\text{o}=5\sigma,8.7\sigma,15\sigma\)) as functions of bending stiffness \(\kappa\). (a) Normalised root-mean-square end-to-end distance, \(R_{\mathrm{e,rms}}/R_{\mathrm{e,rms}}^{0}\). The horizontal dashed line marks the corresponding free-space value \(R_{\mathrm{e,rms}}^{0}\).
    (b) Normalised long-time diffusion coefficient \(D_l/D^\text{Free}_\text{Passive}\), where \(D^\text{Free}_\text{Passive}=0.01\) is the diffusion coefficient of passive polymer in free space.}
    \label{appfig:k-r-relation}
\end{figure}

Figure~\ref{appfig:k-r-relation} (a) shows the normalised root-mean-square end-to-end distance,
\(R_{\mathrm{e,rms}}/R_{\mathrm{e,rms}}^{0}\), as a function of \(\kappa\), where \(R_{\mathrm{e,rms}}^{0}\) is the corresponding free-space value.

For all three obstacle sizes, the strongest confinement-induced extension in random media occurs at an intermediate bending stiffness, rather than in the highly flexible or semiflexible limits. Highly flexible polymers remain compact because they can collapse into pore spaces and undergo trapping-and-hopping dynamics. Conversely, very stiff polymers resist bending and therefore cannot easily adapt to the tortuous pathways of the random medium. Between these limits, moderately flexible polymers can bend around obstacles while maintaining a finite end-to-end extension, producing a maximum in \(R_{\mathrm{e,rms}}/R_{\mathrm{e,rms}}^{0}\). The position of this maximum depends on obstacle size. Within the sampled values of \(\kappa\), the maximum occurs near \(\kappa\simeq 4\) for \(R_\text{o}=5\), near \(\kappa\simeq 5\) for \(R_\text{o}=8.7\), and \(\kappa\simeq 6\) for \(R_\text{o}=15\). This shift indicates that the moderate-flexibility regime is not controlled by bending stiffness alone, but by a geometric matching between polymer flexibility and the characteristic length scale of the pore space. 

The corresponding diffusion coefficient data in Fig.~\ref{appfig:k-r-relation}(b) show a broadly similar trend. with the transport enhancement is likewise concentrated at intermediate \(\kappa\). The diffusion coefficient has larger statistical uncertainty than \(R_{\text{e,rms}}\), so small features should not be overinterpreted without further evidence. Nevertheless, the enhancement of \(D_l\) remains concentrated at intermediate \(\kappa\), and the relative peak becomes more pronounced as the obstacle radius increases.

Together, the conformational and transport trends support the interpretation that enhanced diffusion in this regime arises when the polymer can exploit confinement-induced shape fluctuations while avoiding both compact trapping and excessive stiffness against tortuous pathways.

\bibliographystyle{ieeetr}
\bibliography{references}

@article{wurthner2026geometry,
  title = {Geometry of disordered porous environments regulates cell migration},
  author = {W\"urthner, Laeschkir and Graw, Frederik},
  journal = {Phys. Rev. E},
  volume = {113},
  issue = {1},
  pages = {014407},
  numpages = {15},
  year = {2026},
  month = {Jan},
  publisher = {American Physical Society},
  doi = {10.1103/hvtd-qwp1},
  url = {https://link.aps.org/doi/10.1103/hvtd-qwp1}
}

@article{bhattacharjee2019bacterial,
  title={Bacterial hopping and trapping in porous media},
  author={Bhattacharjee, Tapomoy and Datta, Sujit S},
  journal={Nature communications},
  volume={10},
  number={1},
  pages={2075},
  year={2019},
  publisher={Nature Publishing Group UK London}
}

@article{kurzthaler2021geometric,
  title={A geometric criterion for the optimal spreading of active polymers in porous media},
  author={Kurzthaler, Christina and Mandal, Suvendu and Bhattacharjee, Tapomoy and L{\"o}wen, Hartmut and Datta, Sujit S and Stone, Howard A},
  journal={Nature communications},
  volume={12},
  number={1},
  pages={7088},
  year={2021},
  publisher={Nature Publishing Group UK London}
}

@article{wu2021wechanisms,
  author  = {Wu, Haichao and Greydanus, Benjamin and Schwartz, Daniel K.},
  title   = {Mechanisms of transport enhancement for self-propelled nanoswimmers in a porous matrix},
  journal = {Proceedings of the National Academy of Sciences},
  year    = {2021},
  volume  = {118},
  number  = {27},
  pages   = {e2101807118},
  doi     = {10.1073/pnas.2101807118}
}

@article{licata2016diffusion,
  author  = {Licata, Nicholas A. and Mohari, Bijan and Fuqua, Clay and Setayeshgar, Sima},
  title   = {Diffusion of Bacterial Cells in Porous Media},
  journal = {Biophysical Journal},
  year    = {2016},
  volume  = {110},
  number  = {1},
  pages   = {247--257},
  doi     = {10.1016/j.bpj.2015.09.035}
}

@article{mattingly2025coarse,
  author  = {Mattingly, Henry H.},
  title   = {Coarse-graining bacterial diffusion in disordered media to surface states},
  journal = {Proceedings of the National Academy of Sciences},
  year    = {2025},
  volume  = {122},
  number  = {12},
  pages   = {e2407313122},
  doi     = {10.1073/pnas.2407313122}
}

@article{bertrand2018optimized,
  author  = {Bertrand, Thibault and Zhao, Yongfeng and B{\'e}nichou, Olivier and Tailleur, Julien and Voituriez, Rapha{\"e}l},
  title   = {Optimized Diffusion of Run-and-Tumble Particles in Crowded Environments},
  journal = {Physical Review Letters},
  year    = {2018},
  volume  = {120},
  number  = {19},
  pages   = {198103},
  doi     = {10.1103/PhysRevLett.120.198103}
}

@article{frangipane2019invariance,
  author  = {Frangipane, Giacomo and Vizsnyiczai, Gaszton and Maggi, Claudio and Savo, Romolo and Sciortino, Andrea and Gigan, Sylvain and Di Leonardo, Roberto},
  title   = {Invariance properties of bacterial random walks in complex structures},
  journal = {Nature Communications},
  year    = {2019},
  volume  = {10},
  number  = {1},
  pages   = {2442},
  doi     = {10.1038/s41467-019-10455-y}
}

@article{pietrangeli2025universal,
  author  = {Pietrangeli, T. and Foffi, R. and Stocker, R. and Ybert, C. and Cottin-Bizonne, C. and Detcheverry, F.},
  title   = {Universal Law for the Dispersal of Motile Microorganisms in Porous Media},
  journal = {Physical Review Letters},
  year    = {2025},
  volume  = {134},
  number  = {18},
  pages   = {188303},
  doi     = {10.1103/PhysRevLett.134.188303}
}

@article{volpe2017topography,
  author  = {Volpe, Giorgio and Volpe, Giovanni},
  title   = {The topography of the environment alters the optimal search strategy for active particles},
  journal = {Proceedings of the National Academy of Sciences},
  year    = {2017},
  volume  = {114},
  number  = {43},
  pages   = {11350--11355},
  doi     = {10.1073/pnas.1711371114}
}

@article{sreepadmanabh2025physical,
  author  = {Sreepadmanabh, Mahadev and Dey, Saheli and Kundu, Sayan and Arun, Ashitha B. and Koushika, Sandhya P. and Thutupalli, Shashi and Hewitt, Duncan and Bhattacharjee, Tapomoy},
  title   = {Physical Confinement Regulates Transitions in Nematode Motility},
  journal = {PRX Life},
  year    = {2025},
  volume  = {3},
  number  = {4},
  pages   = {043014},
  doi     = {10.1103/sdhy-5g9n}
}

@article{zeitz2017active,
  title={Active Brownian particles moving in a random Lorentz gas},
  author={Zeitz, Maria and Wolff, Katrin and Stark, Holger},
  journal={The European Physical Journal E},
  volume={40},
  number={2},
  pages={23},
  year={2017},
  publisher={Springer}
}

@article{bechinger2016active,
  title={Active particles in complex and crowded environments},
  author={Bechinger, Clemens and Di Leonardo, Roberto and L{\"o}wen, Hartmut and Reichhardt, Charles and Volpe, Giorgio and Volpe, Giovanni},
  journal={Reviews of modern physics},
  volume={88},
  number={4},
  pages={045006},
  year={2016},
  publisher={APS}
}

@article{winkler2020physics,
  title={The physics of active polymers and filaments},
  author={Winkler, Roland G and Gompper, Gerhard},
  journal={The journal of chemical physics},
  volume={153},
  number={4},
  year={2020},
  publisher={AIP Publishing}
}

@article{wen2011polymer,
  title={Polymer physics of the cytoskeleton},
  author={Wen, Qi and Janmey, Paul A},
  journal={Current Opinion in Solid State and Materials Science},
  volume={15},
  number={5},
  pages={177--182},
  year={2011},
  publisher={Elsevier}
}

@article{teske2006genera,
  title={The genera beggiatoa and thioploca},
  author={Teske, Andreas and Nelson, Douglas C},
  journal={Prokaryotes},
  volume={6},
  pages={784--810},
  year={2006},
  publisher={Springer New York}
}

@article{kuei2017strings,
  title={From strings to coils: Rotational dynamics of DNA-linked colloidal chains},
  author={Kuei, Steve and Garza, Burke and Biswal, Sibani Lisa},
  journal={Physical Review Fluids},
  volume={2},
  number={10},
  pages={104102},
  year={2017},
  publisher={APS}
}

@inproceedings{nakagaki2016chainform,
  title={ChainFORM: a linear integrated modular hardware system for shape changing interfaces},
  author={Nakagaki, Ken and Dementyev, Artem and Follmer, Sean and Paradiso, Joseph A and Ishii, Hiroshi},
  booktitle={Proceedings of the 29th Annual Symposium on User Interface Software and Technology},
  pages={87--96},
  year={2016}
}

@article{martinez2023active,
  title={Active transport in complex environments},
  author={Mart{\'\i}nez-Calvo, Alejandro and Trenado-Yuste, Carolina and Datta, Sujit S},
  year={2023}
}

@article{kudrolli2019burrowing,
  title={Burrowing dynamics of aquatic worms in soft sediments},
  author={Kudrolli, Arshad and Ramirez, Bernny},
  journal={Proceedings of the National Academy of Sciences},
  volume={116},
  number={51},
  pages={25569--25574},
  year={2019},
  publisher={National Acad Sciences}
}

@article{cabrales2017multivalent,
  title={Multivalent cross-linking of actin filaments and microtubules through the microtubule-associated protein Tau},
  author={Cabrales Fontela, Yunior and Kadavath, Harindranath and Biernat, Jacek and Riedel, Dietmar and Mandelkow, Eckhard and Zweckstetter, Markus},
  journal={Nature communications},
  volume={8},
  number={1},
  pages={1981},
  year={2017},
  publisher={Nature Publishing Group UK London}
}

@article{fazelzadeh2023active,
  title={Active motion of tangentially driven polymers in periodic array of obstacles},
  author={Fazelzadeh, Mohammad and Di, Qingyi and Irani, Ehsan and Mokhtari, Zahra and Jabbari-Farouji, Sara},
  journal={The Journal of Chemical Physics},
  volume={159},
  number={22},
  year={2023},
  publisher={AIP Publishing}
}

@article{theeyancheri2023active,
  title={Active dynamics of linear chains and rings in porous media},
  author={Theeyancheri, Ligesh and Chaki, Subhasish and Bhattacharjee, Tapomoy and Chakrabarti, Rajarshi},
  journal={The Journal of Chemical Physics},
  volume={159},
  number={1},
  year={2023},
  publisher={AIP Publishing}
}

@article{philipps2022tangentially,
  title={Tangentially driven active polar linear polymers—An analytical study},
  author={Philipps, Christian A and Gompper, Gerhard and Winkler, Roland G},
  journal={The Journal of Chemical Physics},
  volume={157},
  number={19},
  year={2022},
  publisher={AIP Publishing}
}

@article{mokhtari2019dynamics,
  title={Dynamics of active filaments in porous media},
  author={Mokhtari, Zahra and Zippelius, Annette},
  journal={Physical review letters},
  volume={123},
  number={2},
  pages={028001},
  year={2019},
  publisher={APS}
}

@article{wu2022facilitated,
    author = {Song Wu and Jia-Xiang Li and Qun-Li Lei},
    title = {Facilitated dynamics of an active polymer in 2D crowded environments with obstacles},
    journal = {Soft Matter},
    year = {2022},
    volume = {18},
    number = {48},
    pages = {9263-9272}
}

@article{juarez2010motility,
  title={Motility of small nematodes in wet granular media},
  author={Juarez, Gabriel and Lu, Kevin and Sznitman, Josue and Arratia, Paulo E},
  journal={Europhysics Letters},
  volume={92},
  number={4},
  pages={44002},
  year={2010},
  publisher={IOP Publishing}
}

@article{van2022role,
  title={The role of disorder in the motion of chiral active particles in the presence of obstacles},
  author={Van Roon, Danne M and Volpe, Giorgio and da Gama, Margarida M Telo and Ara{\'u}jo, Nuno AM},
  journal={Soft Matter},
  volume={18},
  number={36},
  pages={6899--6906},
  year={2022},
  publisher={Royal Society of Chemistry}
}

@article{fazelzadeh2023effects,
  title={Effects of inertia on conformation and dynamics of tangentially driven active filaments},
  author={Fazelzadeh, Mohammad and Irani, Ehsan and Mokhtari, Zahra and Jabbari-Farouji, Sara},
  journal={Physical Review E},
  volume={108},
  number={2},
  pages={024606},
  year={2023},
  publisher={APS}
}

@article{alonso2019transport,
  title={Transport and dispersion of active particles in periodic porous media},
  author={Alonso-Matilla, Roberto and Chakrabarti, Brato and Saintillan, David},
  journal={Physical Review Fluids},
  volume={4},
  number={4},
  pages={043101},
  year={2019},
  publisher={APS}
}

@article{isele2015self,
  title={Self-propelled worm-like filaments: spontaneous spiral formation, structure, and dynamics},
  author={Isele-Holder, Rolf E and Elgeti, Jens and Gompper, Gerhard},
  journal={Soft matter},
  volume={11},
  number={36},
  pages={7181--7190},
  year={2015},
  publisher={Royal Society of Chemistry}
}

@article{heddergott2012trypanosome,
  title={Trypanosome motion represents an adaptation to the crowded environment of the vertebrate bloodstream},
  author={Heddergott, Niko and Kr{\"u}ger, Timothy and Babu, Sujin B and Wei, Ai and Stellamanns, Erik and Uppaluri, Sravanti and Pfohl, Thomas and Stark, Holger and Engstler, Markus},
  journal={PLoS pathogens},
  volume={8},
  number={11},
  pages={e1003023},
  year={2012},
  publisher={Public Library of Science San Francisco, USA}
}

@article{majmudar2012experiments,
  title={Experiments and theory of undulatory locomotion in a simple structured medium},
  author={Majmudar, Trushant and Keaveny, Eric E and Zhang, Jun and Shelley, Michael J},
  journal={Journal of the Royal Society Interface},
  volume={9},
  number={73},
  pages={1809--1823},
  year={2012},
  publisher={The Royal Society}
}

@article{karan2024inertia,
  title={Inertia and Activity: Spiral transitions in semi-flexible, self-avoiding polymers},
  author={Karan, Chitrak and Chaudhuri, Abhishek and Chaudhuri, Debasish},
  journal={arXiv preprint arXiv:2404.15748},
  year={2024}
}

@article{peterson2020statistical,
  title={Statistical properties of a tangentially driven active filament},
  author={Peterson, Matthew SE and Hagan, Michael F and Baskaran, Aparna},
  journal={Journal of Statistical Mechanics: Theory and Experiment},
  volume={2020},
  number={1},
  pages={013216},
  year={2020},
  publisher={IOP Publishing}
}

@article{phillips2011pseudo,
  title={Pseudo-random number generation for Brownian Dynamics and Dissipative Particle Dynamics simulations on GPU devices},
  author={Phillips, Carolyn L and Anderson, Joshua A and Glotzer, Sharon C},
  journal={Journal of Computational Physics},
  volume={230},
  number={19},
  pages={7191--7201},
  year={2011},
  publisher={Elsevier}
}

@article{gostick2019porespy,
  title={PoreSpy: A python toolkit for quantitative analysis of porous media images},
  author={Gostick, Jeff T and Khan, Zohaib A and Tranter, Thomas G and Kok, Matthew DR and Agnaou, Mehrez and Sadeghi, Mohammadamin and Jervis, Rhodri},
  journal={Journal of Open Source Software},
  volume={4},
  number={37},
  pages={1296},
  year={2019}
}

@article{lu1993chord,
  title={Chord-length and free-path distribution functions for many-body systems},
  author={Lu, Binglin and Torquato, S},
  journal={The Journal of chemical physics},
  volume={98},
  number={8},
  pages={6472--6482},
  year={1993},
  publisher={American Institute of Physics}
}

@article{muthinja2017microstructured,
  title={Microstructured blood vessel surrogates reveal structural tropism of motile malaria parasites},
  author={Muthinja, Mendi J and Ripp, Johanna and Hellmann, Janina K and Haraszti, Tamas and Dahan, Noa and Lemgruber, Leandro and Battista, Anna and Sch{\"u}tz, Lucas and Fackler, Oliver T and Schwarz, Ulrich S and others},
  journal={Advanced Healthcare Materials},
  volume={6},
  number={6},
  pages={1601178},
  year={2017},
  publisher={Wiley Online Library}
}

@article{sinaasappel2025locomotion,
  title={Locomotion of active polymerlike worms in porous media},
  author={Sinaasappel, Rosa and Fazelzadeh, Mohammad and Hooijschuur, Twan and Di, Qingyi and Jabbari-Farouji, Sara and Deblais, Antoine},
  journal={Physical Review Letters},
  volume={134},
  number={12},
  pages={128303},
  year={2025},
  publisher={APS}
}

@article{dehkharghani2023self,
  title={Self-transport of swimming bacteria is impaired by porous microstructure},
  author={Dehkharghani, Amin and Waisbord, Nicolas and Guasto, Jeffrey S},
  journal={Communications Physics},
  volume={6},
  number={1},
  pages={18},
  year={2023},
  publisher={Nature Publishing Group UK London}
}

@article{ghanbarian2013tortuosity,
  title={Tortuosity in porous media: a critical review},
  author={Ghanbarian, Behzad and Hunt, Allen G and Ewing, Robert P and Sahimi, Muhammad},
  journal={Soil science society of America journal},
  volume={77},
  number={5},
  pages={1461--1477},
  year={2013},
  publisher={Wiley Online Library}
}

@article{wei2025autonomous,
  title={Autonomous life-like behavior emerging in active and flexible microstructures},
  author={Wei, Mengshi and Kraft, Daniela J},
  journal={arXiv preprint arXiv:2506.15198},
  year={2025}
}

@article{fukushima2016gliding,
  title={Gliding motility driven by individual cell-surface movements in a multicellular filamentous bacterium Chloroflexus aggregans},
  author={Fukushima, Shun-ichi and Morohoshi, Sho and Hanada, Satoshi and Matsuura, Katsumi and Haruta, Shin},
  journal={FEMS Microbiology Letters},
  volume={363},
  number={8},
  pages={fnw056},
  year={2016},
  publisher={Oxford University Press}
}

@article{yan2023conformation,
  title={Conformation and dynamics of an active filament in crowded media},
  author={Yan, Ran and Tan, Fei and Wang, Jingli and Zhao, Nanrong},
  journal={The Journal of Chemical Physics},
  volume={158},
  number={11},
  year={2023},
  publisher={AIP Publishing}
}

@article{hiratsuka2001controlling,
  title={Controlling the direction of kinesin-driven microtubule movements along microlithographic tracks},
  author={Hiratsuka, Yuichi and Tada, Tetsuya and Oiwa, Kazuhiro and Kanayama, Toshihiko and Uyeda, Taro QP},
  journal={Biophysical Journal},
  volume={81},
  number={3},
  pages={1555--1561},
  year={2001},
  publisher={Elsevier}
}

@article{zhang2026bacterial,
  title={Bacterial motility patterns vary smoothly with spatial confinement and disorder},
  author={Zhang, Haibei and Wetherington, Miles T and Ko, Hungtang and FitzGerald, Cody E and Luzzatto, Leone V and Kov{\'a}cs, Istv{\'a}n A and Munro, Edwin M and Nirody, Jasmine A},
  journal={PRX Life},
  volume={4},
  number={1},
  pages={013002},
  year={2026},
  publisher={APS}
}

@article{brown2016swimming,
  title={Swimming in a crystal},
  author={Brown, Aidan T and Vladescu, Ioana D and Dawson, Angela and Vissers, Teun and Schwarz-Linek, Jana and Lintuvuori, Juho S and Poon, Wilson CK},
  journal={Soft matter},
  volume={12},
  number={1},
  pages={131--140},
  year={2016},
  publisher={Royal Society of Chemistry}
}

@article{makarchuk2019enhanced,
  title={Enhanced propagation of motile bacteria on surfaces due to forward scattering},
  author={Makarchuk, Stanislaw and Braz, Vasco C and Ara{\'u}jo, Nuno AM and Ciric, Lena and Volpe, Giorgio},
  journal={Nature Communications},
  volume={10},
  number={1},
  pages={4110},
  year={2019},
  publisher={Nature Publishing Group UK London}
}

\end{document}